\documentclass[journal=jctcce, manuscript=article]{achemso}

\usepackage{amsmath}
\usepackage{mathtools}
\usepackage{braket}
\usepackage{multirow}
\usepackage{graphicx}
\usepackage{times}
\usepackage{xcolor}
\usepackage{algorithm}
\usepackage{algpseudocode} 
\usepackage[T1]{fontenc}
\usepackage{booktabs}      
\usepackage{multirow}      
\usepackage{threeparttable} 

\algdef{SE}[INDENT]{Indent}{EndIndent}[1]{#1}{}
\algdef{SE}[BLOCK]{Block}{EndBlock}[1]{\textbf{In} CUDA block with #1\textbf{:}}{\textbf{end block}}
\algdef{SE}[PROC]{Proc}{EndProc}[1]{\textbf{In} the first MPI process for the #1\textbf{:}}{}

\algtext*{EndIf}
\algtext*{EndFor}
\algtext*{EndWhile}
\algtext*{EndIndent}
\algtext*{EndBlock}
\algtext*{EndProc}

\newcommand{\BlockComments}[1]{\State \textcolor{gray}{‘‘‘ #1 '''}}

\makeatother
\makeatletter

\usepackage[colorlinks=true, linkcolor=blue, citecolor=blue]{hyperref}

\author{Qiujiang Liang}
\affiliation[]
{Department of Chemistry, The University of Hong Kong, 
Hong Kong 999077, P.R. China}
\alsoaffiliation[]
{Hong Kong Quantum AI Lab Limited, Hong Kong 999077, P.R. China}
\author{Jun Yang}
\affiliation[]
{Department of Chemistry, The University of Hong Kong, 
Hong Kong 999077, P.R. China}
\alsoaffiliation[]
{Hong Kong Quantum AI Lab Limited, Hong Kong 999077, P.R. China}
\alsoaffiliation[]
{HKU-CAS Joint Laboratory on New Materials, The University of Hong Kong, Hong Kong 999077, P.R. China}
\email{juny@hku.hk}

\title[]{Multi-GPU MBE(3)-OSV-MP2 for Performant Large-Scale \textit{ab initio} Calculations}

\graphicspath{{.}}

\begin{document}

\begin{tocentry}
\centering
\includegraphics[width=1.0\textwidth]{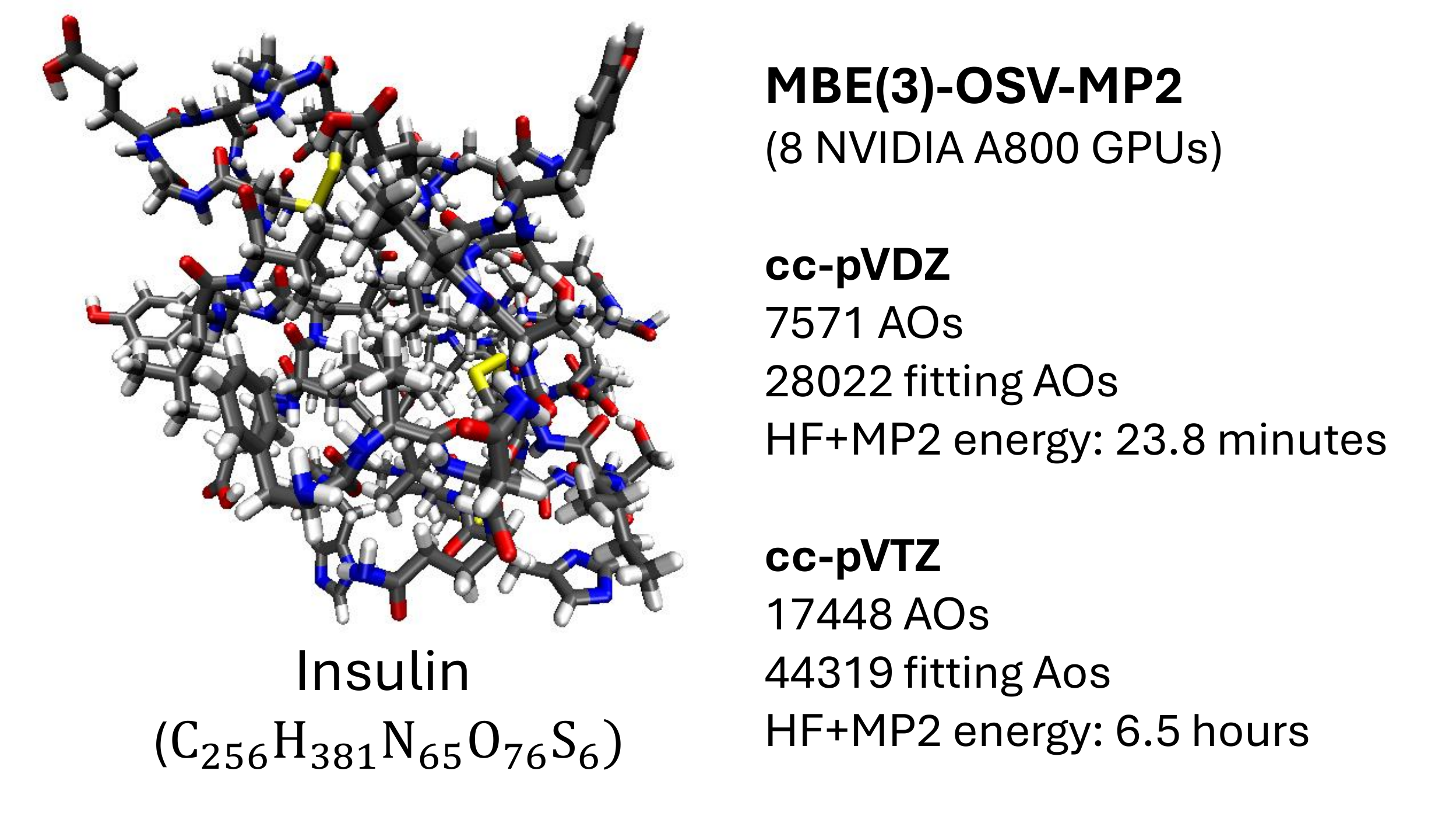}
\end{tocentry}

\begin{abstract}
The computational acceleration of orbital-invariant local correlation methods on graphics processing units (GPUs) has remained largely unexplored due to substantial algorithmic complexities. The runtime efficiency of GPU-implemented local correlation theories can be significantly constrained by the degree of parallelism in the orbital localization procedure, the iterative solution of the local wave function, and the adaptation of CUDA kernels to inherently local or sparse operations. Using the second-order M{\o}ller-Plesset perturbation (MP2) theory, we present a multi-GPU implementation for large-scale third-order many-body expansion orbital-specific virtual MP2 (MBE(3)-OSV-MP2) energy calculations. Accordingly, our algorithms and implementation address the GPU parallelization to maximize peak utilization and the parallelism of local MP2 computation in several aspects, including Jacobi-Pipek-Mezey localization, randomized OSV generation, direct MP2 integral regeneration, and CUDA kernel adaptation for local operations. The GPU-based MBE(3)-OSV-MP2 energy computation achieves an effective empirical $N^{1.9}$ scaling up to $(\text{Gly})_{40}$/def2-TZVP and 84\% parallel efficiency without localization up to 24 GPUs distributed on multiple nodes. The present implementation delivers 40-fold wall-time speedup of the canonical RI-MP2 and 10-fold speedup of the CPU-based MBE(3)-OSV-MP2  for (H$_2$O)$_{128}$/cc-pVDZ and (H$_2$O)$_{190}$/cc-pVDZ, respectively. A large scale computation of a 784-atom insulin peptide yields the full MBE(3)-OSV-MP2 energies in 24 minutes with cc-pVDZ (7571 basis functions) and in 6.4 hours with cc-pVTZ (17448 basis functions) on 8 NVIDIA A800 GPUs. Our work opens new possibilities for fast GPU-based local correlation calculations on real-world macromolecules.
\end{abstract}

\section{Introduction}

\label{sec:introduction}

Accurate \textit{ab initio} electronic structure prediction of macromolecules is  computationally intensive and requires tremendous computing time and resources, compared to classical computations. Significant advancements have been made in the past decade to extend the applicability and efficiency of post-Hartree-Fock calculations that provide systematically controllable accuracy.  Second-order M{\o}ller-Plesset perturbation theory (MP2) is the simplest wave function-based method, but of high importance in computational chemistry and materials, as it captures a large portion of the correlation energy as well as covalent, ionic, and van der Waals interactions more accurately than many popular density functional theory (DFT) approximations. The MP2 correlation energy is an essential component of the fifth-rung double-hybrid functionals\cite{zhang2011doubly, goerigk2014double}, which helps make one of the most robust and accurate DFT approaches. MP2 presently offers  the most practical wave function method to optimize geometries of large main group compounds\cite{schutz2004analytical, lochan2007quartic, pinski2018communication,pinski2019analytical,zhou2019complete,liang2021third}. It is now feasible to apply MP2 to describe dynamic electron correlations in rather complex systems \cite{mochizuki2008large,doser2009linear,nagy2016integral,kjaergaard2017massively,snowdon2024efficient}, up to benchmark demonstrations for biological structure \cite{barca2021enabling} and liquid \cite{barca2022scaling} of  tens to hundreds of thousands of atoms. These developments have been largely driven by implementing novel MP2 methods that reduce the steep $\mathcal{O}(N^5)$ scaling ($N$: molecular size) and that enable massive parallelization on modern computing platforms, including central processing units (CPUs) and graphics processing units (GPUs). 

There have been many MP2 reformulations aiming to overcome the canonical $\mathcal{O}(N^5)$ barrier. The Laplace-transformed MP2 computes the canonical correlation energy on $\mathcal{O}(N^4)$  by contracting molecular orbital (MO) indices to Laplace integration variables implemented for both molecules \cite{haser1992laplace} and solids \cite{schafer2017quartic}, and the relevant linear scaling models have also been developed \cite{ayala1999linear, kobayashi2006implementation,doser2009linear}. The Scaled-Opposite-Spin MP2 (SOS-MP2) gives similar $\mathcal{O}(N^4)$ using Laplace transformation\cite{jung2004scaled} and even lower $\mathcal{O}{(N^3)}$ using atomic orbitals\cite{maurer2014communication}. Another important domain of the scaling-reduced MP2 is a range of local correlation methods that exploit the locality of electrons \cite{pulay1983localizability,kohn1996density} in both occupied and virtual local MOs. A variety of local full-system MP2 ans\"atzs has been defined to  select a subset of compact cluster operators in different ways \cite{maslen1998non, el1998analytical, werner2003fast, doser2009linear, yang2011tensor, werner2015scalable, pavovsevic2016sparsemaps} that are also mutually related.  Moreover, a bottom-up fragmentation approach attempts to solve many smaller MP2 equations of the molecular fragments and synthesize these sub-system solutions to approximate the super-system\cite{li2004divide,kobayashi2006second,gordon2012fragmentation,nagy2016integral,liu2019energy,herbert2019fantasy,liang2021third,barca2021enabling, barca2022scaling}.  In particular, by combining the best of both streams, the present authors have developed a third-order many-body expansion of the wave function amplitudes in an orbital-specific-virtual MP2, coined MBE(3)-OSV-MP2\cite{liang2021third}, to reduce the scalings in occupied and virtual LMOs. The resulting MBE(3)-OSV-MP2 energy and analytical gradient computations achieve empirical $N^2$ and $N^{2\sim3}$ costs\cite{liang2021third}, respectively. The ability of these algorithms for reducing the $\mathcal{O}(N^5)$ complexity emerges at a manageable balance with the exactness of MP2 correlation energy, achieving about 10--100$\times$ speedups. 

Canonical and scaling-reduced MP2 models compute at various $c\mathcal{O}(N^x)$ ($x\approx1-5$; $c$: prefactor) scalings. For large molecules, multiple $\mathcal{O}(N^x)$ throughputs are invoked, resulting in significant prefactors $c$, which is another crucial source of computational bottlenecks besides the formal scaling. For enabling enormous acceleration of MP2 simulations of chemically and biologically relevant macromolecules,  the large prefactor $c$ must be mitigated using accelerators. With the advent of modern terascale high-performance computing platforms, such as the widely used CPUs and GPUs, efficient parallel computational designs are spurred and have been implemented to simultaneously execute MP2 $\mathcal{O}(N^x)$ tasks. In recent decades, GPU-based computations have gained substantial improvements and increasing popularity for quantum chemistry, as compared to CPU-only computation. The high GPU memory bandwidth and large-scale floating-point throughput enable superior linear algebra operations by launching thousands of threads simultaneously. However, this unique GPU architecture necessitates entirely new designs and optimizations for all major computational steps in Hartree-Fock and MP2: Paramount advancements in the past have been made to enhance GPU parallel efficiencies for four-center electron repulsion integrals (ERIs)\cite{yasuda2008two,ufimtsev2008quantum,miao2013acceleration,miao2015acceleration,barca2020high,alkan2024liberi,li2025introducing,palethorpe2024advanced}, three-center density fitting (DF) integrals\cite{kalinowski2017arbitrary,snowdon2024efficient,stocks2024multi,alkan2024liberi}, two-center hypercontraction integrals\cite{song2016atomic,song2017atomic}, Fock construction\cite{asadchev2012new,kussmann2013pre,barca2020high,barca2021faster,qi2023hybrid,stocks2024multi}, canonical MP2\cite{tomlinson2016new,pototschnig2021implementation,huang2025multi}, DF-MP2 \cite{vogt2008accelerating,kwack2019performance,barca2021enabling,stocks2024high,snowdon2024efficient,guo2025byteqc}, and SOS-MP2 models\cite{maurer2014communication, song2016atomic,song2017atomic}.

The GPU-enabled local MP2 correlation calculations are still largely limited to fragmentation methods \cite{bykov2017gpu,kjaergaard2017massively,pham2023porting,kazemian2024high,bykov2017gpu,feng2025efficient,stocks2024breaking}. Each fragment’s DF-MP2 can be ported to GPUs to allow massive parallelization, and the aforementioned molecular GPU-enabled DF-MP2 parallelisms are readily available for implementing fragmentation MP2. However, the local correlation approximation to the entire system wave function introduces additional complexities compared to canonical DF-MP2 on GPUs. Firstly, the localization of occupied and virtual MOs involves an overall $\mathcal{O}(N^3)$-$\mathcal{O}(N^4)$ scaling, which quickly becomes formidably intensive and a significant bottleneck, thereby elevating the overall scaling of local MP2 for large molecules. The parallel scalability of the localization function is poor when the number of parallel processes increases. Second, the Fock matrix in the LMO representation contains significant off-diagonal elements, leading to an iterative procedure where coupled residual equations must be solved for local MP2 amplitudes and energies. This poses a significant challenge to maintaining data transmission balance and memory management within the GPU architecture. Finally, the local MP2 is represented in compact orbital spaces and relies heavily on linear algebra operations for processing many small matrices. These operations are inherently latency- and memory-bound on GPUs, resulting in a considerable degradation in the performance of NVIDIA’s \texttt{cuBLAS} and \texttt{cuSolver} libraries when launching these library functions to execute such high-throughput small operations. 

This work tackles these critical issues by employing our previously developed MBE(3)-OSV-MP2 local correlation model. We adopted a fully integral-direct generator for MP2 half-transformed 3-center-2-electron (3c2e) coefficients, making use of high GPU memory bandwidth and floating point operations to avoid excessive I/O overheads. Additionally, we implemented the following localization methods on GPU thread blocks: the Pipek-Mezey (PM) localization algorithm using Jacobi pairwise orbital rotations~\cite{raffenetti1993efficient} for preparing occupied LMOs at $N^2$ scaling, and the randomized diagonalization scheme for preparing virtual LMOs (i.e., OSVs) at sub-quadratic scaling. Moreover, we developed custom CUDA kernel functions to implement linear algebra transformations as needed throughout the OSV construction, OSV integrals, and OSV-MP2 residual equations.

This paper is structured as follows. Section \ref{sec:mbe} provides an overview of the notation and theory behind the MBE(3)-OSV-MP2 method. Section \ref{sec:implementation} presents our detailed algorithms and GPU implementation for key steps in local MP2 calculations, including orbital localization, OSV generation, computation of local OSV-MP2 intermediates, and residual equations. Section \ref{sec:results} presents benchmark results and illustrative applications to demonstrate the GPU’s run-time efficiency and parallel capabilities for large molecules. Section \ref{sec:conclusions} summarizes this work.

\section{MBE(3)-OSV-MP2 Method}
\label{sec:mbe}

We present a concise overview of the MBE(3)-OSV-MP2  \cite{liang2021third}, adopting the following orbital notation convention: $i, j, k, \dots$ denote occupied molecular orbitals (MOs), either canonical or localized (LMOs), while $a, b, c, \dots$ represent canonical virtual MOs. The indices $\bar{\mu}_k, \bar{\nu}_k, \bar{\xi}_k, \dots$ indicate OSVs associated with the occupied LMO $k$. The indices $p, q, r, \dots$ and $\alpha, \beta, \dots$ refer to generic MOs and atomic orbitals (AOs), respectively. The auxiliary fitting AOs are denoted by $A, B, \dots$. The matrix trace operation is expressed using bra-ket notation $\langle \cdots \rangle$. Tensors and matrices are denoted in boldface, with their elements in italics.

For closed-shell systems, the canonical MP2 wave function amplitudes $\mathbf{T}_{ij}$ are composed of elements by
\begin{equation}
T^{ab}_{ij} = \frac{K^{ab}_{ij}}{f_{ii} + f_{jj} - f_{aa} - f_{bb}},
\label{eq:mp2_tmat}
\end{equation}
where $f_{pp}$ denotes the diagonal elements of the Fock matrix and $K_{ij}^{ab}=(ia|jb)$ the MP2 exchange integral.
However, the local OSV-MP2 correlation amplitudes must be solved iteratively in a set of nonlinear residual equations $\mathbf{R_{(ij,ij)}}$,
\begin{equation}
\begin{aligned}
\mathbf{R}_{(ij,ij)} = {} & \mathbf{K}_{(ij,ij)} + \sum_{k} \Big\{ \mathbf{S}_{(ij,ik)} \mathbf{T}_{(ik,ik)} \left[ \delta_{kj} \mathbf{F}_{(ik,ij)} - f_{kj} \mathbf{S}_{(ik,ij)} \right] \\
{} & + \left[ \delta_{ik} \mathbf{F}_{(ij,kj)} - f_{ik} \mathbf{S}_{(ij,kj)} \right] \mathbf{T}_{(kj,kj)} \mathbf{S}_{(kj,ij)} \Big\},
\end{aligned}
\label{eq:res_close}
\end{equation}
and thus minimize the Hylleraas energy functional for orbital invariance.
Above, all needed matrices in the form of $\mathbf{A}_{(ij,kl)}$ are formulated in the OSV basis $\mathbf{Q}_i$ for an LMO $i$. The OSV basis diagonalizes the semi-canonical MP2 diagonal amplitudes ($\mathbf{T}_{ii}$),
\begin{equation}
\left[ \mathbf{Q}^{\dagger}_{i} \mathbf{T}_{ii} \mathbf{Q}_{i} \right]_{\bar{\mu}_{i}\bar{\nu}_{i}} = \omega_{\bar{\mu}_{i}} \delta_{\bar{\mu}\bar{\nu}}, 
\label{eq:osv_gen}
\end{equation}
with the orthonormality condition $\mathbf{Q}^{\dagger}_{i} \mathbf{Q}_{i} = \mathbf{1}$. The eigenvalues $\omega_{\bar{\mu}_{i}}$ reflect the importance of the corresponding OSV space, allowing selection of OSVs based on a truncation parameter $l_{\text{osv}}$.
In eq~\ref{eq:res_close}, $\mathbf{A}_{(ij,kl)}$ generally represents a 4-block full OSV-based matrix including submatrices corresponding to direct excitation ($i\rightarrow \bar{\mu}_{i}$) and exchange excitation ($i\rightarrow \bar{\nu}_{j}$),
\begin{equation}
\mathbf{A}_{(ij,kl)} 
= \begin{pmatrix} \mathbf{Q}^{\dagger}_{i} \\ \mathbf{Q}^{\dagger}_{j} \end{pmatrix} \mathbf{A} \begin{pmatrix} \mathbf{Q}_{k} & \mathbf{Q}_{l} \end{pmatrix}
= \begin{bmatrix}
\mathbf{A}_{(i,k)} & \mathbf{A}_{(i,l)} \\
\mathbf{A}_{(j,k)} & \mathbf{A}_{(j,l)}
\end{bmatrix}.
\label{eq:4block_osv}
\end{equation}
Here, $\mathbf{A}$ represents canonical one-electron or two-electron molecular integrals, e.g., exchange ($\mathbf{K}$), overlap ($\mathbf{S}$), or Fock ($\mathbf{F}$).


The OSV-MP2 correlation energy is computed using $E_c= \langle\mathbf{K}_{(ij,ij)}[2\mathbf{T}_{(ij,ij)} - \mathbf{T}_{(ij,ij)}^{\dagger}]\rangle$, where the OSV-MP2 amplitudes $\mathbf{T}_{(ij,ij)}$ are approximately decomposed in the third-order many-body expansion, leading to MBE(3)-OSV-MP2 method~\cite{liang2021third}. Different from fragment-based approaches\cite{kitaura1999pair,ziolkowski2010linear,stocks2024breaking} that apply MBE in real space, the MBE(3)-OSV-MP2 expansion is formulated in the space of localized occupied orbitals and their compact OSV domains. The MBE(3)-OSV-MP2 avoids the full couplings between three or more LMOs, enabling massive parallelization of solving all residual equations. The diagonal amplitudes $\mathbf{T}_{(ii,ii)}$ are obtained in the MBE(3) form,
\begin{equation}
\mathbf{T}_{(ii, ii)} = \mathbf{T}^i_{(ii,ii)}  + \sum_{k}\Delta\mathbf{T}^{i,k}_{(ii,ii)} 
                        + \sum_{k>l}\Delta\mathbf{T}^{i,k,l}_{(ii,ii)}, 
\label{eq:tmbediag}
\end{equation}
where $\mathbf{T}^i_{(ii,ii)}$ refers to the solution to 1-body clusters, while the 2-body $\Delta\mathbf{{T}}^{i,k}_{(ii,ii)}$ and 3-body $\Delta\mathbf{T}^{i,k,l}_{(ii,ii)}$ corrections are,
\begin{equation}
\begin{split}
\Delta\mathbf{{T}}^{i,k}_{(ii,ii)}  ={}& \mathbf{{T}}^{i,k}_{(ii,ii)}-\mathbf{T}^i_{(ii,ii)}\\
\Delta\mathbf{T}^{i,k,l}_{(ii,ii)}  ={}& \mathbf{{T}}^{i,k,l}_{(ii,ii)}
 -\Delta \mathbf{T}^{i,k}_{(ii,ii)} - \Delta \mathbf{T}^{i,l}_{(ii,ii)}
 - \mathbf{T}^i_{(ii,ii)}.
 \end{split}
\end{equation}
Similarly, the  off-diagonal pair amplitudes $\mathbf{T}_{(ij,ij)}$ ($i \neq j$) are expanded in MBE(3) as
\begin{equation}
\mathbf{T}_{(ij,ij)} = \mathbf{T}^{i,j}_{(ij,ij)}+\sum_{k}\Delta\mathbf{T}^{i,j,k}_{(ij,ij)}, 
\label{eq:tmbeoff}
\end{equation}
with the 3b correction given by
\begin{equation}
\Delta\mathbf{T}^{i,j,k}_{(ij,ij)} = \mathbf{T}^{i,j,k}_{(ij,ij)}-\mathbf{T}^{i,j}_{(ij,ij)}.
\label{eq:deltaoff}
\end{equation}
Above, an $n$-body ($n$b) cluster includes $n$ LMOs and their associated OSVs. Each 1b cluster is composed of a single LMO. 
A 2b cluster combines two 1b clusters, describing correlated LMO pairs $ij$ within the excitation path $(i,j) \to \bar{\mu}_i \cup \bar{\nu}_j$, that is, a double excitation from $ij$ LMO pairs to their joint OSV subspace $\bar{\mu}_i \cup \bar{\nu}_j$. Similarly, a 3b cluster integrates three 1b clusters, with excitations $(i,j,k) \to \bar{\mu}_i \cup \bar{\nu}_j \cup \bar{\sigma}_k$. The OSV-MP2 amplitudes $\mathbf{T}_{(ij,ij)}$ explicitly depend on 2b interactions and implicitly on 3b interactions through the second term in eq~\ref{eq:res_close}, which can be further simplified by collecting the contributions from MBE clusters without much accuracy loss. 

The MBE(3) ans{\"a}tz further exploits the intrinsic sparsity of explicit third-order contributions to automate selection of only important 2b and 3b interactions, saving substantial computing costs. The sparsity of 2b clusters arises from the shortsightedness of electron correlations, estimated by the average square norm of OSV overlaps before solving MBE(3)-OSV-MP2 equations,
\begin{equation}
s^{\text{2b}}_{ij} = \frac{\sum_{\mu\nu} \langle \bar{\mu}_i | \bar{\nu}_j \rangle^2}{\sqrt{n^{\text{osv}}_i n^{\text{osv}}_j}}.
\label{eq:s2b}
\end{equation}
 Extremely distant 2b clusters can be discarded, while weak ones are efficiently recovered via direct excitation treatment (without exchange blocks) at negligible cost, ensuring linear scaling of strong 2b clusters with system size. The important 3b clusters are determined according to 
\begin{equation}
s^{\text{3b}}_{ijk} = \frac{1}{3} \left( s^{\text{2b}}_{ij} + s^{\text{2b}}_{ik} + s^{\text{2b}}_{jk} \right).
\label{eq:s3b}
\end{equation}
Important 3b clusters are identified to give similar accuracy to the original OSV-MP2 calculation. The number of resulting 3b clusters  grows linearly. The ability of MBE(3)-OSV-MP2 in independently solving the residual equations for each cluster enhances data locality, eliminates repeated host-device data transfer of intermediates, and significantly lowers inter-process communication and synchronization overheads for updating $\mathbf{T}_{(ij,ij)}$ between parallel tasks. Overall, we see a potentially more efficient GPU-based accelerator to enable massive parallelism on low-scaling MBE(3)-OSV-MP2 than canonical methods.

\section{Implementation for Multi-GPU Computing}
\label{sec:implementation}

In data-intensive GPU-based quantum chemistry computations, inefficient non-CUDA operations, such as CPU-bound I/O (e.g., disk or host memory access) and host-device data transfers, often dominate the total runtime. For example, the evaluation of RI-MP2 correlation energy entails $\mathcal{O}(N^{5})$ data movement volume, posing significant overheads and poor scalability.\cite{barca2021enabling, snowdon2024efficient, stocks2024high, guo2025byteqc} The compact local orbital space of MBE(3)‑OSV‑MP2 greatly facilitates an optimization of the algorithmic workflow, which further reduces the data transfer cost of MBE(3)‑OSV‑MP2 calculation to $\mathcal{O}(N^{2})$, as detailed below.

\begin{figure}[ht]
\includegraphics[width=\textwidth]{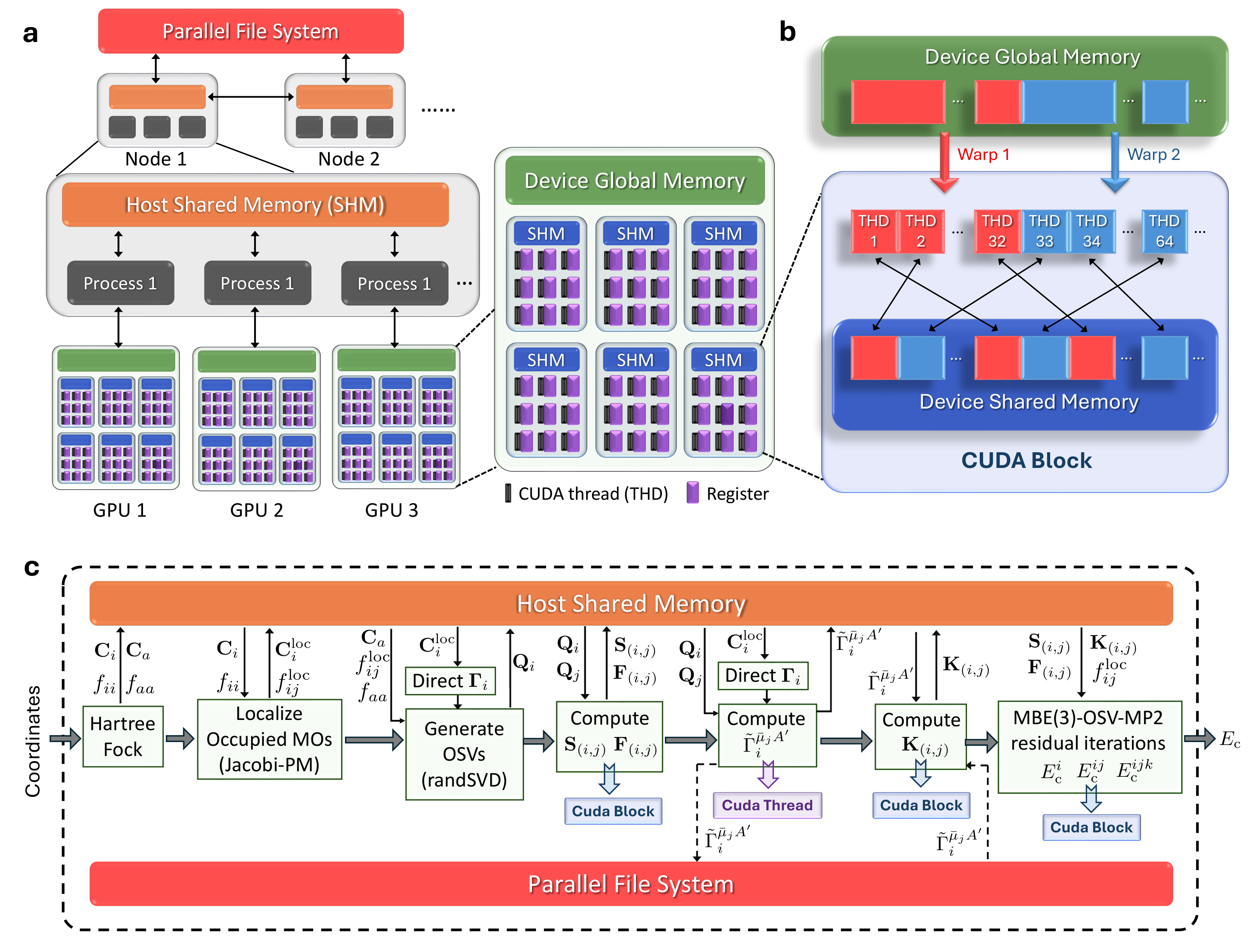}
\centering
\caption{\textbf{a}. Multi-GPU parallel architecture. \textbf{b}. Coalesced GPU memory access. \textbf{c}. GPU parallel scheme for MBE(3)-OSV-MP2.}
\label{fig:arch}
\end{figure}

Atomic orbital integrals and Fock operator are constructed using CUDA kernels in \texttt{GPU4PySCF} package\cite{li2025introducing, wu2025enhancing}. In general, NVIDIA's \texttt{cuBLAS} and \texttt{cuSolver} libraries provide highly optimized linear algebra routines customized to high-performance operations on large matrices, such as AO-to-MO integral transformations and canonical MP2 calculations. However, for low-rank matrices, such as OSV-based tensors, these libraries do not necessarily outperform custom CUDA kernels. In particular, when handling small matrices, the overhead of repeatedly invoking \texttt{cuBLAS} or \texttt{cuSolver} routines can markedly exceed the actual cost of matrix computations. To overcome these limitations, in contrast to many GPU-accelerated canonical MP2 methods that rely on \texttt{cuBLAS}, we developed CUDA kernels by mapping LMO pairs or MBE clusters to CUDA thread blocks, explicitly made for MBE(3)-OSV-MP2, to enable efficient parallelization on GPUs and eliminate the excessive overhead of launching kernel functions. 

The access to GPU global memory is typically limited by low bandwidth and high latency compared to registers and device shared memory. Coalesced access to global memory maximizes effective use of the full bandwidth by allowing warp threads to fetch contiguous addresses in a single transaction, while scattered accesses cause fragmented transactions and latency. As illustrated in Figure \ref{fig:arch}\textbf{b}, we implemented coalesced offloading mechanisms for OSV-based submatrices and their efficient reuse via the device shared memory accessible to all threads within the block, substantially enhancing kernel performance while reducing the global memory traffic. 

The CPU parallelism is implemented via the Message Passing Interface (MPI), which orchestrates the concurrent execution of multiple GPUs by coordinating tasks distributed across CPU nodes. For better intra-node data communication, the MPI-3 shared memory is utilized to reduce the local memory duplication on all processes and achieve near-zero data exchange latency within the same node. Inter-node data communication is enhanced through passive remote memory access (RMA), which  effectively reduces memory copies and synchronization latency, as compared to conventional point-to-point data transmission.  


Figure \ref{fig:arch}\textbf{c} illustrates the MBE(3)-OSV-MP2 computational steps and their parallelization implementation. The GPU calculation begins with Hartree-Fock to determine canonical MO coefficients and energies,  followed by the localization of occupied MOs. 
The density fitting 3c2e coefficient tensor $\mathbf{\Gamma}_{i}$ (eq~\ref{eq:fitting}) is generated on-the-fly and then immediately contracted into quantities of much smaller dimension, which avoids huge storage and data movement overheads due to $\mathbf{\Gamma}_i$ size. 
With $\mathbf{\Gamma}_{i}$ generator, the randomized SVD is applied to generate and select OSV basis vectors $\mathbf{Q}_{i}$ according to a single parameter $l_\text{osv}=10^{-4}$. Subsequently, the OSV-based overlap $\mathbf{S}_{(i, j)}$ and Fock $\mathbf{F}_{(i, j)}$ matrices are computed for $(i,j)$ orbital pairs, alongside the calculation of 2b selection metric $s^{\text{2b}}_{ij}$ that is tuned by the same parameter $l_\text{osv}$. Following this, we compute the OSV-based exchange integrals $\mathbf{K}_{(i,j)}$ in two steps: transforming $\mathbf{\Gamma}_{i}$ into the OSV-based intermediate $\tilde{\mathbf{\Gamma}}^{\bar{\mu}_{j}A'}_{i}$, followed by its contraction to $\mathbf{K}_{(i,j)}$. The preconditioning step is then performed for amplitude updates. Finally, residual equations for MBE clusters are iteratively solved to converge the MP2 correlation energy. Due to compact OSV orbitals and tremendous pair screening, most OSV-based data arrays, including $\mathbf{Q}_{i}$, $\mathbf{S}_{(i, j)}$, $\mathbf{F}_{(i, j)}$, $\mathbf{K}_{(i, j)}$ and $\mathbf{T}_{(i, j)}$, are lightweight and thus stored in the host shared memory, which consume, for instance, only a few dozen GB for protein hormone insulin (C$_{256}$H$_{381}$N$_{65}$O$_{76}$S$_6$) with cc-pVTZ basis. The $\tilde{\mathbf{\Gamma}}^{\bar{\mu}_{j}A'}_{i}$ tensor forms a major storage bottleneck, e.g.,  $\sim$197 GB for insulin/cc-pVTZ, and can be cached in host shared memory when possible or otherwise flushed to disk.

\subsection{Direct Density Fitting  Integrals}

Transformations of two-electron integrals are fundamental and expensive in both Hartree-Fock and MP2 calculations. The utilization of density fitting integrals has been very successful in CPU-based Fock construction, but its application to GPU calculations is disadvantaged by large intermediate 3c2e tensors, creating memory and computation bottlenecks. In contrast, the non-DF Hartree-Fock methods achieve much better GPU parallel efficiency\cite{seritan2021terachem,palethorpe2024advanced,li2025introducing,alkan2024liberi}.



However, density fitting integrals still substantially reduce MP2 computational costs. Existing GPU-accelerated RI-MP2 implementations either cached the fundamental MO-based 3c2e integrals $(ia|A)$ in memory\cite{barca2021enabling,snowdon2024efficient,stocks2024high}, or on disk\cite{guo2025byteqc,huang2025multi}. Nevertheless, generating this tensor set requires $\sim$2$\times$ its size in temporary storage, for instance, over 17 TB for insulin/cc-pVTZ, imposing severe memory/disk pressure on large systems. Moreover, repeatedly loading $(ia|A)$ and $(jb|A)$ to device memory for each pair incurs massive CPU I/O and host–device transfer overheads. To address these issues, we implemented a direct half-transform integral generator for $\mathbf{\Gamma}_i$ that processes MO batches on-the-fly (see Algorithm~\ref{alg:gamma_i}). By fully exploiting integral sparsity and permutational symmetry, GPU-based generation of the 3c2e ERIs $(\alpha\beta|A)$ achieves approximately two orders of magnitude higher throughput than the subsequent transformation. Therefore, we regenerate ERIs for each MO batch $\{i\}$, construct $\mathbf{\Gamma}_{i}$ segments
\begin{equation}
\begin{split}
    (i\alpha|B) ={}& \sum_{\beta} C_{\beta i}(\beta\alpha|B)\\
    \mathbf{\Gamma}_i : \Gamma^{\alpha A}_{i} ={}& \sum_B (i\alpha|B)V^{-1/2}_{AB},
\end{split}
\label{eq:fitting}
\end{equation}
and transform them directly to low-rank OSV-based arrays (e.g., $\Gamma^{\alpha A}_{i} \to \tilde{\mathbf{\Gamma}}^{\bar{\mu}_{j}A'}_{i}$). This strategy eliminates large intermediate storage and reduces data movement amount to $\mathcal{O}(N^{2})$. Moreover, as the localization of occupied orbitals induces sparsity in $\mathbf{\Gamma}_i$, a compact subset of the LMO-specific $\mathbf{\Gamma}_i$ is selected when the sum of squared $\Gamma^{\alpha A'}_{i}$ exceeds a threshold $l_{\text{fit}}$: 
\begin{equation}
    s^{\text{fit}}_{iA'} = \sum_{\alpha}\left(\Gamma^{\alpha A'}_{i}\right)^{2} > l_{\text{fit}}\text{, }A'\in D_{i},
\label{eq:sparse_fit_metric}
\end{equation}
where only auxiliary orbitals belonging to the domain $D_i$ close to an LMO $i$ are selected.

\begin{algorithm}
\caption{Direct $\mathbf{\Gamma}_{i}$ generator}
\label{alg:gamma_i}
\begin{algorithmic}[1]
\State Compute Coulomb kernel 2c2e integrals $V_{AB}=(A\vert B)$ $\cdots\cdots N^{2}_{\text{aux}}$
\State Device memory: Cholesky decomposition of $V_{AB}$ for $V^{-1/2}_{AB}$ $\cdots\cdots N^{3}_{\text{aux}}$
\State Transfer MOs $\mathbf{C}$ to device memory
\For{MO chunk $I$ in $\{i\}$ on the GPU}
    \State Initialize $(I\alpha|B)$ with zeros
	\For {Sparse AO pair chunk $\beta\alpha$}
        \For {Fitting orbital chunk $B\in D_I$}
		      \State Construct $(\beta\alpha|B)$ $\cdots\cdots N_{\text{aoPair}}N^{\text{chunk}}_{\text{aux}}$
		      \State Half-transformation: $(I\alpha|B) \mathrel{+}= \sum_{\beta}C_{\beta I}(\beta\alpha | B) $ $\cdots\cdots N^{\text{chunk}}_{\text{occ}}N_{\text{aoPair}}N^{\text{chunk}}_{\text{aux}}$
	\EndFor
    \EndFor
    \State Fitting: $\Gamma^{\alpha A}_{I} = \sum_{B\in D_I}(I\alpha|B)V^{-1/2}_{AB}$ $\cdots\cdots N^{\text{chunk}}_{\text{occ}}N_{\text{ao}}N^{2}_{\text{aux}}$
	\State Perform OSV generation or $\tilde{\mathbf{\Gamma}}^{\bar{\mu}_{j}A'}_{I}$ computation.
\EndFor
\end{algorithmic}
\end{algorithm}

\subsection{Occupied Orbital Localization}

We implemented the PM localization algorithm using Jacobi sweeps~\cite{raffenetti1993efficient} and L{\"o}wdin charges on GPUs, a stable and efficient pairwise-rotation method widely adopted in quantum chemistry programs\cite{werner2012molpro,neese2012orca,furche2014turbomole}. As shown in Algorithm~\ref{alg:loc_omo}, the L{\"o}wdin charge tensor $P^{\mathcal{A}}_{ij}$ is initialized for an atom $A$
based on the symmetrically orthonormalized and PM localized MOs $\mathbf{\tilde{C}} = \mathbf{S}^{1/2}\mathbf{C}\mathbf{U}$ using a custom CUDA kernel on GPUs.
The resulting local MOs maximize the PM charges on  each atom,
\begin{equation}
    L = \sum_{i\mathcal{A}} \left(P^{\mathcal{A}}_{ii}\right)^{2}.
\end{equation}
For each $(i, j)$ pair with $i < j$, the rotation angle $\theta$ can be determined with CUDA kernels using
\begin{equation}
    a_{ij} = \sum_{\mathcal{A}} \left(P^{\mathcal{A}}_{ij}\right)^{2} - \frac{1}{4}\left(P^{\mathcal{A}}_{ii}-P^{\mathcal{A}}_{jj}\right)^{2},\quad
    b_{ij} =  \sum_{\mathcal{A}} P^{\mathcal{A}}_{ij} 
    \left(P^{\mathcal{A}}_{ii}-P^{\mathcal{A}}_{jj}\right),\quad
    \theta = \frac{1}{4}\text{arctan}\left(\frac{b_{ij}}{-a_{ij}}\right).
\label{eq:loc_theta}
\end{equation}
The orbital rotation $U_{ij}$ can be computed in parallel by distributing $k$ rows on CUDA threads, which updates 
$P^{\mathcal{A}}_{ij}$ pair-wisely for each $ij$ at a time,
\begin{equation}
     \left(P^{\mathcal{A}}_{ki}\right)^{\text{new}} = P^{\mathcal{A}}_{ki} \text{cos}\theta + P^{\mathcal{A}}_{kj} \text{sin}\theta,\quad\quad
     \left(P^{\mathcal{A}}_{kj}\right)^{\text{new}} = P^{\mathcal{A}}_{kj} \text{cos}\theta - P^{\mathcal{A}}_{ki} \text{sin}\theta.
\end{equation}
The localization procedure completes when the incremental functional falls below a localization threshold
\begin{equation}
    \Delta L = \sum_{ij}\sqrt{a^{2}_{ij}+b^{2}_{ij}} (1 - \cos 4\theta) < l_{\text{loc}},
\end{equation}
where typically $l_{\text{loc}}=10^{-3}$. 

\begin{algorithm}
\caption{Occupied orbital localization with Jacobi sweeps}
\label{alg:loc_omo}
\begin{algorithmic}[1]
\State Initialize identity matrix $U_{ij}$ on GPU
\State Initialize canonical $P^{\mathcal{A}}_{ij} = \sum_{\alpha\in\mathcal{A}}\tilde{C}_{\alpha i}\tilde{C}_{\alpha j}$ on GPU $\cdots\cdots N^{2}_{\text{occ}}N_{\text{ao}}$

\While{$\Delta L > l_{\text{loc}}$}
    \State $\Delta L$ = 0.0
	\For{$(i, j)$ pair if $i< j$}
		\State Compute $a_{ij}$, $b_{ij}$ and $\theta$ using equations in eq~\ref{eq:loc_theta} $\cdots\cdots N_{\text{atom}}$
        \If{$|\text{sin}\theta| \ge 10^{-10}$}
            \State $\Delta L \mathrel{+}= \sqrt{a^{2}_{ij}+b^{2}_{ij}} (1 - \cos 4\theta)$ 
            \State Rotate columns $U_{ki}$ and $U_{kj}$ $\cdots\cdots 2N_{\text{occ}}$
            \State Rotate first MO index $P^{\mathcal{A}}_{ik}$ and $P^{\mathcal{A}}_{jk}$ $\cdots\cdots 2N_{\text{occ}}N_{\text{atom}}$
            \State Rotate second MO index $P^{\mathcal{A}}_{ki}$ and $P^{\mathcal{A}}_{kj}$ $\cdots\cdots 2N_{\text{occ}}N_{\text{atom}}$
            
        \EndIf
	\EndFor
\EndWhile
\State Transfer $U_{ij}$ to host memory
\end{algorithmic}
\end{algorithm}

\subsection{Randomized OSV generation}

The diagonalization of the semi-canonical MP2 diagonal amplitudes for generating OSVs in eq~\ref{eq:osv_gen} has a formal scaling of $\mathcal{O}(N^{4})$ over all LMOs, posing a bottleneck for computing macromolecules. We previously reduced this scaling to $\mathcal{O}(N^{2})$ via  interpolative decomposition (ID). However, the cost of ID-OSV generation increases rapidly by expanding the column subset of the $\mathbf{T}_{ii}$ matrix as needed for better accuracy.  
Alternatively, random projections effectively approximate the dominant row subspace of $\mathbf{T}_{ii}$\cite{halko2011finding}, 
\begin{equation}
\mathbf{Y} = \mathbf{G}\mathbf{T}_{ii}, 
\end{equation}
where $\mathbf{G}$ is $N_{\text{rosv}}\times N_{\text{vir}}$ random Gaussian matrix, with $N_{\text{vir}}$ denoting the number of canonical virtual orbitals and $N_{\text{rosv}}$ the number of the sampled rows of $\mathbf{Y}$. Typically, there is $N_{\text{rosv}} \ll N_{\text{vir}}$ due to the rapid decay of $\mathbf{T}_{ii}$ singular values, which ensures a rather low OSV generation cost that grows only linearly with $N_\text{vir}$. In our implementation, the $N_\text{rosv}$ rows of $\mathbf{Y}$ are incrementally sampled by fulfilling the following condition for every $r=10$ rows,
\begin{equation}
\max \left\{ \left\| \mathbf{y}^{(s+1)} \right\|, \left\| \mathbf{y}^{(s+2)} \right\|, \ldots, \left\| \mathbf{y}^{(s+r)} \right\| \right\} \le l_{\text{osv}} / (10 \sqrt{2/\pi}).
\end{equation}
Subsequently, the selected rows are orthonormalized in eq~\ref{eq:rsvd_z} to transform the amplitude in eq~\ref{eq:tsub}.
\begin{eqnarray}
\mathbf{Z} &=& \text{orth}(\mathbf{Y})
\label{eq:rsvd_z} \\
\tilde{\mathbf{T}}_{ii} &=& \mathbf{Z}\mathbf{T}_{ii}
\label{eq:tsub}
\end{eqnarray}
The transformed amplitude is further decomposed via singular value decomposition in eq~\ref{eq:tii_rsvd}, followed by a back-transformation to obtain the OSV basis vector $\mathbf{Q}_i$ in eq~\ref{eq:back_q}.
\begin{eqnarray}
\tilde{\mathbf{T}}_{ii} &=& \tilde{\mathbf{Q}}_{i}\omega_{i} \tilde{\mathbf{V}}_{i}^{\dagger},
\label{eq:tii_rsvd}\\
\mathbf{Q}_{i} &=& \mathbf{Z}^{\dagger}\tilde{\mathbf{Q}}_{i}.
\label{eq:back_q}
\end{eqnarray}

\begin{algorithm}
\caption{Randomized OSV generation}
\label{alg:rand_osv}
\begin{algorithmic}[1]
\State Transfer $f_{aa}$, $C_{\beta a}$, $C_{\beta i}$ and $V^{-1/2}_{PQ}$ to device memory

\For{chunk $I$  in $\{i\}$ on the GPU}
    \State Generate $(I\alpha|B)$ using Algorithm \ref{alg:gamma_i}
    \For{$i$ in chunk $I$ }
        \State Compute $\Gamma^{\alpha A'}_{i}=\sum_{B}(i\alpha|B)V^{-1/2}_{BA'}, A'\in D_{i}$ 
        $\cdots\cdots N_{\text{ao}}N_{\text{aux}}N_{\text{sFit}}$
        \State Compute $T^{ab}_{ii}$ from $\Gamma^{\alpha A'}_{i}$ using eq~\ref{eq:mp2_tkk}. $\cdots\cdots N_{\text{vir}}(N_{\text{vir}}+N_{\text{ao}})N_{\text{sFit}}$
        \Statex
        \BlockComments{Adaptive randomized range finder}
        \State Initialize selected rows $s = 0$
        \State Generate standard Gaussian matrices $\mathbf{G}$ and $\bar{\mathbf{G}}$
        \State Generate $\mathbf{Y} = \mathbf{G}\mathbf{T}_{ii}$ $\cdots\cdots rN^{2}_{\text{vir}}$
        \State Generate $\bar{\mathbf{Y}} = \bar{\mathbf{G}}\mathbf{T}_{ii}$
        $\cdots\cdots N^{\text{max}}_{\text{rosv}}N^{2}_{\text{vir}}$ 
        \State Pre-allocate $N^{\text{max}}_{\text{rosv}} \times N_{\text{vir}} $ array to initialize $\mathbf{Z}^{(0)} = [\;]$
        \While{$\max \left\{ \left\| \mathbf{y}^{(s+1)} \right\|, \left\| \mathbf{y}^{(s+2)} \right\|, \ldots, \left\| \mathbf{y}^{(s+r)} \right\| \right\} > l_{\text{osv}} / (10 \sqrt{2/\pi})$}
            \State $s \mathrel{+}= 1$
            \State Compute $\mathbf{z}^{(s)}$ orthonormal to $\mathbf{Z}^{(s-1)}$ from $\mathbf{y}^{(s)}$ using eq~\ref{eq:rsvd_za}
            $\cdots\cdots 2N_{\text{rosv}}N_{\text{vir}}$
            \State Appending: $\mathbf{Z}^{(s)} = 
                \begin{bmatrix}
                    \mathbf{Z}^{(s-1)} \\
                    \mathbf{z}^{(s)}
                \end{bmatrix}$
            \State Shift block $\mathbf{y}^{(t)} \gets \mathbf{y}^{(t+1)}$, $t = s, \ldots, s+r-1$
            \State Compute $\mathbf{y}^{(s+r)} = \bar{\mathbf{y}}^{(s)}[\mathbf{I} - \left(\mathbf{Z}^{(s-1)}\right)^{\dagger}\mathbf{Z}^{(s-1)}]$
            $\cdots\cdots 2N_{\text{rosv}}N_{\text{vir}}$
            \State Compute $\mathbf{Y} \mathrel{-}= \mathbf{Y}\left(\mathbf{z}^{(s)}\right)^{\dagger}\mathbf{z}^{(s)}$
            $\cdots\cdots 2rN_{\text{rosv}}N_{\text{vir}}$
        \EndWhile
        \State Compute $\tilde{\mathbf{T}}_{ii} = \mathbf{Z}\mathbf{T}_{ii}$
        $\cdots\cdots N_{\text{rosv}}N^{2}_{\text{vir}}$
        \State Low-rank SVD: $\tilde{\mathbf{T}}_{ii} = \tilde{\mathbf{Q}}_{i}\omega_{i}\tilde{\mathbf{V}}_{i}^{\dagger}$
        $\cdots\cdots N^{2}_{\text{rosv}}N_{\text{vir}}$
        \State Compute OSV matrix: $\mathbf{Q}_{i}= \mathbf{Z}^{\dagger}\tilde{\mathbf{Q}}_{i}.$
        $\cdots\cdots N_{\text{vir}N^{2}_{\text{rosv}}}$
        \State Select $\mathbf{Q}_{i}$ vectors with $\omega_{i} \ge l_{\text{osv}}$
        \State Transfer $\mathbf{Q}_{i}$ $(N_{\text{vir}}, N_{\text{osv}})$ to host shared memory
    \EndFor
\EndFor
\end{algorithmic}
\end{algorithm}

The randomized OSV (rOSV) generator is implemented as Algorithm~\ref{alg:rand_osv}. The calculation of the semi-canonical diagonal MP2 amplitudes is substantially expedited through the sparse fitting of eq~\ref{eq:sparse_fit_metric}, 
\begin{equation}
    \tilde{\Gamma}^{a A'}_{i} = \sum_{\alpha}C_{\alpha a} \Gamma^{\alpha A'}_{i}, A'\in D_{i},\quad\quad
    T^{ab}_{ii} =  \sum_{A'\in D_{i}}\frac{\tilde{\Gamma}^{a A'}_{i}\tilde{\Gamma}^{b A'}_{i}}{f_{aa}+f_{bb}-2f_{ii}}.
\label{eq:mp2_tkk}
\end{equation}
The low-rank randomized projection matrix $\mathbf{Z}$ (defined in eq~\ref{eq:rsvd_z}) is built incrementally using the adaptive randomized range finder~\cite{halko2011finding}. A row-based implementation is employed to improve memory efficiency. The range finder uses Gram-Schmidt orthogonalization to ensure the new row $\mathbf{z}^{(s)}$ is orthonormal to all prior rows $\mathbf{Z}^{(s-1)}$:
\begin{equation}
    \tilde{\mathbf{z}}^{(s)} = \mathbf{y}^{(s)}
    \left[\mathbf{I} - \left(\mathbf{Z}^{(s-1)}\right)^{\dagger}\mathbf{Z}^{(s-1)}\right],\quad\quad
    \mathbf{z}^{(s)} = \tilde{\mathbf{z}}^{(s)} / \lVert \tilde{\mathbf{z}}^{(s)} \rVert,
\label{eq:rsvd_za}
\end{equation}
where $\mathbf{y}^{(s)} = \mathbf{g}^{(s)}\mathbf{T}_{ii}$ and $\mathbf{g}^{(s)}$ is a $1 \times N_{\text{vir}}$ Gaussian random vector at the row $s$. The $r \times N_{\text{vir}}$ matrix $\mathbf{Y}$ is iteratively accumulated. To reduce computational latency from frequent small-matrix multiplications and random vector generation, the new rows $\mathbf{g}^{(s+r)}\mathbf{T}_{ii}$ are precomputed. The rOSV rank $N_{\text{rosv}}$ is oversampled for enhancing accuracy, and the resulting $\mathbf{Q}_{i}$ in eq~\ref{eq:back_q} can be further pruned by removing columns corresponding to eigenvalues $\omega_{i} < l_{\text{osv}}$. In this compact OSV representation, the $\mathbf{Q}_{i}$ vectors are small enough ($\sim$9 GB for insulin/cc-pVTZ) to be stored in host shared memory and can be efficiently fetched with near-zero latency in the following pairwise calculations.

\subsection{OSV Overlap and Fock Computation}

In our previous CPU-based implementation for computing OSV overlap $\mathbf{S}_{(i, j)}$ and Fock $\mathbf{F}_{(i, j)}$ matrices,\cite{liang2021third} $\mathbf{Q}_{i}$ and $\mathbf{Q}_{j}$ are loaded for every $(i, j)$ pair in parallel via remote memory access, which is problematic to GPUs due to an excessive data transfer volume as $(N_{\text{occ}}+1)N_{\text{occ}}N_{\text{vir}}N_{\text{osv}}$ from the host to the device. To address this, we revamp the data transfer mechanism from a pair-wise to an LMO-wise implementation, by batching LMOs out of all pairs, and the batch length is determined by the available device global memory. The number of batches is minimized by greedily selecting sorted pairs that group no more than $N_i^\text{batch}$ unique LMOs, as detailed in Algorithm \ref{alg:min_batch}. All $\mathbf{Q}_{i}$ vectors specific to LMOs belonging to each batch are transferred to device global memory. The computations of pair $\mathbf{S}_{(i, j)}$ and $\mathbf{F}_{(i, j)}$ are mapped to CUDA blocks for extensive parallelization. To enable coalesced access, $f_{aa}$ and $\mathbf{Q}_{i}$ are further loaded to device shared memory along virtual orbital slices according to the pre-allocated shared memory per CUDA block. After computing $\mathbf{S}_{(i, j)}$ and $\mathbf{F}_{(i, j)}$,  the 2b cluster criteria $s^{\text{2b}}_{ij}$ are evaluated by performing parallel reduction from threads within the block. Based on the 2b and 3b selection metrics defined in eqs~\ref{eq:s2b} and \ref{eq:s3b}, 2b clusters are classified into 3 categories for different treatments: distant ($s^{\text{2b}}_{ij} < 10^{-7}$), weak ($10^{-7} \le s^{\text{2b}}_{ij} < 10^{-2}$) and close ($s^{\text{2b}}_{ij} \ge 10^{-2}$). Unimportant 3b clusters with $s^{\text{3b}}_{ijk} < 0.2$ are discarded. Only the $\mathbf{S}_{(i, j)}$ and $\mathbf{F}_{(i, j)}$ for non-distant pairs are stored in the host shared memory.

\begin{algorithm}
\caption{Minimal Batches of LMOs for Pairs}
\label{alg:min_batch}
\begin{algorithmic}[1]
\State{\textbf{Inputs:} pairs, $N^{\text{batch}}_{i}$}
\State Extract all unique LMOs from pairs
\For{LMO $i$}
    \State Count the total number of occurrences of $i$ among all pairs ($n_{i}$)
\EndFor
\State Sort pairs in the descending order of $n_i+n_j$ for each $ij$ pair
\State Generate LMO batches among all sorted pairs: each batch contains pairs that possess at most $N^{\text{batch}}_{i}$ LMOs
\State \textbf{Outputs:} batches of LMOs and pairs 
\end{algorithmic}
\end{algorithm}

\begin{algorithm}
\caption{$\mathbf{S}_{(i, j)}$ and $\mathbf{F}_{(i, j)}$ Computation}
\label{alg:osv_sf}
\begin{algorithmic}[1]
\State Transfer $f_{aa}$ to device global memory
\State Generate LMO batches using Algorithm~\ref{alg:min_batch}
\For {batch LMOs $K$}
    \State Transfer all $\mathbf{Q}_{K}$ vectors to device global memory
    \State Map pairs to CUDA blocks belonging to batch $K$
    \Block{pair $(i, j)$}
        \State Batch virtual orbitals to fit the device shared memory
        \For{chunk $a$ in full virtual $\{a\}$}
            \State Coalesced load of $f_{aa}$, $Q^{a\bar{\mu}_{i}}_{i}$ and $Q^{a\bar{\nu}_{j}}_{j}$ to device shared memory
            \State Compute $S^{\bar{\mu}_{i}\bar{\nu}_{j}}_{ij} \mathrel{+}= \sum_{a}Q^{a\bar{\mu}}_{i}Q^{a\bar{\nu}}_{j}$
            $\cdots\cdots N^{2}_{\text{osv}}N_{\text{vir}}$
            \State Compute $F^{\bar{\mu}_{i}\bar{\nu}_{j}}_{ij} \mathrel{+}= \sum_{a}Q^{a\bar{\mu}}_{i}f_{aa}Q^{a\bar{\nu}}_{j}$
            $\cdots\cdots N^{2}_{\text{osv}}N_{\text{vir}}$
        \EndFor
        \State Parallel reduction: $s^{\text{2b}}_{ij} = \sum_{\mu\nu} \left(S^{\bar{\mu}_{i}\bar{\nu}_{j}}_{ij}\right)^2/\sqrt{n_i n_j}$
        $\cdots\cdots N^{2}_{\text{osv}}$
    \EndBlock
    \State Transfer $\mathbf{S}_{(i, j)}$, $\mathbf{F}_{(i, j)}$ and $s^{\text{2b}}_{ij}$ back to host memory
    \State Perform pair classification based on $s^{\text{2b}}_{ij}$
    \State Store $\mathbf{S}_{(i, j)}$ and $\mathbf{F}_{(i, j)}$, each sized $(N^{\text{batch}}_{\text{kPair}},N_{\text{osv}}, N_{\text{osv}})$, in host shared memory for kept pairs
\EndFor
\end{algorithmic}
\end{algorithm}

\subsection{OSV Exchange Integrals}

\begin{algorithm}
\caption{OSV Exchange Integral Evaluation}
\label{alg:osv_k}
\begin{algorithmic}[1]
\BlockComments{$\tilde{\Gamma}_i^{\bar{\mu}_j A^\prime}$ evaluation}
\State Sort process LMOs $\{i\}$ and their close $\{j\}$
\State Transfer first batch of $\mathbf{Q}_{j}$ to device global memory
\For{LMO chunk $I$ on the GPU}
    \State Generate $\mathbf{\Gamma}_{I}$ using Algorithm \ref{alg:gamma_i}
   \For{$i$ in chunk $I$}
    \State Update $\mathbf{Q}_{j}$ for $j$ close to $i$
    \State Map $\tilde{\Gamma}_i^{\bar{\mu}_j A^\prime}$ elements to CUDA threads
    \Indent{\textbf{In} each CUDA block\textbf{:}}
        \State Batch atomic orbitals by shared memory availability
        \State Fetch $\Gamma^{\alpha A'}_{i}$ and $Q^{a\bar{\mu}_{j}}_{j}$ to device shared memory for the first AO batch
        \For{chunk $\alpha$ in $\{\alpha\}$}
            \State Fetch $\Gamma^{\alpha A'}_{i}$ and $Q^{a\bar{\mu}_{j}}_{j}$ for the next AO batch
            \State Compute $\tilde{\Gamma}_i^{\bar{\mu}_j A^\prime} \mathrel{+}= \sum_{\alpha} Q^{\alpha\bar{\mu}_{j}}_{j} \Gamma^{\alpha A'}_{i}$
            $\cdots\cdots N_{\text{osv}}N_{\text{ao}}N_{\text{sFit}}$
        \EndFor
        \State Write $\tilde{\Gamma}_i^{\bar{\mu}_j A^\prime}$ to global memory
    \EndIndent
    \State Store $\tilde{\Gamma}_i^{\bar{\mu}_j A^\prime}$ ($N_{j},N_{\text{osv}},N_{\text{sFit}}$) in host shared memory or on disk 

  \EndFor  
\EndFor
\Statex

\BlockComments{$\tilde{\Gamma}_i^{\bar{\mu}_j A^\prime}$ contraction}
\State Generate LMO batches using Algorithm~\ref{alg:min_batch}
\For{batch LMOs $K$}
    \State Transfer diagonal pair $\tilde{\Gamma}_K^{\bar{\mu}_K A^\prime}$ to device global memory
    \State Make close pair batches to fit device global memory
    \For{batch close pairs $(i,j)$}
        \State Transfer $\tilde{\Gamma}_i^{\bar{\mu}_j A^\prime}$ and $\tilde{\Gamma}_j^{\bar{\mu}_i A^\prime}$ to device global memory
        \Block{close pair $(i, j)$}
            \State Batch fitting AOs to fit device shared memory
            \For{chunk $A'$ in $D_{i}\cup D_{j}$}
                \State Fetch $\tilde{\Gamma}_i^{\bar{\mu}_i A^\prime}$, $\tilde{\Gamma}_j^{\bar{\mu}_j A^\prime}$, $\tilde{\Gamma}_i^{\bar{\mu}_j A^\prime}$ and $\tilde{\Gamma}_j^{\bar{\mu}_i A^\prime}$ to device shared memory
                \State Compute $\mathbf{K}_{(ij,ij)}$ using eq~\ref{eq:osv_close_k}
                $\cdots\cdots 4N^{2}_{\text{osv}}N_{\text{sFit}}$
            \EndFor
        \EndBlock
    \EndFor
    \For{batch weak pairs $(i,j)$}
    \Block{weak pair $(i, j)$}
        \State Batch fitting AOs to fit device shared memory
        \For{chunk $A'$ in $D_{i}\cup D_{j}$}
            \State Fetch $\tilde{\Gamma}_i^{\bar{\mu}_i A^\prime}$ and $\tilde{\Gamma}_j^{\bar{\mu}_j A^\prime}$ to device shared memory
            \State Compute $\mathbf{K}_{(i,j)}$ using eq~(\ref{eq:osv_weak_k})
            $\cdots\cdots N^{2}_{\text{osv}}N_{\text{sFit}}$
        \EndFor
    \EndBlock
    \EndFor
    \State Transfer $\mathbf{K}_{(ij,ij)}$ ($N^{\text{batch}}_{\text{cPair}}, 2N_{\text{osv}}, 2N_{\text{osv}}$) and $\mathbf{K}_{(ij)}$ ($N^{\text{batch}}_{\text{wPair}}, N_{\text{osv}}, N_{\text{osv}}$) to host shared memory
\EndFor

\end{algorithmic}
\end{algorithm}

The OSV exchange integrals $\mathbf{K}_{(ij.ij)}$ are formulated in eq~\ref{eq:4block_osv}
where the AO-to-OSV transformation matrices $\mathbf{Q}_{i}$ are cached in host shared memory. Due to the compact OSV space, we can avoid loading large $\mathbf{\Gamma}_i$ and $\mathbf{\Gamma}_j$ for every LMO pair that is a common implementation for RI-MP2 GPU algorithms\cite{guo2025byteqc,katouda2016massively,snowdon2024efficient,barca2021enabling,stocks2024high,martinez2017gpu}. Here, we adopt a two-step workflow by first computing the OSV-based half-transformed integrals as
\begin{equation}
    \tilde{\Gamma}_i^{\bar{\mu}_j A^\prime} = \sum_{\alpha} Q^{\alpha\bar{\mu}_{j}}_{j} \Gamma^{\alpha A'}_{i}, A'\in D_{i}\cup D_{j}
\label{eq:osv_imua}
\end{equation}
where the sparse fitting basis $A'$ belongs to the  union of the domains $D_i$ and $D_j$. $\mathbf{\Gamma}_{i}$ in eq~\ref{eq:osv_imua} is generated on-the-fly, eliminating storage and I/O overhead. Small $\mathbf{Q}_{j}$ tensors can be reused across several LMOs to avoid excessive data transmissions. The OSV-based $\tilde{\mathbf\Gamma}_i$ is substantially reduced compared to $\mathbf{\Gamma}_i$, which therefore can be stored in memory or on disk. The second step makes rapid contraction of $\tilde{\Gamma}_i^{\bar{\mu}_j A^\prime}$ for close and weak 2b clusters in eqs~\ref{eq:osv_close_k} and \ref{eq:osv_weak_k}, respectively.
\begin{equation}
    \begin{pmatrix}
K^{\bar{\mu}_i \bar{\mu}_i}_{ij} & K^{\bar{\mu}_i \bar{\nu}_j}_{ij} \\
K^{\bar{\nu}_j \bar{\mu}_i}_{ij} & K^{\bar{\nu}_j \bar{\nu}_j}_{ij}
\end{pmatrix} = \sum_{A'}
\begin{bmatrix}
\tilde{\Gamma}_i^{\bar{\mu}_i A^\prime}  \\
\tilde{\Gamma}_i^{\bar{\nu}_j A^\prime}
\end{bmatrix}
\begin{bmatrix}
\tilde{\Gamma}_j^{\bar{\mu}_i A^\prime}  &
\tilde{\Gamma}_i^{\bar{\nu}_j A^\prime}
\end{bmatrix}, A' \in D_{i}\cup D_{j},
\label{eq:osv_close_k}
\end{equation}
\begin{equation}
K^{\bar{\mu}_i \bar{\nu}_j}_{ij}  = \sum_{A'}
\tilde{\Gamma}_i^{\bar{\mu}_i A^\prime}  
\tilde{\Gamma}_j^{\bar{\nu}_j A^\prime}, A' \in D_{i}\cup D_{j}.
\label{eq:osv_weak_k}
\end{equation}

Custom CUDA kernels are implemented for eqs~\ref{eq:osv_imua}-\ref{eq:osv_weak_k} in Algorithm~\ref{alg:osv_k}, which slices $\mathbf{\Gamma}_i$ and $\mathbf{Q}_j$ into AO batches to fit the device shared memory and bandwidth consumption. In addition, the double buffering is used to overlap processes of data fetching and computation during the construction of $\tilde{\mathbf\Gamma}_i$, with each element assigned to a CUDA thread and accumulated in registers. The $\tilde{\mathbf\Gamma}_i$ contractions in eqs~\ref{eq:osv_close_k} and \ref{eq:osv_weak_k} are mapped pair-wise to CUDA blocks, which are placed in device shared memory to reduce otherwise repeated access to global memory.

\subsection{MBE(3)-OSV-MP2 Residual Equations}

\begin{algorithm}
\caption{Residual Iterations}
\label{alg:osv_res}
\begin{algorithmic}[1]
\BlockComments{For close MBE clusters}
\For{batch MBE cluster}
    \State Get unique LMO pairs from the batch clusters
    \State Transfer $f_{ij}$, $\mathbf{S}_{(i,j)}$, $\mathbf{F}_{(i,j)}$, $\mathbf{K}_{(ij,ij)}$, $\mathbf{X}_{(ij,ij)}$, $\tilde{\mathbf{E}}_{(ij,ij)}$ and $\mathbf{T}_{(i,j)}$ to device global memory
    \State Map MBE clusters to CUDA blocks
    \Block{MBE cluster}
        \State Initialize $E^{\text{current}}_{\text{c}}$, $E^{\text{last}}_{\text{c}}$ and $\mathbf{T}_{(ij,ij)}$
        \While{$(E^{\text{current}}_{\text{c}}-E^{\text{last}}_{\text{c}}) \ge l_{\text{mp2e}}$}
            \State $E^{\text{last}}_{\text{c}}=E^{\text{current}}_{\text{c}}$
            \State $E^{\text{current}}_{\text{c}} = 0.0$
            \For{pair $(i,j)$ from cluster}
                \State Compute $\mathbf{R}_{(ij,ij)}$ using eq~(\ref{eq:res_close})
                $\cdots\cdots 32N_{k}N^{3}_{\text{osv}}$
                
                \State Transform $\tilde{\mathbf{R}}_{(ij,ij)} = \mathbf{X}^{\dagger}_{(ij,ij)}\mathbf{R}_{(ij,ij)}\mathbf{X}_{(ij,ij)}$
                $\cdots\cdots 16N^{3}_{\text{osv}}$
                \State Update $\mathbf{T}_{(ij,ij)} \mathrel{+}= \mathbf{X}_{(ij,ij)} (\tilde{\mathbf{R}}_{(ij,ij)}/\tilde{\mathbf{E}}_{(ij,ij)}) \mathbf{X}^{\dagger}_{(ij,ij)}$
                $\cdots\cdots 16N^{3}_{\text{osv}}$
                \State $E^{\text{current}}_{\text{c}}\mathrel{+}=\langle \mathbf{K}_{(ij,ij)} (2\mathbf{T}_{(ij,ij)}-\mathbf{T}^{\dagger}_{(ij,ij)}) \rangle$
                $\cdots\cdots 16N^{2}_{\text{osv}}$
            \EndFor
            \State 
        \EndWhile
    \EndBlock
    \State Accumulate cluster $\mathbf{T}_{(ij,ij)}$ ($N^{\text{batch}}_{\text{cPair}}, 2N_{\text{osv}}, 2N_{\text{osv}}$) to host shared memory
    \State Accumulate $E^{\text{current}}_{\text{c}}$ to total $E_{\text{c}}$
\EndFor
\Statex
\BlockComments{For weak pairs}
\State Generate LMO batches using Algorithm~\ref{alg:min_batch}
\For{batch $k$}
    \State Transfer $\mathbf{T}_{(k,k)}$, $\mathbf{X}_{(k,k)}$ and $\tilde{\mathbf{E}}_{(k,k)}$ to device global memory
    \State Transfer $f_{ij}$, $\mathbf{S}_{(i,j)}$, $\mathbf{F}_{(i,j)}$ and $\mathbf{K}_{(i,j)}$ to device global memory

    \State Map weak pairs formed by $k$ to CUDA blocks
    \Block{weak pair $(i,j)$}
        \State Compute $\mathbf{R}_{(i,j)}$ using eq~(\ref{eq:res_weak})
        $\cdots\cdots 4N^{3}_{\text{osv}}$
        \State Transform $\tilde{\mathbf{R}}_{(i,j)} = \mathbf{X}^{\dagger}_{(i,i)}\mathbf{R}_{(i,j)}\mathbf{X}_{(j,j)}$
        $\cdots\cdots 2N^{3}_{\text{osv}}$
        \State Compute $\mathbf{T}_{(i,j)} = \mathbf{X}_{(i,i)} \left[\tilde{\mathbf{R}}_{(i,j)}/(f_{ii}+f_{jj}-\tilde{\mathbf{E}}_{(i,i)}-\tilde{\mathbf{E}}_{(j,j)})\right] \mathbf{X}^{\dagger}_{(j,j)}$
        $\cdots\cdots 2N^{3}_{\text{osv}}$
        \State Compute $E^{ij}_{\text{c}} = 2\left< \mathbf{K}_{(i,j)} \mathbf{T}_{(ij,ij)} \right>$
        $\cdots\cdots 2N^{2}_{\text{osv}}$
    \EndBlock
    \State Accumulate $E^{ij}_{\text{c}}$ to total $E_{\text{c}}$
\EndFor
\end{algorithmic}
\end{algorithm}

Our GPU implementation for solving residual equations is described in Algorithm \ref{alg:osv_res}. All intermediates in the OSV basis corresponding to unique LMO pairs are loaded into the device global memory only for each batch of MBE clusters, which naturally reduces both memory usage and data transfer. An individual residual equation for a given cluster is solved independently on each CUDA block. Evaluating the 4-block residual matrix $\mathbf{R}_{(ij,ij)}$ involves the composite matrix $\mathbf{S}_{(ij,kl)}$ and $\mathbf{F}_{(ij,kl)}$, whose sub-blocks are scattered across non-contiguous memory regions. To avoid conditional branching and thread divergence in a unified loop, each sub-block of $\mathbf{R}_{(ij,ij)}$ is computed separately. 

Among the tremendous number of weak pairs, the exchange excitations ($i\rightarrow \bar{\mu}_{j},~j\rightarrow \bar{\nu}_{i}$) are negligible for such long-range interactions. Therefore single-block residual equations $\mathbf{R}_{(i,j)}$, which are a drastic simplification of the full residual equations $\mathbf{R}_{(ij,ij)}$, are sufficiently accurate,
\begin{equation}
    \mathbf{R}_{(i,j)} = \mathbf{K}_{(i,j)} + \mathbf{T}_{(i,j)} \mathbf{F}_{(j,i)} + \mathbf{F}_{(i,i)} \mathbf{T}_{(i,j)}- (f_{ij} + f_{ii}) \mathbf{T}_{(i,j)}
        - f_{ij} \left[ 
                \mathbf{T}_{(i,i)} \mathbf{S}_{(i,j)} 
                + \mathbf{S}_{(i,i)} \mathbf{T}_{(j,j)} 
                \right].
\label{eq:res_weak}
\end{equation}
Since weak-pair amplitudes $\mathbf{T}_{(i,j)}$ couple exclusively to the pre-computed diagonal-pair amplitudes, they can be determined directly without iteration. This feature, along with the one-block residual formulation, makes the evaluation cost negligible for long-range correlation energies, as compared to that of solving close pair residual equations.

The update of $\mathbf{T}_{(ij,ij)}$ amplitudes constitutes another computational bottleneck in solving close pair residual equations, as the generalized eigenvalue problem is targeted for each close pair $(i,j)$,
\begin{equation}
\mathbf{F}_{(ij,ij)} \mathbf{X}_{(ij,ij)} = \mathbf{S}_{(ij,ij)} \mathbf{X}_{(ij,ij)}\mathbf{\tilde{E}}_{(i,j)},
\label{eq:prec_eig}
\end{equation}
where $\mathbf{X}_{(ij,ij)}$ and $\mathbf{\tilde{E}}_{(i,j)}$ denote the eigenvectors and eigenvalues, respectively. On CPUs, this equation can be conveniently solved for each pair assigned to an MPI process via Cholesky decomposition. GPU implementations, however, suffer from significant overhead: the repeated invocation of \texttt{cuSolver}'s eigenvalue solver is much more expensive than the computation itself, due to the small size of OSV-based matrices. We eliminate this overhead by constructing a CUDA kernel to enable pair-wise diagonalizations over blocks. 
Furthermore, in the CUDA kernel, we implement a device function to carry out block-wise Cholesky factorization $\mathbf{S}_{(ij,ij)} = \mathbf{L}_{(ij)}\mathbf{L}^{\dagger}_{(ij)}$, where the elements of the lower-triangular factor $\mathbf{L}_{(ij)}$ are determined in parallel across successive column iterations. A Jacobi diagonalization routine is adapted for small OSV-based matrices, where all eigenvalue updates and eigenvector rotations are executed concurrently on threads within each block, similar to Algorithm \ref{alg:loc_omo}. We develop another device function that solves triangular eigen-systems involving $\mathbf{L}_{(ij)}$, thereby avoiding explicit computation of the numerically unstable inverse $\mathbf{L}^{-1}_{(ij)}$. The pre-allocated device shared memory in each thread block is reserved across device functions to cache intermediate quantities and accumulate partial results, significantly reducing global memory accesses and enhancing overall computational performance.

\section{Results}
\label{sec:results}

In all MBE(3)-OSV-MP2 calculations presented in this work, we adopt the truncation thresholds for determining various local orbital spaces that were previously optimized for an optimal efficiency–accuracy trade-off\cite{liang2021third}. The sparsity identified by these cutoff thresholds is demonstrated with insulin/cc-pVTZ in Table \ref{tab:sparsity}. Numerical benchmarks demonstrate that the correlation energy accuracy from our GPU-accelerated MBE(3)-OSV-MP2 implementation is comparable to the previously reported CPU-based results, as shown in Table S1.

To assess the computational performance and scalability of the GPU acceleration, we evaluate the capability of our implementation for efficiently handling large molecules and provide timing comparisons with other GPU-accelerated RI-MP2 implementations\cite{guo2025byteqc,snowdon2024efficient} as well as our own CPU-based algorithm\cite{liang2021third}. All computations in this study were performed on the Tianhe-2 supercomputer at the National Supercomputing Center.  Our GPU computations utilized 1--3 nodes, each equipped with 8$\times$ NVIDIA A800 (80 GB) GPUs connected via PCIe 4.0 $\times$16. The A800 (80 GB PCIe) version delivers identical performance to the A100 (80 GB PCIe) in all other aspects, including compute throughput, memory bandwidth (HBM2e), and single-GPU operation. The Lustre parallel file system enables the sharing of large datasets across processes via files. CPU calculations employed 64 cores on Intel Xeon Platinum 8358P (2.60 GHz). 

\begin{table}[H]
\centering
    \caption{The chosen cutoff parameters for predefining orbital and MBE cluster spaces in MBE(3)-OSV-MP2 calculations and the resulting sparsity (demonstrated with insulin/cc-pVTZ).}
\begin{tabular}{lrccc}
\hline \midrule
&	Threshold	&	Full	&	Kept	&	Sparsity (\%)	\\
\cmidrule(lr){2-5}
AO pairs	&	$10^{-10}$	&	304432704	&	25267990	&	91.70\\
Average MP2 sparse fitting	&	$10^{-6}$	&	44319	&	5957	&	86.56\\
Average OSVs	&	$10^{-4}$	&	15910	&	70	&	99.56\\
Close 2-body clusters	&	$10^{-2}$	&	617716	&	21204	&	96.57\\
3-body clusters	&	0.2	&	227938315	&	59456	&	99.97\\ 
\hline\hline
\end{tabular}
\label{tab:sparsity}
\end{table}

\subsection{GPU Parallel Scalability with Molecular Sizes}

\begin{figure}[H]
\includegraphics[width=0.7\linewidth]{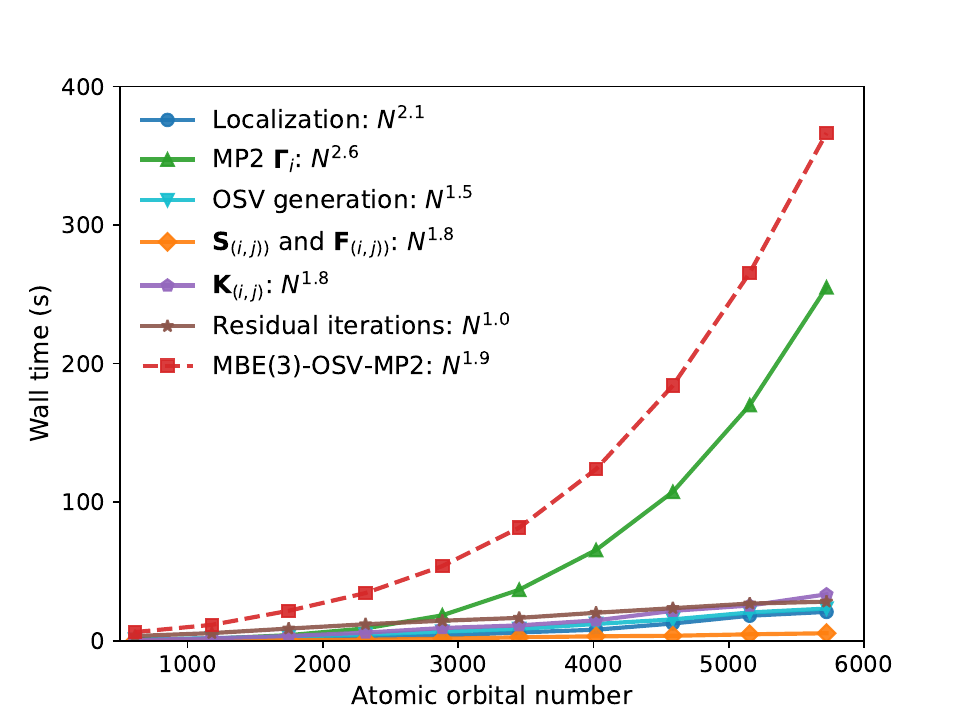}
\centering
\caption{The comparison of the total Wall time (seconds) as a function of the number of atomic orbitals for MBE(3)-OSV-MP2 and its individual computing components using one NVIDIA A800 (80 GB) GPU. All timings of polyglycines $(\text{Gly})_{n}$ ($n=4,8,...,40$) were obtained using def2-TZVP/def2-TZVP-RIFit basis sets. The empirical scaling exponents were obtained by fitting the polynomial curve of the computation time versus number of atomic orbitals with $(\text{Gly})_{4}$--$(\text{Gly})_{40}$.}
\label{fig:scal_mol}
\end{figure}

Figure~\ref{fig:scal_mol} compares the wall time and scaling performance of individual MBE(3)-OSV-MP2 steps on a single NVIDIA A800 (80 GB) GPU. The MBE(3)-OSV-MP2  energy calculation was completed within about 400 s for the longest polymer $(\text{Gly})_{40}$ with def2-TZVP basis set, having 5723 atomic basis functions. Overall, the GPU MBE(3)-OSV-MP2 energy calculation scales as $N^{1.9}$ empirically up to $(\text{Gly})_{40}$/def2-TZVP. 
The GPU calculation is dominated by the on-the-fly construction of $\mathbf{\Gamma}_{i}$ for which both the scaling complexity of $N^{2.6}$ and timing fraction are relatively high compared to other steps. Excluding the orbital localization and $\mathbf{\Gamma}_{i}$ build, all of other MBE(3)-OSV-MP2 computational steps scale only as $N^{1.3}$, much more favorable than the $N^{2.1}$ scaling of the CPU computation ~\cite{liang2021third}. This scaling improvement arises from substantial enhancements in the parallel algorithms, as well as the flatter runtime curve observed for small molecules, which was also similarly observed for DFT scaling on GPUs\cite{kussmann2021highly}. On small systems, the computations underutilize the GPU’s massive parallel resources, with the total runtime heavily constrained by repeated overheads such as kernel launches and synchronizations.

Several other CPU computational bottlenecks are now alleviated  because of their algorithmic adaptation to the GPU device. The orbital localization on GPU becomes only a  minor fraction of the total computational cost with an empirical scaling $N^{2.1}$, which is much reduced from the formal $N^{3}$ complexity. The GPU-based rOSV generation scales only as $N^{1.5}$, making a substantial improvement over the ID-OSV scaling of $N^{2.5}$\cite{liang2021third}. The sparsity exploration in LMO pairs, OSVs, and auxiliary fitting space considerably accelerates the GPU  evaluations of OSV overlap, Fock, and exchange tensors, which add up to only a small portion of the total runtime. Finally, both the preconditioning and residual iterations exhibit a linear scaling, managed by effective truncations of OSVs and MBE 2b/3b clusters.

\subsection{Strong Scaling with Multiple GPUs}

\begin{figure}[H]
\includegraphics[width=0.7\linewidth]{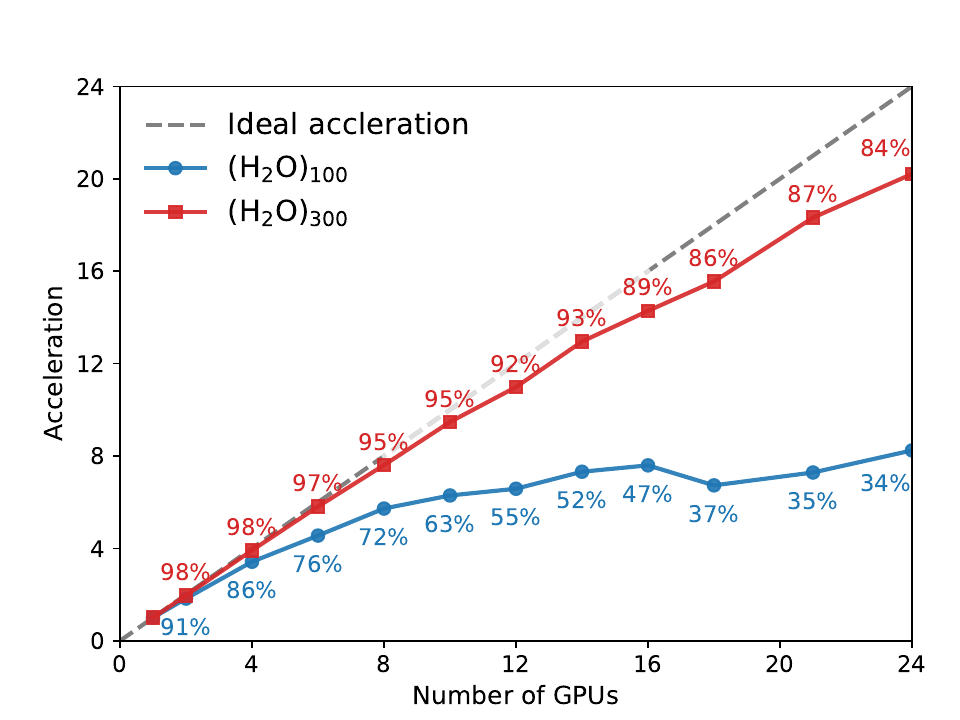}
\centering
\caption{Strong scaling performance of MBE(3)-OSV-MP2 without localization with respect to the number of GPUs (A800/80 GB). Each computing node is equipped with 8 GPUs. The parallel efficiency refers to the percentage of the actual acceleration to the ideal acceleration fold. Calculations were performed using cc-pVDZ/cc-pVDZ-RIFit basis sets for both (H$_2$O)$_{100}$ and (H$_2$O)$_{300}$ clusters.}
\label{fig:scal_node}
\end{figure}

The GPU parallel speedup performance of MBE(3)-OSV-MP2 computation is demonstrated up to 24 A800 (80 GB) GPUs across 3 nodes for (H$_2$O)$_{100}$ and (H$_2$O)$_{300}$ which are optimized with ChargeNN model\cite{liang2025polarizable}, as shown in Figure~\ref{fig:scal_node}. The localization is excluded from MBE(3)-OSV-MP2 parallel tests, as it was performed on only one GPU, and the multi-GPU localization has not been implemented. 
The MBE(3)-OSV-MP2 calculation shows scalable acceleration for the large (H$_2$O)$_{300}$ cluster with the number of GPUs. The parallel efficiency is still maintained by $\sim$90\% up to 16 GPUs and only moderately drops to 84\% on 24 GPUs. For the smaller (H$_2$O)$_{100}$ cluster, however, the efficiency descends markedly to 47\% on 16 GPUs and 34\% on 24 GPUs, because the relatively low computational intensity per task for small systems is insufficient to effectively mask non-kernel overheads, such as communication, synchronization, and host–device transfers.


The overall end-to-end runtime efficiency of the MBE(3)-OSV-MP2 calculation depends on many factors including the size of molecules, basis sets, the number of GPUs and device memory.The MBE(3)-OSV-MP2 performance remains only moderate when the HF and localization costs become dominant. 
This situation occurs when the speedup of post-HF correlation calculations is substantially superior to HF and localization steps executed on many GPUs. Consequently, the post-HF contribution to the total wall time is effectively and continuously reduced as the number of GPUs increases, as illustrated in Figure S1\textbf{b} for (H$_2$O)$_{300}$. 
On the other hand, for very large molecules such as insulin/cc-pVTZ (Table 3), the localization cost is in fact negligibly small compared to post-HF steps on 8 A800 GPUs, e.g., only 3.1\% of the total runtime. For smaller systems on a single GPU, the localization timings also constitute only minor fraction of the total runtime, for instance, only 2\% and 8\% for (H$_2$O)$_{190}$ and C$_{60}$@Catcher using cc-pVTZ basis, respectively, as seen in Figure S2.

The HF cost arises primarily from significant serial overheads in the McMurchie–Davidson (for $J_{\alpha\beta}$) and Rys quadrature (for $K_{\alpha\beta}$) kernels in \texttt{GPU4PySCF}. Switching to the Head–Gordon–Pople algorithm could substantially mitigate these bottlenecks.\cite{palethorpe2024advanced} For Jacobi localization, the 2-by-2 orbital rotation depends on the result of the previous rotation, which yields poor strong scaling, as the orbital data communication quickly overwhelms the calculation.

Overall, while the post-HF components demonstrate excellent strong scaling, the parallel inefficiencies of the Hartree–Fock reference and the high cost of the single-GPU localization currently constrain efficient deployment of the end-to-end MBE(3)-OSV-MP2 method on a large number of GPUs.

\subsection{Large Molecules}

As demonstrated in Table S1, MBE(3)-OSV-MP2 yields sufficiently accurate correlation energy for a series of molecular systems, e.g., recovering 99.97\% of the RI-MP2 correlation energy for (H$_2$O)$_{32}$/aug-cc-pVTZ. Furthermore, we validated the accuracy of MBE(3)-OSV-MP2 for large molecular systems by computing the protein–ligand interaction energies of 6GQ5 (392 atoms) and 4NAG (420 atoms) using the cc-pVTZ basis set. As shown in Table~\ref{tab:pro_inter}, relative to the reference RI-MP2 values, MBE(3)-OSV-MP2 yields chemically accurate deviations for 6GQ5 and 4NAG, with errors of only 1.02 and 0.77 kcal/mol at the normal OSV truncation threshold ($10^{-4}$).  Tightening the threshold to $10^{-5}$ reduces these errors to 0.61 and 0.37 kcal/mol, comparable to the reduced CIM-RI-MP2 ($\mu=9.75\text{ \AA}$) values of 0.45 and 0.36 kcal/mol. Moreover, the interaction energy computation (including protein, ligand, and complex correlation energies) on MBE(3)-OSV-MP2 required approximately 1 hour at the normal threshold and 2.5 hours at the tight threshold using two A800 (80 GB) GPUs. In comparison, the reduced CIM-RI-MP2 required approximately 17.4 hours on four V100 (32 GB) GPUs. Consequently, GPU-accelerated MBE(3)-OSV-MP2 retains high accuracy for large molecular complexes while delivering excellent computational efficiency.

\begin{table}[H]
\centering
\caption{Comparison of protein–ligand interaction energies (kcal/mol) and wall-clock times (hours) for MBE(3)-OSV-MP2 versus RI-MP2 and reduced CIM-RI-MP2 on the 6GQ5 and 4NAG systems. The cc-pVTZ basis set was used. The geometries were taken from Ref. \citenum{feng2025efficient}.}
\begin{tabular}{llcccc}
\hline \midrule
 & & \multicolumn{2}{c}{6GQ5} & \multicolumn{2}{c}{4NAG} \\
\cmidrule(lr){3-4} \cmidrule(lr){5-6}
 & Cutoff & Eint & Time & Eint & Time \\
 & & (kcal/mol) & (hours) & (kcal/mol) & (hours) \\
\midrule
RI-MP2$^a$ & & -9.08 & & -10.52 & \\
CIM-RI-MP2$^b$
 & $\mu=9.75\text{ \AA}$ & -8.63 & 17.49 & -10.16 & 17.33 \\
\multirow{2}{*}{MBE(3)-OSV-MP2$^c$} & $l_{\text{osv}}=10^{-4}$ & -8.06 & 0.81 & -9.75 & 1.08 \\
 & $l_{\text{osv}}=10^{-5}$ & -8.47 & 2.39 & -10.15 & 2.73 \\
\hline\hline
\end{tabular}
\begin{tablenotes}
\item$^{a}$ From Ref. \citenum{feng2025efficient}.
\item$^{b}$ From Ref. \citenum{feng2025efficient}, computed with 4$\times$NVIDIA V100 (32 GB).
\item$^{c}$ This work, computed with 2$\times$NVIDIA A800 (80 GB).
\end{tablenotes}
\label{tab:pro_inter}
\end{table}

\begin{figure}[H]
\includegraphics[width=0.7\linewidth]{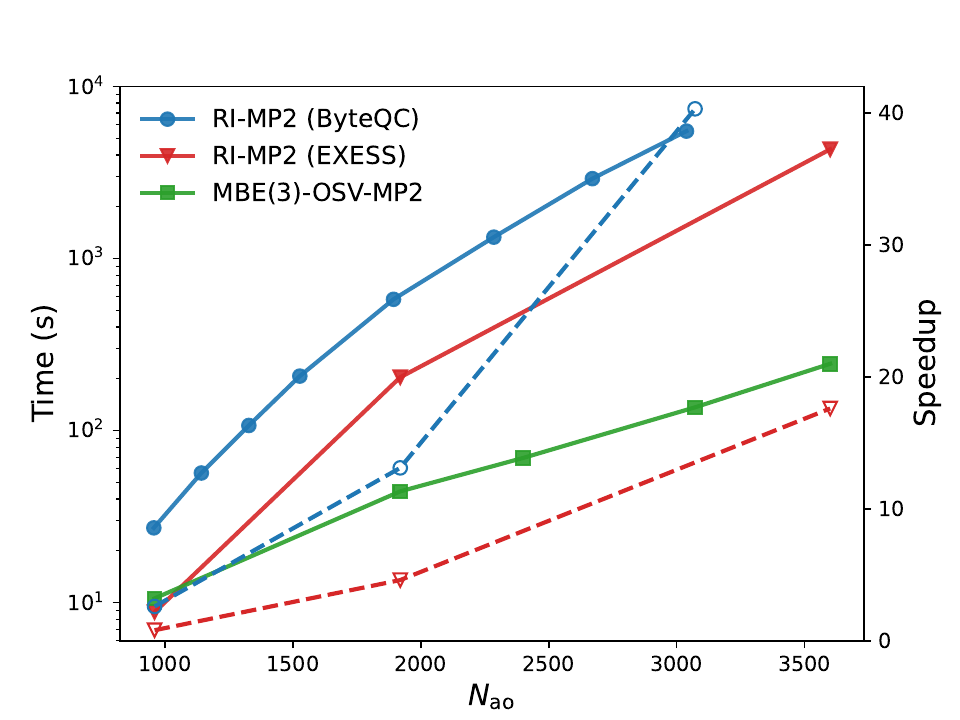}
\centering
\caption{Wall time (seconds) comparison between GPU-accelerated MBE(3)-OSV-MP2, \texttt{ByteQC}'s RI-MP2\cite{guo2025byteqc}, as well as computational time (seconds) of \texttt{EXESS}'s RI-MP2\cite{snowdon2024efficient}. Dashed curves indicate the speedup of MBE(3)‑OSV‑MP2 relative to RI‑MP2. RI-MP2 calculations were performed on an A100 (80 GB) GPU\cite{guo2025byteqc,snowdon2024efficient}, while MBE(3)-OSV-MP2 calculations were carried out on an A800 (80 GB) GPU. A800 (80 GB, PCIe) exhibits the same performance as A100 (80 GB, PCIe). For MBE(3)-OSV-MP2 calculations, the structures of water clusters were optimized using the QM-polarizable water model ChargeNN~\cite{liang2025polarizable}.}
\label{fig:vs_rimp2}
\end{figure}

Figure~\ref{fig:vs_rimp2} compares the single-GPU performance of RI-MP2 timings in \texttt{ByteQC}\cite{guo2025byteqc} and \texttt{EXESS}\cite{snowdon2024efficient} with the current MBE(3)-OSV-MP2 GPU implementation on water clusters using cc-pVDZ/cc-pVDZ-RIFit basis sets. 
Overall, the MBE(3)-OSV-MP2 significantly outperforms RI-MP2 due to both the low-scaling algorithm and GPU parallelism, which systematically exploit the locality in LMOs, OSVs, and sparse fitting AOs. Compared to \texttt{ByteQC}'s RI-MP2, the MBE(3)-OSV-MP2 calculation achieves 2.6$\times$, 13.1$\times$ and 40.3$\times$ accelerations for (H$_2$O)$_{40}$, (H$_2$O)$_{80}$ and (H$_2$O)$_{128}$, respectively. Even CPU I/O and host–device transfer costs are excluded from the \texttt{EXESS}'s RI-MP2 timings, the MBE(3)-OSV-MP2 is only slightly slower for (H$_2$O)$_{40}$, but yields 4.6$\times$ and 17.6$\times$ folds of acceleration for (H$_2$O)$_{80}$ and (H$_2$O)$_{128}$, respectively. The greater acceleration of MBE(3)-OSV-MP2 relative to RI-MP2 for larger systems originates from its dramatically improved scaling, which drops from the formal $\mathcal{O}(N^{5})$ complexity for RI-MP2 to an empirical sub-quadratic scaling. Moreover, the on-the-fly regeneration of $\mathbf{\Gamma}_{i}$ in MBE(3)-OSV-MP2  avoids $N_{\text{occ}}(N_{\text{ao}}+N_{\text{vir}})N_{\text{aux}}$ storage and $N_{\text{occ}}(N_{\text{occ}}+1)N_{\text{vir}}N_{\text{aux}}$ data movement that plague the canonical RI-MP2 GPU calculation for large molecules. This used to be an important bottleneck that limited disk I/O and host–device data transfers arising from low-bandwidth operations, which are now significantly alleviated in the present GPU algorithm. 

Moreover, we demonstrate the GPU acceleration of MBE(3)-OSV-MP2 energy calculation compared to our previous CPU-based implementation~\cite{liang2021third} for large molecules, as shown in Figure~\ref{fig:vs_cpu}.  The localization was carried out using 1 CPU core, and all subsequent steps employed 48 cores for (H$_2$O)$_{190}$/cc-pVTZ due to memory limitation and 64 cores for all other tested systems. 
In the CPU implementation, Figure \ref{fig:vs_cpu}a shows that the single-core PM localization using meta-L{\"o}wdin charges together with the co-iterative augmented Hessian (CIAH) rotation method\cite{sun2016co} requires even longer time than the following parallel MBE(3)-OSV-MP2 steps for all tested systems. Switching to L{\"o}wdin charges and Jacobi optimizer, the localization step on CPU is substantially accelerated, achieving $\sim$19$\times$ of speedup for (H$_2$O)$_{190}$/cc-pVTZ. Using the same charge model and optimizer, the localization step is further expedited on GPU: about 3-fold speedup for C$_{60}$@catcher (374 occupied orbitals) and 22--27-fold for (H$_2$O)$_{190}$ (950 occupied orbitals). The latter speedup is substantially greater due to the larger orbital pairs that better exploit GPU parallelism with higher active thread occupancy for (H$_2$O)$_{190}$ than C$_{60}$@catcher. In addition, the time required for Pipek–Mezey (PM) localization using Jacobi sweeps increases only modestly with larger basis sets. Consequently, the localization step constitutes a significantly smaller fraction of the total computational time when a larger basis set is used. For example, the Figure S2 shows a substantial drop of the localization proportion from 16.0\% with cc-pVDZ to 2.0\% with cc-pVTZ for (H$_2$O)$_{190}$.

\begin{figure}[H]
\includegraphics[width=1\linewidth]{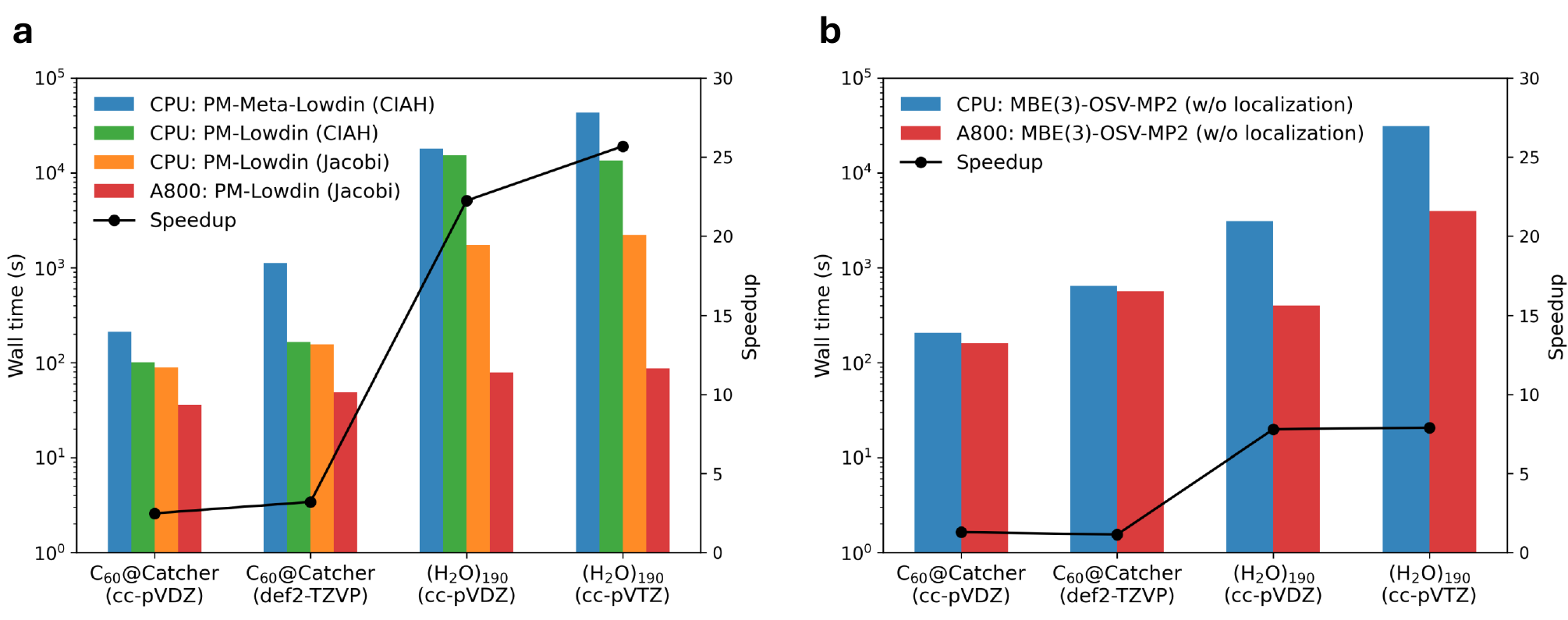}
\centering
\caption{Wall time (seconds) for (\textbf{a}) localization and (\textbf{b}) subsequent MBE(3)-OSV-MP2 processes, measured on a single NVIDIA A800 (80 GB) GPU and 64 Intel Xeon Platinum 8358P (2.60 GHz) CPU cores. The coordinates of C$_{60}$@catcher and $(\text{H}_{2}\text{O})_{190}$ were taken from Refs. \citenum{sure2015comprehensive} and \citenum{watergeo}, respectively.}
\label{fig:vs_cpu}
\end{figure}

 For the subsequent MBE(3)-OSV-MP2 steps after localization, only 1.1--1.3 folds of GPU acceleration is achieved for the conjugated system C$_{60}$@catcher compared to parallel calculations on 64 CPU cores. This minor improvement  stems primarily from two interrelated consequences of the extensive electron delocalization: first, the large OSV space (average 93 for C$_{60}$@catcher/cc-pVTZ) generates heavy global memory traffic, as well as restricts kernel occupancy as fewer pairs or MBE clusters can be processed in each kernel launch; second, the highly nonuniform OSV distribution (standard deviation of 41) leads to severe workload imbalance among thread blocks, each of which is assigned with calculations of an LMO pair or an MBE cluster. Therefore, Figure S2 reveals significantly larger proportions of residual iterations timing out of the total runtime for the conjugated $\text{C}_{60}\text{@Catcher}$ (42.8\%--49.9\%) compared to the localized system $(\text{H}_2\text{O})_{190}$(1.2\%--2.6\%). Furthermore, these factors result in residual iterations for $\text{C}_{60}\text{@Catcher}$ to run approximately 1.7$\times$ slower on the GPU than on the CPU. Specific workflow and kernel optimizations will be necessary to improve efficiency for systems exhibiting similarly extended and heterogeneous virtual orbital spaces. 
 
 About 8.0-fold speedup is obtained for (H$_2$O)$_{190}$, owing to its much more compact and uniform OSV spaces (average size of 55 with a standard deviation of 7 for (H$_2$O)$_{190}$/cc-pVTZ). In the corresponding CPU calculations, the 3c2e coefficient tensor $\mathbf{\Gamma}_{i}$ is precomputed and stored on disk, requiring 3.6 TB of storage. Although multi-process MPI parallelization enables effective overlap of I/O operations to access $\mathbf{\Gamma}_{i}$ with computation, pure CPU processing still accounts for only about 40\% of the total elapsed time for the (H$_2$O)$_{190}$/cc-pVTZ system. The GPU implementation eliminates both the high storage demand and the dominant CPU I/O bottleneck by generating the $\mathbf{\Gamma}_{i}$ intermediates directly on-the-fly, thereby enabling markedly better scalability for large-scale calculations.

 Overall, GPU implementation yields 1.3$\times$--1.5$\times$ accelerations for C$_{60}$@catcher and 8.3$\times$--10.2$\times$ for (H$_2$O)$_{190}$ in MBE(3)-OSV-MP2 compared to CPU-only execution. Based on the hourly rate of the Tianhe-2 platform, each CPU node costs 64 core hours per hour, while each A800 card consumes resources economically equivalent to 80 core hours per hour. Consequently, the GPU-accelerated MBE(3)-OSV-MP2 offers substantial reductions in computation time and cost, which translate directly into significant real-world savings in both monetary and energy consumption for large-scale computing tasks.

Finally, we showcase the performance of our multi-GPU implementation on a large biomolecule: the 784-atom human insulin peptide (a key hormone comprising 51 amino acids). Calculations were carried out using both cc-pVDZ and cc-pVTZ basis sets, together with their corresponding RIFit auxiliary basis sets, on 8 NVIDIA A800 GPUs. The memory usage and wall time are significantly reduced for all important MBE(3)-OSV-MP2 processes as summarized in Table~\ref{tab:insulin}. The efficient MBE(3)-OSV-MP2 computations are promoted due to the highly compact LMOs, OSVs, and fitting space. The peak memory storage remains modest at 88 GB for cc-pVDZ and 241 GB for cc-pVTZ, which fit well with available host memory without incurring expensive disk I/O operations.
For cc-pVDZ basis ($N_{\text{ao}}$=7571, $N_{\text{aux}}$=28022), the MBE(3)-OSV-MP2 energy calculation was completed in 1429.7 s, including 406.7 s for Hartree–Fock and 1023.0 s for MP2 correlation energy (including localization). With a larger cc-pVTZ basis ($N_{\text{ao}}$=17448, $N_{\text{aux}}$=44319), the total wall time rises to 6.5 hours, consisting of 1.7 hours for Hartree-Fock and 4.8 hours for MP2 correlation (including localization). 

\begin{table}[H]
\centering
    \caption{Molecular size, memory usage (GB), and wall time (s) for MBE(3)-OSV-MP2 calculations of insulin performed on 8 NVIDIA A800 GPUs, using cc-pVDZ and cc-pVTZ basis sets with corresponding auxiliary basis sets. The structure of insulin was taken from Ref. \citenum{RCSB3I40}.}
\begin{tabular}{lrrcrr}
\hline \hline
\multicolumn{6}{c}{Molecular size}									\\
Basis set	&	cc-pVDZ	&		&&	cc-pVTZ	&		\\
Atoms	&	784	&		&&	784	&		\\
Occupied orbitals	&	1538	&		&&	1538	&		\\
Orbital basis	&	7571	&		&&	17448	&		\\
MP2 fitting basis	&	28022	&		&&	44319	&		\\
\multicolumn{6}{c}{Memory usage (GB)}									\\
$\mathbf{Q}_{i}$	&	1.9	&		&&	9.3	&		\\
$\tilde{\mathbf{\Gamma}}^{\bar{\mu}_{j}A'}_{i}$	&	77.3	&		&&	196.7	&		\\
OSV S matrix	&	3.2	&		&&	12.2	&		\\
OSV K matrix	&	4.1	&		&&	14.5	&		\\
Max memory usage	&	88.0	&		&&	240.6	&		\\
\multicolumn{6}{c}{Wall time}									\\
	&	Wall time (s)	&	Fraction (\%)	&&	Wall time (s)	&	Fraction (\%)	\\
Hartree-Fock 	&	406.7	&		&&	6226.7	&		\\
Localization	&	470.4	&	46.0	&&	536.0	&	3.1	\\
$\mathbf{\Gamma}_{i}$ (ERI)	&	125.1	&	12.2	&&	11272.8	&	65.9	\\
$\mathbf{\Gamma}_{i}$ (transformation)	&	332.0	&	32.4	&&	4597.0	&	26.9	\\
rSVD	&	30.7	&	3.0	&&	331.6	&	1.9	\\
OSV S/F	&	4.4	&	0.4	&&	16.1	&	0.1	\\
OSV K matrix	&	47.8	&	4.7	&&	316.5	&	1.9	\\
Residual iteration	&	12.8	&	1.2	&&	37.7	&	0.2	\\
MBE(3)-OSV-MP2	&	1023.0	&	100	&&	17107.8	&	100	\\
Total 	&	1429.7	&	 	&&	23334.5	&		\\
\hline\hline
\end{tabular}
\label{tab:insulin}
\end{table}

We further analyze the computational costs of MBE(3)-OSV-MP2 individual steps for  insulin, as shown in Table~\ref{tab:insulin}. 
For the cc-pVDZ basis, the localization carried out on a single GPU forms a major task as compared to other multi-GPU steps in correlation computation. It is expected that the multi-GPU localization implementation would shorten its single-GPU wall time significantly and become unimportant.
The primary bottlenecks are the repeated construction of the two-electron intermediate tensor $\mathbf{\Gamma}_{i}$ on-the-fly, which accounts for 44.6\% of the total time with cc-pVDZ and increases to to 92.8\% with cc-pVTZ. This strong basis-set dependence arises because the present implementation of $\mathbf{\Gamma}_{i}$ generation is subject to the availability of limited GPU memory: a small capacity forces excessive recomputation of the 3c2e integrals $(\alpha\beta|A)$. The number of such regenerations ranges from only 13 for cc-pVDZ to 207 for cc-pVTZ on each A800 (80 GB) GPU, thereby increasing their contribution to the total wall time from 12.2\% to 65.9\%. The development of a $\mathbf{\Gamma}_{i}$ construction scheme with reduced sensitivity to memory constraints will be crucial for enabling the application of the method to substantially larger systems in future studies. In addition to the substantial speedup achieved by exploiting sparsity in the localized orbital spaces, the remaining computational steps are significantly accelerated through the use of highly optimized CUDA kernels specifically designed for the massively parallel evaluation of thousands of LMO pairs (or MBE clusters). This strategy leads to a very low overall computational overhead, amounting to only 9.3\% for the cc-pVDZ basis set and 4.1\% for cc-pVTZ.


\subsection{Illustrative Application}
We further demonstrate the capability of our GPU-accelerated MBE(3)-OSV-MP2 method for tackling realistic biochemical systems. As a representative case, we investigated the enzymatic Friedel--Crafts acylation of monoacetylphloroglucinol (MAPG) catalyzed by Pseudomonas protegens (PpATase), a highly promising enzyme for biocatalysis applications.\cite{hayashi2012molecular,schmidt2017biocatalytic} Following the protocol by Sheng et al., an extensive 413-atom quantum mechanical (QM) region was employed to describe the long-range proton transfer chain ($>$10~\AA) involved in substrate protonation.\cite{sheng2019mechanism} The optimized geometries, solvation energies, and zero-point energy corrections were adopted from Ref. \citenum{sheng2019mechanism}, where they were computed at the B3LYP-D3/6-31G(d,p) level of theory. Electronic energies were computed with GPU-accelerated MBE(3)-OSV-MP2 the triple-$\zeta$ 6-311+G(2d,2p) basis set. Our results reveal that the first half-reaction is limited by substrate protonation ($\text{Int3} \rightarrow [\text{TS2}] \rightarrow \text{Int4}$) with a barrier of 16.6~kJ/mol, while the second half-reaction is governed by C--C bond formation between the substrate and the acyl group on a cysteine residue ($\text{Int8} \rightarrow [\text{TS4}] \rightarrow \text{Int9}$), with a barrier of 15.5~kJ/mol. The overall computed free-energy barrier of 16.6~kJ/mol is in excellent agreement with the experimental value of approximately 17~kJ/mol. Notably, a full single-point MBE(3)-OSV-MP2 energy calculation (including the Hartree--Fock contribution) on this 413-atom system requires only 1.7~hours on two A800 GPUs. This combination of high accuracy and practical computational efficiency highlights an important potential of our GPU-accelerated implementation for high-level \textit{ab initio} simulations of complex biomolecular processes.

\begin{figure}[H]
\includegraphics[width=1\linewidth]{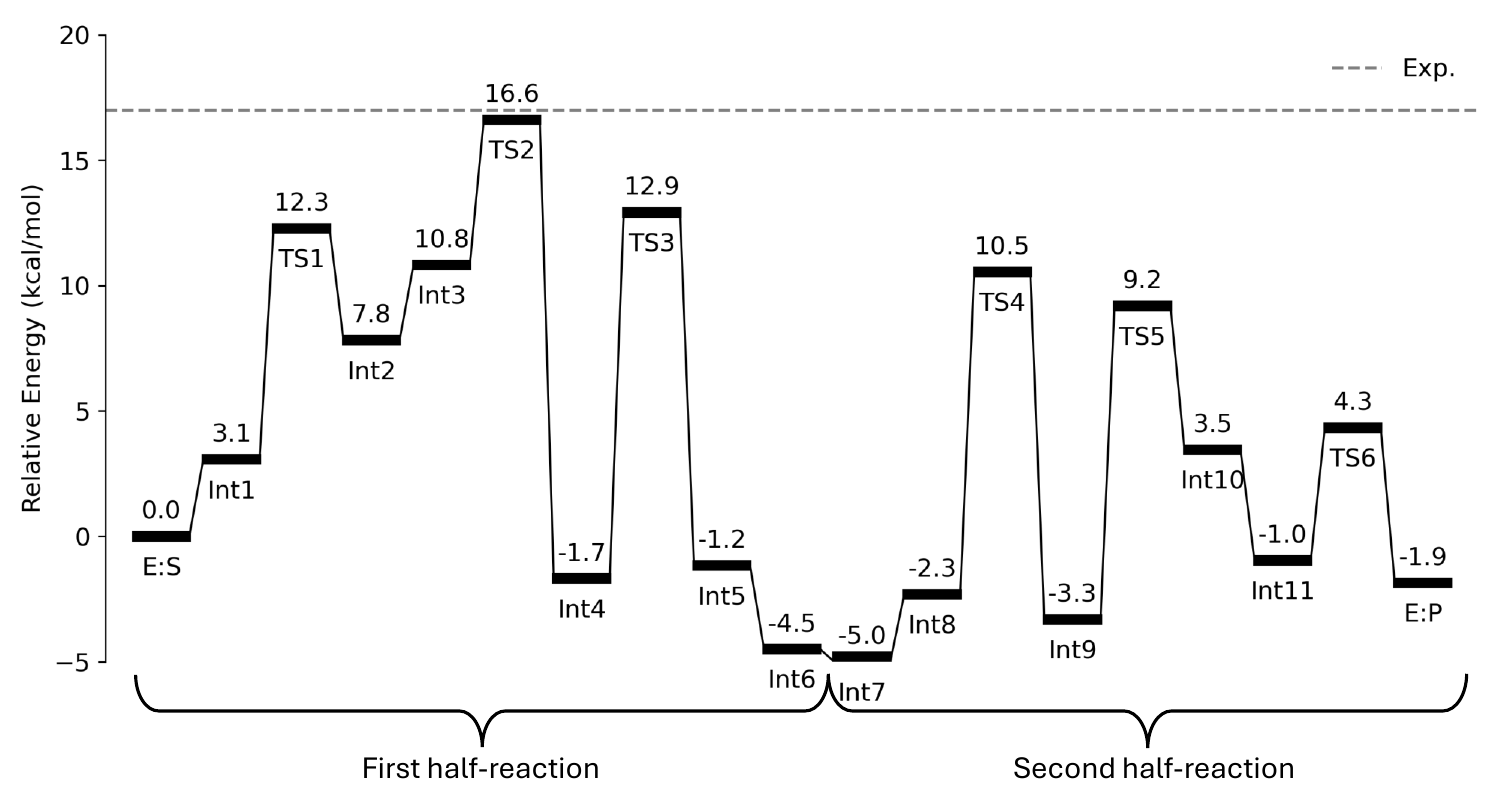}
\centering
\caption{Computed energy diagram for the mechanism of acylation of MAPG catalysed by PpATase. Optimized structures were taken from Ref. \citenum{sheng2019mechanism}.}
\label{fig:vs_cpu}
\end{figure}
\section{Conclusions}
\label{sec:conclusions}
In this work, we introduce a novel multi-GPU implementation to enable efficient large-scale  MBE(3)-OSV-MP2 energy computation across distributed compute nodes. Key advancements include robust scaling-reduced GPU algorithms such as Jacobi-Pipek–Mezey orbital localization and randomized OSV generation. We also implemented a direct $\mathbf{\Gamma}_{i}$ generator to eliminate I/O bottlenecks. In addition, we engineered highly specialized CUDA kernels to maximize GPU parallel efficiency and arithmetic intensity. As a result, the GPU-accelerated MBE(3)-OSV-MP2 achieves empirical sub-$N^{2}$ emperical scalings with molecular size up to (Gly)$_{40}$/def2-TZVP and 84\% parallel efficiency without localization across 24 GPUs distributed on multiple nodes. Compared to cutting-edge canonical RI-MP2 implementations, our GPU-accelerated MBE(3)-OSV-MP2 delivers 40$\times$ speedup in wall-clock time for the (H$_2$O)$_{128}$ cluster in the cc-pVDZ basis set. Relative to our previous CPU-based MBE(3)-OSV-MP2 implementation, the present GPU version provides up to 10$\times$ acceleration, highlighting the substantial computational gain on GPU platforms. Finally, we demonstrate the practical applicability of this implementation to treating large biochemical systems: a full MBE(3)-OSV-MP2 energy calculation of the human insulin peptide with 784 atoms completes in only 24 minutes using cc-pVDZ basis set and in 6.4 hours with cc-pVTZ basis set.

There are several limitations in the current implementation. The GPU parallel localization has been implemented, but can only function on a single GPU. In addition, the direct 3c2e generator of $\mathbf{\Gamma}_{i}$ heavily relies on the availability of the often short device memory, which limits the application to molecular systems consisting of thousands of atoms in a single MP2 calculation without using system fragmentation. Furthermore, advanced kernel optimizations can be conducted to promote the GPU parallel efficiency, including tuning specialized variants/template pools to OSV compactness, better shared memory use for contracting intermediates with reduced bank conflicts, occupancy-aware block/grid configurations for higher warp occupancy, mixed precision, and Tensor Core acceleration at acceptable precision. 

For very large systems, memory constraints will limit linear algebra operations on uncompressed matrices, such as Cholesky decomposition of auxiliary Coulomb integrals and fitting of half‑transformed integrals. Such limitations can be mitigated by implementing matrix tiling or exploiting multi-GPU libraries like cuBLASXt and cuSolverMg. Moreover, the MBE(3)-OSV-MP2 can be integrated with the state-of-the-art fragmentation schemes to handle systems over a million electrons\cite{stocks2024breaking}, which would considerably accelerate the calculations of the individual
fragments relative to the current RI-MP2-based fragmentation methods. 

The current GPU-accelerated implementation provides a promising tool for \textit{ab initio} calculations on large molecular systems. Future work will be extended to GPU-supported coupled-cluster model, periodic systems, MP2 analytical energy gradients, and machine learning frameworks.

\begin{acknowledgement}

The authors acknowledge financial support from the Hong Kong Research Grant Council through General Research Funds (17309020, 17310922, 17305724), the Hong Kong Quantum AI Lab through Research Talent Hub from Innovation and Technology Fund of Hong Kong Innovation and Technology Commission, and the Hung Hing Ying Physical Sciences Research Fund of the University of Hong Kong. The authors also thank the National Supercomputing Center in Guangzhou for providing high-performance computational resources. Q. L. thanks Dr. Qiming Sun for helpful discussions.

\end{acknowledgement}

\section*{Data Availability}
The source code is available under the CC BY-NC license at https://github.com/QCLabHKU/OSVMP2

\begin{suppinfo}

The Supporting Information is available free of charge.

\begin{quote}
Results of correlation energy accuracy of GPU-based MBE(3)-OSV-MP2 compared to CPU implementation across various molecular systems, a list of custom kernels implemented in this work, and coordinates of water clusters optimized with ChargeNN.
\end{quote}
\end{suppinfo}

\bibliography{manuscript}

\end{document}


\section{Accuracy}
\begin{table}[H]
\centering
    \caption{Percentage of canonical RI-MP2 correlation energy recovered by GPU-based and CPU-based\cite{liang2021third} MBE(3)-OSV-MP2 for various molecular systems.}
\begin{tabular}{lrccc}
\hline \midrule
&&&\multicolumn{2}{c}{$E_{\text{c}}^{\text{MBE(3)-OSV-MP2}}/E_{\text{c}}^{\text{RI-MP2}}$}\\
\cmidrule(lr){4-5}
	&	basis	&	$N_{\text{ao}}$	&	CPU\cite{liang2021third} (\%)	&	GPU	(\%) \\
\cmidrule(lr){2-5}
DIAD	&	def2-TZVP	&	1392	&	99.67&	99.68\\
BHS	&	def2-TZVP	&	1586	&	99.80&	99.81\\
Nonactin	&	def2-TZVP	&	1996	&	99.81&	99.86\\
YIVNOG	&	def2-TZVP	&	2046	&	99.86&	99.87\\
FLP	&	def2-TZVP	&	2059	&	99.69&	99.71\\
(H$_{2}$O)$_{32}$	&	aug-cc-pVTZ	&	2944	&	99.92&	99.97\\
\hline\hline
\end{tabular}
\label{tab:sparsity}
\end{table}

\begin{figure}[H]
\includegraphics[width=\linewidth]{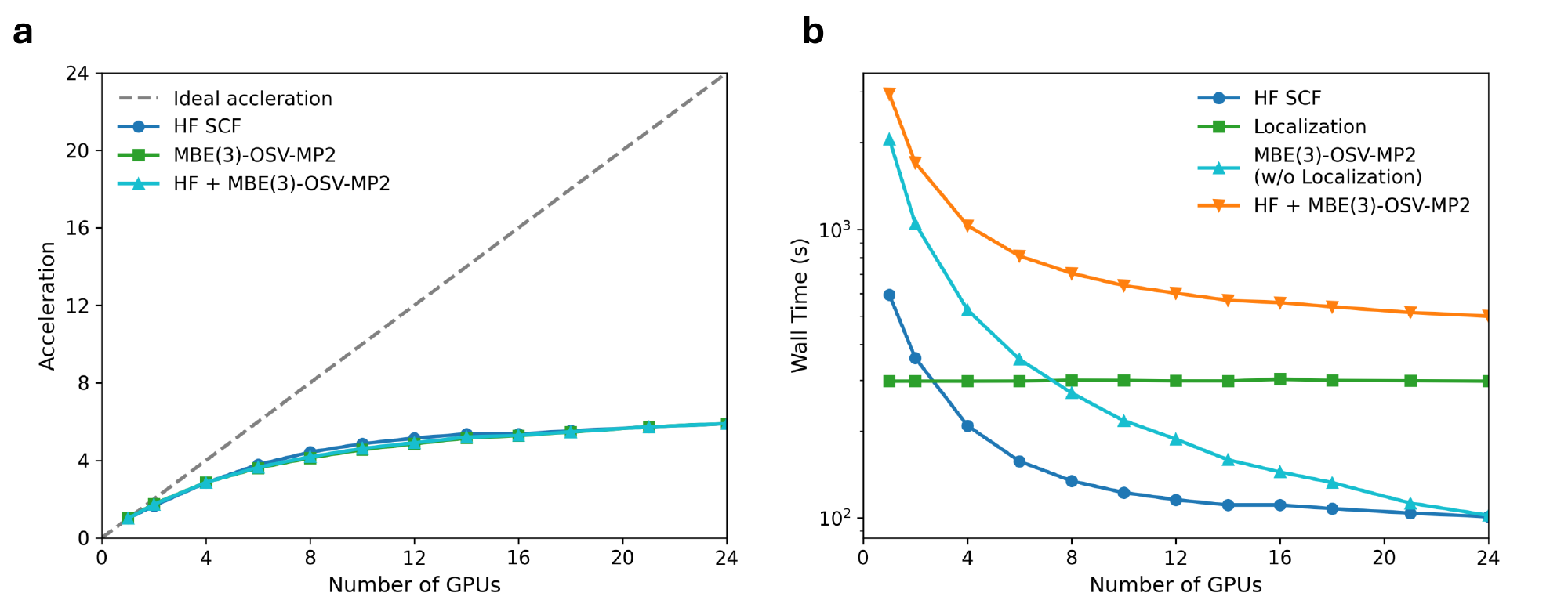}
\centering
\caption{Parallel performance of the GPU based HF SCF and MBE(3)-OSV-MP2 calculations.
(a) Acceleration relative to a single GPU as a function of the number of GPUs.
(b) Wall time for key steps as a function of the number of GPUs. Calculations were performed using a (H$_2$O)$_{300}$ cluster with cc-pVDZ/cc-pVDZ-RIFit basis sets.}
\label{fig:scal_ngpu_step}
\end{figure}

\begin{figure}[H]
\includegraphics[width=\linewidth]{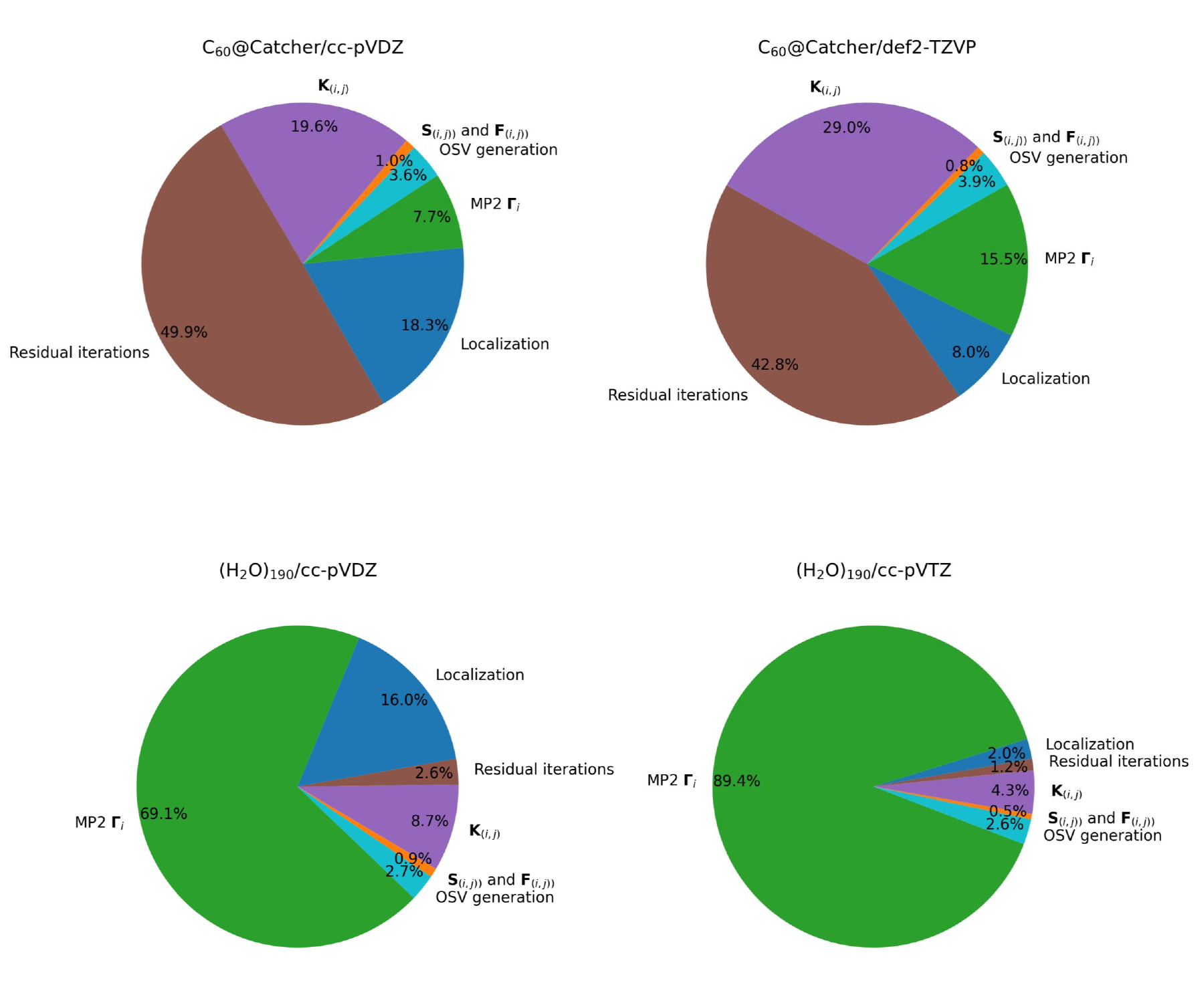}
\centering
\caption{Percentage breakdown of total wall time for the GPU-accelerated MBE(3)-OSV-MP2 calculations on C$_{60}$@Catcher and (H$_2$O)$_{190}$ with various basis sets.}
\label{fig:pie_large}
\end{figure}

\begin{table}[h]
\centering
\begin{tabular}{lcc}
\hline\hline
(H$_2$O)$_{190}$/cc-pVDZ & DLPNO-MP2 & MBE(3)-OSV-MP2 \\
\hline
Hardware & 64 $\times$ 8358P cores & 1 $\times$ A800 (80GB) \\
$E_c$ / Hartree & -39.96629748 & -39.98306756 \\
Localization time / s & 1203.8 & 86.3 \\
MP2 time (w/o loc.) / s & 16756.4 & 3960.6 \\
Total MP2 / s & 17960.2 & 4046.9 \\
\hline\hline
\end{tabular}
\caption{Computational performance comparison for (H$_2$O)$_{190}$/cc-pVDZ between CPU-parallelized DLPNO-MP2 (64 cores, Intel Xeon Platinum 8358P @ 2.60 GHz) in ORCA 6.1.1\cite{ORCA6,pinski2020domain} and GPU-accelerated MBE(3)-OSV-MP2 (single NVIDIA A800, 80 GB).}
\label{tab:vs_dlpno}
\end{table}

\begin{figure}[H]
\includegraphics[width=\linewidth]{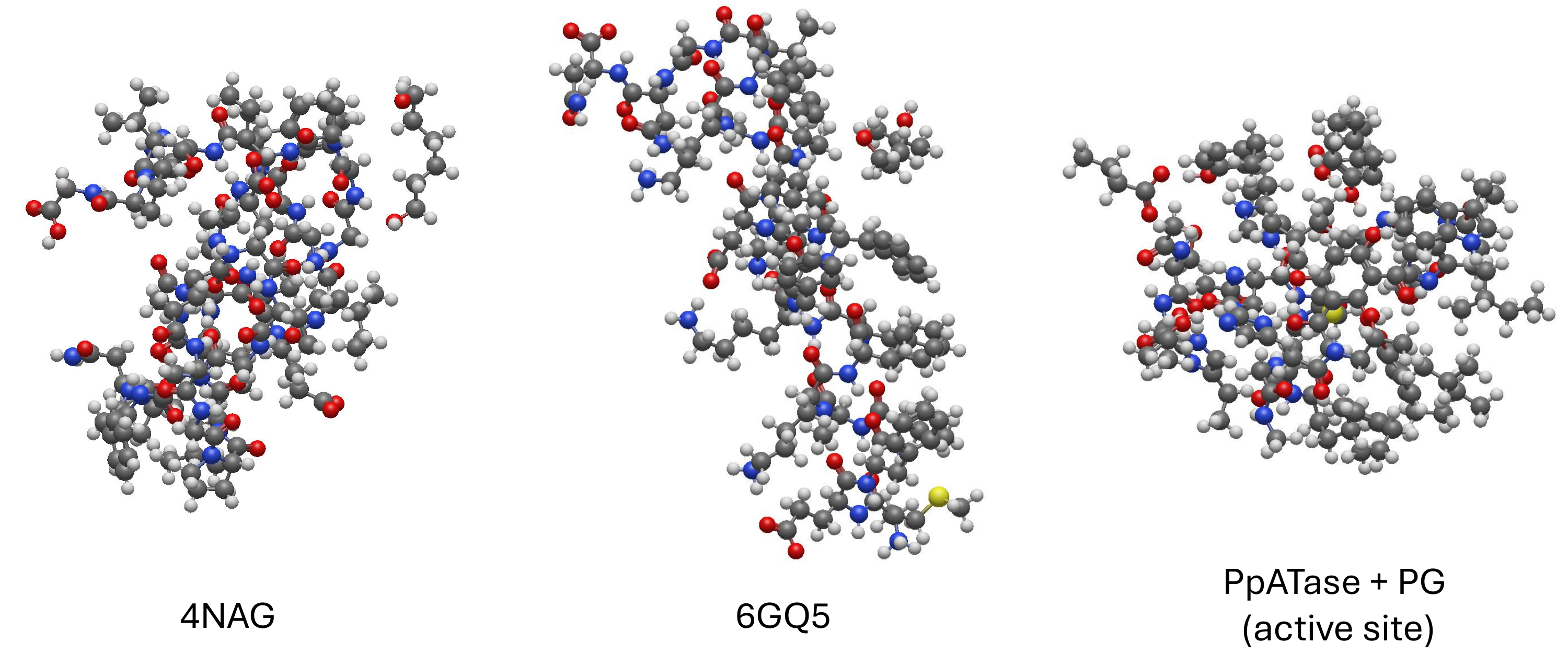}
\centering
\caption{Protein structures simulated in this work. The structures of 4NAG and 6GQ5 were obtained from Ref. \citenum{feng2025efficient}, while the active site configuration of PpATase + PG was sourced from Ref. \citenum{sheng2019mechanism}.}
\label{fig:proteins}
\end{figure}

\section{Algorithms}
\begin{table}[H]
\centering
\caption{Custom Kernels developed in this work}
\label{tab:gpu_kernels}
\begin{tabular}{ll}
\toprule
\toprule
\textbf{Steps} & \textbf{Kernels} \\
\midrule
$\mathbf{\Gamma}_{i}$ generation
    & sparseHalfTransKernel \\
    
\midrule
\multirow{5}{*}{Localization} 
    & pijaLowdinKernel \\
    & computeRotAnglesKernel \\
    & reduceAijBij1BlockKernel \\
    & rotateUKernel \\
    & rotatePijaRowKernel \\
    
\midrule
\multirow{2}{*}{OSV generation} 
    & normalizeVec1BlockKernel \\
    & batchedDnorm2Kernel \\
    & findMaxValue1BlockKernel \\
\midrule
OSV S/F
    & osvSFKernel \\
    
\midrule
\multirow{4}{*}{OSV K} 
    & imujpKernel \\
    & closeOsvKmatKernel \\
    & remoteOsvKmatKernel \\
    
\midrule
\multirow{5}{*}{OSV Residual} 
    & closeOsvPreconKernel \\
    & remoteOsvPreconKernel \\
    & closeResidualKernel \\
    & remoteResidualKernel \\
    & clusterResidualKernel \\
\bottomrule
\bottomrule
\end{tabular}
\end{table}

\section{Coordinates}

\begin{table}[H]
\footnotesize
\setlength{\tabcolsep}{6pt} 
\renewcommand{\arraystretch}{0.95} 
\begin{minipage}{0.3\textwidth}
\begin{tabular}{lrrr}
\multicolumn{4}{l}{120}\\ 
\multicolumn{4}{l}{w40 optmized with ChargeNN}\\ 
O & 6.195 & 6.913 & 7.494 \\ 
H & 5.771 & 6.046 & 7.335 \\ 
H & 5.530 & 7.589 & 7.269 \\ 
O & 4.614 & 9.046 & 7.007 \\ 
H & 4.747 & 9.412 & 6.114 \\ 
H & 3.673 & 8.795 & 7.077 \\ 
O & 5.085 & 7.063 & 3.937 \\ 
H & 5.162 & 7.986 & 4.238 \\ 
H & 4.826 & 7.124 & 2.992 \\ 
O & 4.857 & 4.626 & 7.051 \\ 
H & 3.894 & 4.527 & 6.904 \\ 
H & 5.258 & 4.390 & 6.191 \\ 
O & 2.445 & 6.683 & 4.992 \\ 
H & 3.365 & 6.791 & 4.700 \\ 
H & 2.297 & 7.287 & 5.740 \\ 
O & 5.815 & 4.413 & 4.563 \\ 
H & 6.782 & 4.344 & 4.449 \\ 
H & 5.557 & 5.300 & 4.260 \\ 
O & 5.274 & 9.741 & 4.467 \\ 
H & 5.219 & 9.957 & 3.522 \\ 
H & 6.128 & 10.112 & 4.765 \\ 
O & 8.429 & 8.072 & 6.425 \\ 
H & 8.768 & 8.482 & 7.246 \\ 
H & 7.613 & 7.604 & 6.690 \\ 
O & 2.134 & 8.074 & 7.371 \\ 
H & 2.342 & 7.454 & 8.091 \\ 
H & 1.734 & 8.836 & 7.827 \\ 
O & 8.404 & 4.060 & 3.736 \\ 
H & 8.375 & 4.943 & 3.332 \\ 
H & 7.949 & 3.470 & 3.104 \\ 
O & 7.681 & 6.576 & 2.900 \\ 
H & 8.360 & 7.219 & 3.158 \\ 
H & 6.840 & 6.872 & 3.284 \\ 
O & 9.613 & 8.200 & 3.931 \\ 
H & 10.542 & 7.959 & 3.907 \\ 
H & 9.324 & 8.128 & 4.857 \\ 
O & 2.800 & 6.772 & 9.709 \\ 
H & 2.986 & 7.612 & 10.163 \\ 
H & 3.399 & 6.118 & 10.113 \\ 
O & 7.095 & 7.596 & 9.943 \\ 
H & 6.689 & 7.198 & 9.147 \\ 
H & 6.519 & 7.390 & 10.699 \\ 
O & 4.351 & 4.945 & 11.063 \\ 
H & 4.899 & 4.400 & 10.467 \\ 
H & 3.563 & 4.377 & 11.170 \\ 
O & 5.820 & 9.974 & 9.279 \\ 
H & 6.355 & 9.203 & 9.535 \\ 
H & 5.389 & 9.750 & 8.433 \\ 
O & 8.142 & 11.043 & 8.080 \\ 
H & 7.316 & 10.902 & 8.572 \\ 
H & 8.802 & 10.454 & 8.477 \\ 
O & 5.132 & 3.107 & 9.248 \\ 
\end{tabular}
\end{minipage}\hfill
\hspace{0.02\textwidth}
\begin{minipage}{0.3\textwidth}
\begin{tabular}{lrrr}
H & 5.088 & 3.640 & 8.430 \\ 
H & 5.075 & 2.178 & 8.942 \\ 
O & 7.645 & 10.542 & 5.463 \\ 
H & 7.975 & 9.665 & 5.720 \\ 
H & 7.769 & 11.077 & 6.266 \\ 
O & 2.230 & 4.389 & 6.534 \\ 
H & 1.650 & 4.543 & 7.296 \\ 
H & 2.138 & 5.145 & 5.930 \\ 
O & 4.609 & 7.556 & 1.372 \\ 
H & 3.669 & 7.558 & 1.101 \\ 
H & 4.925 & 8.472 & 1.307 \\ 
O & 9.883 & 4.022 & 6.085 \\ 
H & 9.447 & 4.025 & 5.216 \\ 
H & 9.607 & 3.199 & 6.522 \\ 
O & 6.467 & 3.013 & 2.139 \\ 
H & 5.892 & 3.301 & 2.862 \\ 
H & 6.393 & 3.691 & 1.442 \\ 
O & 6.330 & 5.392 & 0.732 \\ 
H & 5.583 & 6.008 & 0.742 \\ 
H & 7.009 & 5.801 & 1.286 \\ 
O & 0.892 & 4.958 & 8.897 \\ 
H & 0.000 & 5.101 & 9.215 \\ 
H & 1.442 & 5.688 & 9.232 \\ 
O & 7.922 & 2.844 & 9.517 \\ 
H & 6.955 & 2.926 & 9.540 \\ 
H & 8.292 & 3.707 & 9.781 \\ 
O & 8.151 & 10.315 & 2.788 \\ 
H & 8.757 & 9.568 & 2.894 \\ 
H & 8.043 & 10.675 & 3.684 \\ 
O & 6.726 & 0.916 & 6.050 \\ 
H & 6.653 & 1.319 & 5.182 \\ 
H & 7.569 & 1.212 & 6.435 \\ 
O & 10.149 & 6.209 & 7.626 \\ 
H & 9.563 & 6.690 & 7.025 \\ 
H & 10.264 & 5.340 & 7.195 \\ 
O & 3.741 & 9.110 & 10.844 \\ 
H & 4.221 & 8.566 & 11.487 \\ 
H & 4.435 & 9.568 & 10.331 \\ 
O & 4.925 & 0.732 & 8.066 \\ 
H & 4.330 & 0.000 & 7.906 \\ 
H & 5.571 & 0.738 & 7.332 \\ 
O & 2.484 & 3.168 & 10.380 \\ 
H & 1.875 & 3.583 & 9.751 \\ 
H & 3.249 & 2.844 & 9.878 \\ 
O & 5.655 & 10.128 & 1.776 \\ 
H & 6.611 & 10.171 & 1.979 \\ 
H & 5.452 & 10.919 & 1.273 \\ 
O & 1.565 & 9.950 & 9.288 \\ 
H & 1.157 & 10.780 & 9.536 \\ 
H & 2.264 & 9.767 & 9.938 \\ 
O & 8.870 & 5.391 & 9.938 \\ 
H & 9.437 & 5.576 & 9.169 \\ 
H & 8.269 & 6.141 & 10.032 \\ 
O & 9.435 & 8.665 & 8.892 \\ 
\end{tabular}
\end{minipage}\hfill
\hspace{0.02\textwidth}
\begin{minipage}{0.3\textwidth}
\begin{tabular}{lrrr}
H & 10.045 & 7.931 & 8.717 \\ 
H & 8.752 & 8.310 & 9.485 \\ 
O & 0.884 & 6.106 & 2.806 \\ 
H & 0.686 & 5.171 & 2.863 \\ 
H & 1.377 & 6.340 & 3.612 \\ 
O & 1.984 & 7.541 & 0.779 \\ 
H & 1.537 & 7.072 & 1.510 \\ 
H & 1.440 & 7.424 & 0.000 \\ 
O & 9.001 & 1.735 & 7.358 \\ 
H & 9.669 & 1.119 & 7.661 \\ 
H & 8.608 & 2.132 & 8.165 \\ 
O & 5.370 & 7.186 & 12.056 \\ 
H & 4.996 & 6.320 & 11.794 \\ 
H & 5.699 & 7.079 & 12.951 \\ 
\multicolumn{4}{l}{240}\\ 
\multicolumn{4}{l}{w80 optmized with ChargeNN}\\ 
O & 7.663 & 6.995 & 9.251 \\ 
H & 7.393 & 6.106 & 8.952 \\ 
H & 6.921 & 7.599 & 9.062 \\ 
O & 5.949 & 9.046 & 8.814 \\ 
H & 6.170 & 9.394 & 7.929 \\ 
H & 5.066 & 8.633 & 8.742 \\ 
O & 6.140 & 7.515 & 5.431 \\ 
H & 6.398 & 8.352 & 5.863 \\ 
H & 5.449 & 7.780 & 4.789 \\ 
O & 6.881 & 4.512 & 8.460 \\ 
H & 6.120 & 4.825 & 7.937 \\ 
H & 7.526 & 4.176 & 7.810 \\ 
O & 4.807 & 5.788 & 7.169 \\ 
H & 5.230 & 6.244 & 6.423 \\ 
H & 4.397 & 6.488 & 7.710 \\ 
O & 7.690 & 5.331 & 5.796 \\ 
H & 8.617 & 5.595 & 5.934 \\ 
H & 7.186 & 6.149 & 5.644 \\ 
O & 6.805 & 9.918 & 6.416 \\ 
H & 6.456 & 10.787 & 6.141 \\ 
H & 7.766 & 10.054 & 6.502 \\ 
O & 9.688 & 8.672 & 8.838 \\ 
H & 9.468 & 9.390 & 9.464 \\ 
H & 8.971 & 8.013 & 8.914 \\ 
O & 3.683 & 7.674 & 8.753 \\ 
H & 3.787 & 7.328 & 9.660 \\ 
H & 2.841 & 8.157 & 8.718 \\ 
O & 10.300 & 6.102 & 6.077 \\ 
H & 10.215 & 7.014 & 5.745 \\ 
H & 10.765 & 5.618 & 5.368 \\ 
O & 7.925 & 7.539 & 3.271 \\ 
H & 8.743 & 7.875 & 3.673 \\ 
H & 7.246 & 7.618 & 3.963 \\ 
O & 9.950 & 8.502 & 4.856 \\ 
H & 10.772 & 8.767 & 4.402 \\ 
H & 9.765 & 9.180 & 5.526 \\ 
O & 4.343 & 6.584 & 11.158 \\ 
H & 3.733 & 6.952 & 11.825 \\ 
\end{tabular}
\end{minipage}
\end{table}

\begin{table}[H]
\footnotesize
\setlength{\tabcolsep}{6pt} 
\renewcommand{\arraystretch}{0.95} 
\begin{minipage}{0.3\textwidth}
\begin{tabular}{lrrr}
H & 5.101 & 6.206 & 11.642 \\ 
O & 8.190 & 7.343 & 11.868 \\ 
H & 8.030 & 7.323 & 10.907 \\ 
H & 7.538 & 6.722 & 12.238 \\ 
O & 6.416 & 5.337 & 12.439 \\ 
H & 6.692 & 4.574 & 11.896 \\ 
H & 6.695 & 5.138 & 13.349 \\ 
O & 6.449 & 10.667 & 10.880 \\ 
H & 6.266 & 10.331 & 11.777 \\ 
H & 6.161 & 9.971 & 10.259 \\ 
O & 9.153 & 10.736 & 10.439 \\ 
H & 8.223 & 10.676 & 10.714 \\ 
H & 9.303 & 11.691 & 10.277 \\ 
O & 7.233 & 3.285 & 10.820 \\ 
H & 7.095 & 3.658 & 9.924 \\ 
H & 7.366 & 2.322 & 10.698 \\ 
O & 9.457 & 10.357 & 6.821 \\ 
H & 9.643 & 9.709 & 7.528 \\ 
H & 9.732 & 11.213 & 7.203 \\ 
O & 2.814 & 3.995 & 7.673 \\ 
H & 2.870 & 4.009 & 8.646 \\ 
H & 3.602 & 4.475 & 7.359 \\ 
O & 4.569 & 8.555 & 3.627 \\ 
H & 3.603 & 8.706 & 3.648 \\ 
H & 4.954 & 9.374 & 3.251 \\ 
O & 12.111 & 5.603 & 8.052 \\ 
H & 11.405 & 5.812 & 7.410 \\ 
H & 11.978 & 4.675 & 8.297 \\ 
O & 6.915 & 3.454 & 4.033 \\ 
H & 7.059 & 4.102 & 4.751 \\ 
H & 6.948 & 3.975 & 3.210 \\ 
O & 7.045 & 5.283 & 2.015 \\ 
H & 6.180 & 5.633 & 1.732 \\ 
H & 7.493 & 6.047 & 2.426 \\ 
O & 2.950 & 4.398 & 10.357 \\ 
H & 2.103 & 4.643 & 10.762 \\ 
H & 3.551 & 5.136 & 10.566 \\ 
O & 9.990 & 3.538 & 10.925 \\ 
H & 9.019 & 3.455 & 10.881 \\ 
H & 10.190 & 4.460 & 11.177 \\ 
O & 10.506 & 11.838 & 4.780 \\ 
H & 11.107 & 11.321 & 4.223 \\ 
H & 10.078 & 11.198 & 5.366 \\ 
O & 8.923 & 2.912 & 7.364 \\ 
H & 9.113 & 2.922 & 6.408 \\ 
H & 9.776 & 2.917 & 7.822 \\ 
O & 11.814 & 7.520 & 10.008 \\ 
H & 11.071 & 7.887 & 9.495 \\ 
H & 12.214 & 6.868 & 9.413 \\ 
O & 6.178 & 12.550 & 8.745 \\ 
H & 6.429 & 13.343 & 9.248 \\ 
H & 6.318 & 11.801 & 9.346 \\ 
O & 7.007 & 0.958 & 7.697 \\ 
H & 6.738 & 0.730 & 6.786 \\ 
H & 7.631 & 1.695 & 7.586 \\ 
O & 4.572 & 2.274 & 11.101 \\ 
H & 3.945 & 2.979 & 10.868 \\ 
\end{tabular}
\end{minipage}\hfill
\hspace{0.02\textwidth}
\begin{minipage}{0.3\textwidth}
\begin{tabular}{lrrr}
H & 5.462 & 2.654 & 11.083 \\ 
O & 7.746 & 11.934 & 3.841 \\ 
H & 8.689 & 12.027 & 4.036 \\ 
H & 7.257 & 12.312 & 4.591 \\ 
O & 3.448 & 12.374 & 8.279 \\ 
H & 3.274 & 12.542 & 7.340 \\ 
H & 4.411 & 12.417 & 8.390 \\ 
O & 10.549 & 5.998 & 11.889 \\ 
H & 11.167 & 6.501 & 11.331 \\ 
H & 9.741 & 6.539 & 11.939 \\ 
O & 9.841 & 9.070 & 13.292 \\ 
H & 10.532 & 9.469 & 12.736 \\ 
H & 9.326 & 8.475 & 12.721 \\ 
O & 1.921 & 8.928 & 3.434 \\ 
H & 1.824 & 8.182 & 2.818 \\ 
H & 1.597 & 8.617 & 4.301 \\ 
O & 3.161 & 11.430 & 3.390 \\ 
H & 2.620 & 10.625 & 3.317 \\ 
H & 4.006 & 11.256 & 2.936 \\ 
O & 11.400 & 2.956 & 8.706 \\ 
H & 11.904 & 2.179 & 8.959 \\ 
H & 10.895 & 3.190 & 9.510 \\ 
O & 6.436 & 7.299 & 15.448 \\ 
H & 6.881 & 6.458 & 15.258 \\ 
H & 7.093 & 8.009 & 15.353 \\ 
O & 1.267 & 8.991 & 8.534 \\ 
H & 0.747 & 9.776 & 8.340 \\ 
H & 0.640 & 8.273 & 8.755 \\ 
O & 11.643 & 9.999 & 11.288 \\ 
H & 10.767 & 10.286 & 10.973 \\ 
H & 11.822 & 9.123 & 10.902 \\ 
O & 11.585 & 4.789 & 4.097 \\ 
H & 11.600 & 4.566 & 3.144 \\ 
H & 12.069 & 4.084 & 4.567 \\ 
O & 6.748 & 8.678 & 0.972 \\ 
H & 6.036 & 8.025 & 0.895 \\ 
H & 7.351 & 8.333 & 1.648 \\ 
O & 6.313 & 0.897 & 5.071 \\ 
H & 5.358 & 0.781 & 4.931 \\ 
H & 6.560 & 1.728 & 4.636 \\ 
O & 5.825 & 10.329 & 13.507 \\ 
H & 6.606 & 10.116 & 14.047 \\ 
H & 5.133 & 9.707 & 13.789 \\ 
O & 3.419 & 12.942 & 5.558 \\ 
H & 3.103 & 13.794 & 5.254 \\ 
H & 3.281 & 12.327 & 4.807 \\ 
O & 12.712 & 7.374 & 4.150 \\ 
H & 12.360 & 6.491 & 3.957 \\ 
H & 13.256 & 7.269 & 4.952 \\ 
O & 4.493 & 0.587 & 8.878 \\ 
H & 5.362 & 0.718 & 8.464 \\ 
H & 4.476 & 1.108 & 9.700 \\ 
O & 0.737 & 5.263 & 6.641 \\ 
H & 1.497 & 4.767 & 7.011 \\ 
H & 0.058 & 4.613 & 6.446 \\ 
O & 4.134 & 8.437 & 14.664 \\ 
H & 3.764 & 8.769 & 15.484 \\ 
\end{tabular}
\end{minipage}\hfill
\hspace{0.02\textwidth}
\begin{minipage}{0.3\textwidth}
\begin{tabular}{lrrr}
H & 4.920 & 7.913 & 14.917 \\ 
O & 6.089 & 12.499 & 6.006 \\ 
H & 6.188 & 12.661 & 6.960 \\ 
H & 5.151 & 12.657 & 5.796 \\ 
O & 2.568 & 1.282 & 7.118 \\ 
H & 3.173 & 0.890 & 7.781 \\ 
H & 2.578 & 2.237 & 7.291 \\ 
O & 5.675 & 10.712 & 2.482 \\ 
H & 6.410 & 11.187 & 2.915 \\ 
H & 6.083 & 10.166 & 1.783 \\ 
O & 1.902 & 4.994 & 3.957 \\ 
H & 2.685 & 4.405 & 3.974 \\ 
H & 1.633 & 5.138 & 4.873 \\ 
O & 9.554 & 2.842 & 4.677 \\ 
H & 8.634 & 2.877 & 4.367 \\ 
H & 10.015 & 3.587 & 4.260 \\ 
O & 2.105 & 9.673 & 11.137 \\ 
H & 2.322 & 10.626 & 11.185 \\ 
H & 1.845 & 9.503 & 10.221 \\ 
O & 7.478 & 0.664 & 10.338 \\ 
H & 6.885 & 0.000 & 10.696 \\ 
H & 7.362 & 0.646 & 9.368 \\ 
O & 7.549 & 4.804 & 14.863 \\ 
H & 8.443 & 4.579 & 14.538 \\ 
H & 7.292 & 4.086 & 15.444 \\ 
O & 9.297 & 13.349 & 9.930 \\ 
H & 9.691 & 13.375 & 9.041 \\ 
H & 8.529 & 13.936 & 9.955 \\ 
O & 4.778 & 6.756 & 1.588 \\ 
H & 3.848 & 6.578 & 1.392 \\ 
H & 4.744 & 7.321 & 2.382 \\ 
O & 11.914 & 9.819 & 3.429 \\ 
H & 12.416 & 9.012 & 3.658 \\ 
H & 11.837 & 9.826 & 2.473 \\ 
O & 7.955 & 9.587 & 15.081 \\ 
H & 8.377 & 10.079 & 15.786 \\ 
H & 8.663 & 9.384 & 14.434 \\ 
O & 0.000 & 6.662 & 9.002 \\ 
H & 0.268 & 6.221 & 9.825 \\ 
H & 0.225 & 6.082 & 8.257 \\ 
O & 10.634 & 12.685 & 7.566 \\ 
H & 10.831 & 12.768 & 6.621 \\ 
H & 11.440 & 12.365 & 8.014 \\ 
O & 4.069 & 3.381 & 4.015 \\ 
H & 3.857 & 2.456 & 4.233 \\ 
H & 5.035 & 3.446 & 4.005 \\ 
O & 2.375 & 7.579 & 12.741 \\ 
H & 2.219 & 8.395 & 12.218 \\ 
H & 2.926 & 7.853 & 13.494 \\ 
O & 14.054 & 6.757 & 6.461 \\ 
H & 13.411 & 6.360 & 7.073 \\ 
H & 14.927 & 6.503 & 6.762 \\ 
O & 1.994 & 6.627 & 1.859 \\ 
H & 1.326 & 6.345 & 1.235 \\ 
H & 1.956 & 5.997 & 2.612 \\ 
O & 12.045 & 2.712 & 5.890 \\ 
H & 11.203 & 2.421 & 5.503 \\ 
\end{tabular}
\end{minipage}
\end{table}

\begin{table}[H]
\footnotesize
\setlength{\tabcolsep}{6pt} 
\renewcommand{\arraystretch}{0.95} 
\begin{minipage}{0.3\textwidth}
\begin{tabular}{lrrr}
H & 11.892 & 2.840 & 6.834 \\ 
O & 5.370 & 12.864 & 12.059 \\ 
H & 5.293 & 12.587 & 12.977 \\ 
H & 5.738 & 12.081 & 11.611 \\ 
O & 11.424 & 4.143 & 1.508 \\ 
H & 10.506 & 4.064 & 1.182 \\ 
H & 11.994 & 4.024 & 0.748 \\ 
O & 1.292 & 7.910 & 5.905 \\ 
H & 1.422 & 8.328 & 6.768 \\ 
H & 1.072 & 6.981 & 6.077 \\ 
O & 8.890 & 3.895 & 0.575 \\ 
H & 8.238 & 4.425 & 1.068 \\ 
H & 8.393 & 3.313 & 0.000 \\ 
O & 12.620 & 11.571 & 9.134 \\ 
H & 13.541 & 11.779 & 9.295 \\ 
H & 12.308 & 11.072 & 9.909 \\ 
O & 3.581 & 0.780 & 4.726 \\ 
H & 3.126 & 0.026 & 4.352 \\ 
H & 3.188 & 0.929 & 5.613 \\ 
O & 9.882 & 4.049 & 13.727 \\ 
H & 10.299 & 4.858 & 13.387 \\ 
H & 9.883 & 3.456 & 12.961 \\ 
O & 0.815 & 5.707 & 11.500 \\ 
H & 1.317 & 6.407 & 11.960 \\ 
H & 0.164 & 5.385 & 12.125 \\ 
O & 2.795 & 12.273 & 10.950 \\ 
H & 3.617 & 12.576 & 11.361 \\ 
H & 2.908 & 12.376 & 9.987 \\ 
O & 6.774 & 14.593 & 10.456 \\ 
H & 6.300 & 14.097 & 11.150 \\ 
H & 6.441 & 15.490 & 10.481 \\ 
\multicolumn{4}{l}{300}\\ 
\multicolumn{4}{l}{w100 optmized with ChargeNN}\\ 
O & 11.836 & 9.813 & 15.357 \\ 
H & 11.611 & 10.408 & 16.072 \\ 
H & 11.512 & 10.217 & 14.534 \\ 
O & 4.065 & 14.160 & 11.264 \\ 
H & 4.744 & 13.625 & 11.707 \\ 
H & 4.422 & 14.411 & 10.397 \\ 
O & 8.556 & 12.358 & 13.669 \\ 
H & 8.065 & 11.619 & 14.074 \\ 
H & 9.390 & 11.945 & 13.387 \\ 
O & 3.183 & 12.678 & 3.389 \\ 
H & 2.920 & 11.763 & 3.574 \\ 
H & 4.089 & 12.639 & 3.036 \\ 
O & 2.154 & 10.686 & 11.857 \\ 
H & 1.828 & 11.598 & 11.838 \\ 
H & 2.201 & 10.465 & 10.912 \\ 
O & 11.316 & 12.876 & 4.352 \\ 
H & 12.268 & 12.932 & 4.523 \\ 
H & 11.054 & 11.993 & 4.668 \\ 
O & 11.115 & 6.059 & 0.895 \\ 
H & 10.482 & 6.697 & 1.280 \\ 
H & 10.765 & 5.856 & 0.026 \\ 
O & 5.691 & 12.063 & 8.066 \\ 
H & 6.186 & 11.851 & 7.252 \\ 
H & 5.130 & 11.288 & 8.255 \\ 
\end{tabular}
\end{minipage}\hfill
\hspace{0.02\textwidth}
\begin{minipage}{0.3\textwidth}
\begin{tabular}{lrrr}
O & 10.808 & 7.301 & 15.050 \\ 
H & 11.120 & 7.106 & 14.148 \\ 
H & 11.297 & 8.102 & 15.305 \\ 
O & 9.519 & 12.152 & 6.996 \\ 
H & 9.739 & 11.916 & 7.917 \\ 
H & 9.844 & 13.061 & 6.854 \\ 
O & 4.289 & 11.053 & 13.546 \\ 
H & 3.488 & 10.931 & 13.009 \\ 
H & 4.570 & 10.181 & 13.865 \\ 
O & 13.912 & 5.226 & 12.569 \\ 
H & 13.089 & 5.712 & 12.755 \\ 
H & 14.498 & 5.864 & 12.133 \\ 
O & 9.195 & 9.587 & 3.615 \\ 
H & 9.905 & 9.815 & 4.241 \\ 
H & 8.415 & 9.361 & 4.150 \\ 
O & 8.531 & 9.196 & 9.478 \\ 
H & 8.024 & 8.476 & 9.064 \\ 
H & 7.923 & 9.946 & 9.620 \\ 
O & 8.032 & 6.612 & 5.705 \\ 
H & 8.926 & 6.948 & 5.908 \\ 
H & 7.561 & 7.343 & 5.273 \\ 
O & 8.968 & 8.933 & 12.134 \\ 
H & 8.870 & 9.150 & 11.189 \\ 
H & 8.463 & 8.106 & 12.233 \\ 
O & 7.197 & 7.101 & 8.226 \\ 
H & 6.318 & 7.391 & 7.922 \\ 
H & 7.669 & 6.883 & 7.400 \\ 
O & 7.273 & 15.650 & 10.291 \\ 
H & 7.626 & 15.258 & 11.107 \\ 
H & 7.054 & 16.561 & 10.501 \\ 
O & 4.937 & 8.277 & 11.081 \\ 
H & 4.296 & 8.322 & 11.816 \\ 
H & 5.764 & 7.938 & 11.468 \\ 
O & 0.170 & 8.618 & 9.355 \\ 
H & 0.447 & 7.958 & 10.016 \\ 
H & 0.187 & 8.168 & 8.496 \\ 
O & 5.093 & 9.689 & 2.892 \\ 
H & 4.166 & 9.652 & 3.185 \\ 
H & 5.302 & 10.644 & 2.829 \\ 
O & 13.934 & 12.508 & 5.297 \\ 
H & 13.906 & 11.534 & 5.370 \\ 
H & 14.709 & 12.723 & 4.778 \\ 
O & 5.091 & 4.287 & 11.010 \\ 
H & 4.447 & 4.896 & 10.603 \\ 
H & 5.953 & 4.529 & 10.631 \\ 
O & 5.002 & 3.534 & 4.361 \\ 
H & 5.096 & 2.747 & 3.825 \\ 
H & 5.897 & 3.801 & 4.631 \\ 
O & 11.846 & 4.638 & 9.097 \\ 
H & 12.350 & 3.975 & 9.598 \\ 
H & 11.100 & 4.851 & 9.686 \\ 
O & 15.114 & 7.155 & 10.963 \\ 
H & 14.341 & 7.750 & 10.914 \\ 
H & 15.880 & 7.712 & 11.108 \\ 
O & 6.107 & 12.587 & 12.335 \\ 
H & 6.892 & 12.614 & 12.901 \\ 
H & 5.454 & 12.030 & 12.811 \\ 
\end{tabular}
\end{minipage}\hfill
\hspace{0.02\textwidth}
\begin{minipage}{0.3\textwidth}
\begin{tabular}{lrrr}
O & 9.036 & 13.746 & 2.962 \\ 
H & 9.890 & 13.542 & 3.374 \\ 
H & 8.841 & 13.016 & 2.355 \\ 
O & 6.781 & 8.987 & 4.904 \\ 
H & 6.898 & 9.864 & 5.327 \\ 
H & 6.175 & 9.133 & 4.152 \\ 
O & 7.608 & 1.954 & 13.632 \\ 
H & 6.795 & 2.463 & 13.476 \\ 
H & 7.508 & 1.549 & 14.496 \\ 
O & 11.962 & 11.239 & 10.573 \\ 
H & 11.102 & 11.650 & 10.348 \\ 
H & 12.601 & 11.963 & 10.459 \\ 
O & 12.433 & 4.459 & 6.367 \\ 
H & 12.027 & 3.727 & 5.870 \\ 
H & 12.057 & 4.424 & 7.263 \\ 
O & 12.087 & 7.317 & 8.618 \\ 
H & 11.402 & 7.744 & 8.063 \\ 
H & 11.844 & 6.379 & 8.678 \\ 
O & 7.497 & 5.180 & 10.054 \\ 
H & 7.386 & 5.844 & 9.343 \\ 
H & 7.610 & 4.318 & 9.600 \\ 
O & 2.781 & 14.645 & 5.239 \\ 
H & 2.235 & 15.284 & 4.778 \\ 
H & 2.915 & 13.900 & 4.623 \\ 
O & 14.842 & 5.765 & 6.321 \\ 
H & 14.010 & 5.266 & 6.236 \\ 
H & 15.169 & 5.520 & 7.200 \\ 
O & 2.465 & 14.347 & 8.007 \\ 
H & 2.444 & 14.384 & 7.039 \\ 
H & 3.377 & 14.553 & 8.267 \\ 
O & 13.560 & 3.291 & 10.772 \\ 
H & 13.692 & 3.880 & 11.546 \\ 
H & 14.187 & 3.633 & 10.112 \\ 
O & 11.031 & 10.316 & 5.497 \\ 
H & 11.988 & 10.135 & 5.540 \\ 
H & 10.758 & 10.679 & 6.364 \\ 
O & 13.268 & 13.601 & 10.229 \\ 
H & 12.514 & 14.148 & 10.531 \\ 
H & 14.052 & 14.005 & 10.602 \\ 
O & 4.502 & 9.778 & 8.833 \\ 
H & 4.705 & 9.311 & 9.664 \\ 
H & 3.564 & 10.046 & 8.897 \\ 
O & 14.665 & 5.205 & 9.055 \\ 
H & 14.938 & 5.897 & 9.683 \\ 
H & 13.706 & 5.299 & 8.944 \\ 
O & 0.626 & 7.388 & 6.897 \\ 
H & 1.472 & 6.909 & 7.004 \\ 
H & 0.000 & 6.763 & 6.530 \\ 
O & 3.046 & 6.251 & 7.211 \\ 
H & 3.203 & 6.136 & 8.168 \\ 
H & 3.753 & 6.853 & 6.916 \\ 
O & 7.147 & 6.947 & 12.099 \\ 
H & 7.264 & 6.248 & 11.432 \\ 
H & 6.884 & 6.514 & 12.932 \\ 
O & 12.838 & 8.637 & 10.817 \\ 
H & 12.588 & 9.575 & 10.797 \\ 
H & 12.509 & 8.254 & 9.979 \\ 
\end{tabular}
\end{minipage}
\end{table}

\begin{table}[H]
\footnotesize
\setlength{\tabcolsep}{6pt} 
\renewcommand{\arraystretch}{0.95} 
\begin{minipage}{0.3\textwidth}
\begin{tabular}{lrrr}
O & 9.576 & 14.923 & 8.932 \\ 
H & 9.818 & 15.002 & 7.990 \\ 
H & 8.684 & 15.280 & 9.061 \\ 
O & 9.721 & 12.324 & 9.754 \\ 
H & 8.809 & 12.047 & 9.972 \\ 
H & 9.695 & 13.289 & 9.606 \\ 
O & 4.437 & 1.875 & 9.733 \\ 
H & 4.873 & 1.979 & 8.874 \\ 
H & 4.691 & 2.642 & 10.268 \\ 
O & 3.610 & 6.238 & 9.856 \\ 
H & 2.712 & 6.334 & 10.233 \\ 
H & 4.110 & 7.008 & 10.182 \\ 
O & 9.085 & 0.294 & 11.952 \\ 
H & 8.486 & 0.838 & 12.484 \\ 
H & 9.910 & 0.783 & 11.833 \\ 
O & 11.424 & 2.497 & 4.714 \\ 
H & 11.876 & 1.663 & 4.576 \\ 
H & 11.529 & 3.003 & 3.886 \\ 
O & 13.699 & 6.019 & 1.826 \\ 
H & 12.804 & 6.237 & 1.508 \\ 
H & 14.242 & 5.890 & 1.048 \\ 
O & 8.554 & 8.364 & 16.149 \\ 
H & 8.119 & 7.662 & 16.636 \\ 
H & 9.379 & 7.985 & 15.792 \\ 
O & 13.734 & 9.819 & 5.407 \\ 
H & 13.932 & 9.129 & 4.750 \\ 
H & 13.996 & 9.410 & 6.256 \\ 
O & 7.200 & 11.398 & 5.923 \\ 
H & 7.012 & 12.196 & 5.386 \\ 
H & 8.081 & 11.573 & 6.312 \\ 
O & 11.995 & 3.914 & 2.473 \\ 
H & 12.795 & 4.460 & 2.481 \\ 
H & 11.349 & 4.470 & 2.013 \\ 
O & 5.162 & 8.426 & 14.207 \\ 
H & 5.428 & 7.514 & 14.401 \\ 
H & 5.984 & 8.947 & 14.275 \\ 
O & 5.280 & 3.426 & 13.508 \\ 
H & 5.199 & 3.708 & 12.573 \\ 
H & 4.399 & 3.148 & 13.769 \\ 
O & 6.753 & 9.045 & 0.792 \\ 
H & 6.219 & 8.976 & 0.000 \\ 
H & 6.135 & 9.139 & 1.540 \\ 
O & 5.309 & 15.428 & 6.039 \\ 
H & 5.300 & 15.252 & 6.993 \\ 
H & 4.413 & 15.224 & 5.720 \\ 
O & 11.111 & 14.977 & 11.137 \\ 
H & 10.658 & 15.118 & 10.286 \\ 
H & 10.405 & 14.925 & 11.795 \\ 
O & 0.879 & 4.051 & 11.001 \\ 
H & 0.088 & 3.539 & 11.173 \\ 
H & 1.311 & 3.624 & 10.232 \\ 
O & 8.445 & 0.178 & 9.239 \\ 
H & 7.962 & 1.013 & 9.139 \\ 
H & 8.510 & 0.000 & 10.191 \\ 
O & 14.350 & 7.440 & 4.133 \\ 
H & 14.721 & 6.834 & 4.800 \\ 
H & 14.272 & 6.946 & 3.301 \\ 
\end{tabular}
\end{minipage}\hfill
\hspace{0.02\textwidth}
\begin{minipage}{0.3\textwidth}
\begin{tabular}{lrrr}
O & 6.882 & 13.692 & 4.638 \\ 
H & 7.661 & 14.010 & 4.140 \\ 
H & 6.442 & 14.458 & 5.054 \\ 
O & 10.615 & 14.595 & 6.450 \\ 
H & 10.681 & 14.239 & 5.550 \\ 
H & 11.531 & 14.583 & 6.792 \\ 
O & 8.235 & 11.287 & 1.692 \\ 
H & 7.833 & 10.630 & 1.107 \\ 
H & 8.693 & 10.758 & 2.371 \\ 
O & 1.250 & 6.727 & 11.028 \\ 
H & 1.623 & 7.088 & 11.851 \\ 
H & 1.038 & 5.783 & 11.167 \\ 
O & 9.681 & 2.760 & 6.814 \\ 
H & 10.170 & 2.612 & 5.991 \\ 
H & 10.148 & 2.219 & 7.482 \\ 
O & 10.853 & 1.247 & 8.758 \\ 
H & 11.137 & 1.437 & 9.660 \\ 
H & 10.038 & 0.710 & 8.846 \\ 
O & 8.675 & 14.871 & 12.628 \\ 
H & 8.610 & 13.949 & 12.943 \\ 
H & 8.649 & 15.416 & 13.417 \\ 
O & 5.479 & 2.800 & 7.279 \\ 
H & 4.570 & 3.049 & 7.028 \\ 
H & 6.024 & 3.111 & 6.536 \\ 
O & 7.492 & 9.919 & 14.236 \\ 
H & 7.875 & 9.478 & 15.021 \\ 
H & 7.966 & 9.550 & 13.469 \\ 
O & 1.985 & 2.794 & 8.880 \\ 
H & 2.773 & 2.331 & 9.223 \\ 
H & 2.202 & 3.058 & 7.973 \\ 
O & 11.797 & 6.581 & 4.737 \\ 
H & 12.648 & 6.969 & 4.472 \\ 
H & 12.002 & 5.775 & 5.243 \\ 
O & 5.832 & 12.251 & 2.670 \\ 
H & 6.147 & 12.814 & 3.404 \\ 
H & 6.634 & 12.060 & 2.156 \\ 
O & 2.784 & 8.193 & 12.849 \\ 
H & 2.451 & 9.092 & 12.670 \\ 
H & 3.453 & 8.263 & 13.551 \\ 
O & 1.757 & 9.852 & 6.368 \\ 
H & 1.893 & 10.272 & 7.232 \\ 
H & 1.297 & 9.014 & 6.550 \\ 
O & 5.045 & 8.040 & 6.856 \\ 
H & 5.555 & 8.411 & 6.117 \\ 
H & 4.813 & 8.777 & 7.456 \\ 
O & 11.614 & 1.595 & 11.566 \\ 
H & 12.107 & 0.808 & 11.815 \\ 
H & 12.295 & 2.235 & 11.285 \\ 
O & 13.098 & 14.119 & 7.444 \\ 
H & 13.223 & 13.805 & 8.352 \\ 
H & 13.502 & 13.485 & 6.832 \\ 
O & 7.697 & 2.892 & 8.750 \\ 
H & 6.857 & 2.787 & 8.260 \\ 
H & 8.400 & 2.890 & 8.074 \\ 
O & 1.968 & 10.568 & 9.081 \\ 
H & 1.724 & 11.482 & 9.331 \\ 
H & 1.183 & 9.994 & 9.215 \\ 
\end{tabular}
\end{minipage}\hfill
\hspace{0.02\textwidth}
\begin{minipage}{0.3\textwidth}
\begin{tabular}{lrrr}
O & 1.781 & 12.991 & 10.291 \\ 
H & 2.504 & 13.426 & 10.782 \\ 
H & 1.753 & 13.489 & 9.453 \\ 
O & 10.858 & 9.983 & 8.297 \\ 
H & 11.337 & 10.472 & 8.983 \\ 
H & 10.021 & 9.709 & 8.720 \\ 
O & 7.418 & 3.932 & 5.608 \\ 
H & 7.612 & 4.888 & 5.630 \\ 
H & 8.227 & 3.490 & 5.916 \\ 
O & 11.675 & 6.813 & 12.519 \\ 
H & 12.024 & 7.540 & 11.974 \\ 
H & 11.046 & 6.322 & 11.963 \\ 
O & 9.187 & 7.813 & 1.532 \\ 
H & 8.341 & 8.111 & 1.167 \\ 
H & 9.340 & 8.368 & 2.317 \\ 
O & 5.206 & 14.622 & 8.726 \\ 
H & 5.980 & 14.987 & 9.180 \\ 
H & 5.421 & 13.695 & 8.499 \\ 
O & 10.728 & 10.793 & 13.017 \\ 
H & 11.271 & 10.997 & 12.239 \\ 
H & 10.154 & 10.048 & 12.749 \\ 
O & 9.880 & 3.329 & 12.984 \\ 
H & 10.552 & 2.721 & 12.641 \\ 
H & 9.079 & 2.799 & 13.142 \\ 
O & 14.386 & 8.261 & 7.488 \\ 
H & 13.576 & 8.023 & 7.970 \\ 
H & 14.686 & 7.443 & 7.058 \\ 
O & 8.947 & 5.261 & 14.872 \\ 
H & 9.346 & 4.589 & 14.298 \\ 
H & 9.627 & 5.946 & 15.007 \\ 
O & 3.077 & 3.680 & 6.364 \\ 
H & 3.546 & 3.636 & 5.515 \\ 
H & 3.019 & 4.626 & 6.595 \\ 
O & 10.035 & 5.192 & 11.018 \\ 
H & 9.131 & 5.299 & 10.675 \\ 
H & 9.969 & 4.526 & 11.729 \\ 
O & 10.302 & 7.777 & 6.549 \\ 
H & 10.475 & 8.735 & 6.511 \\ 
H & 10.848 & 7.372 & 5.841 \\ 
O & 7.220 & 11.465 & 10.172 \\ 
H & 6.797 & 11.816 & 10.980 \\ 
H & 6.621 & 11.697 & 9.437 \\ 
O & 2.549 & 10.011 & 3.821 \\ 
H & 1.709 & 9.849 & 3.389 \\ 
H & 2.359 & 9.985 & 4.782 \\ 
O & 6.336 & 5.841 & 14.436 \\ 
H & 7.249 & 5.639 & 14.724 \\ 
H & 5.928 & 4.989 & 14.210 \\ 
\multicolumn{4}{l}{384}\\ 
\multicolumn{4}{l}{w128 optmized with ChargeNN}\\ 
O & 9.711 & 10.825 & 11.117 \\ 
H & 9.372 & 10.049 & 11.597 \\ 
H & 8.953 & 11.401 & 10.908 \\ 
O & 7.893 & 12.696 & 10.323 \\ 
H & 8.065 & 12.953 & 9.399 \\ 
H & 6.973 & 12.363 & 10.371 \\ 
O & 8.123 & 10.640 & 7.273 \\ 
\end{tabular}
\end{minipage}
\end{table}

\begin{table}[H]
\footnotesize
\setlength{\tabcolsep}{6pt} 
\renewcommand{\arraystretch}{0.95} 
\begin{minipage}{0.3\textwidth}
\begin{tabular}{lrrr}
H & 8.370 & 11.504 & 7.657 \\ 
H & 7.558 & 10.872 & 6.509 \\ 
O & 7.235 & 7.400 & 9.871 \\ 
H & 6.739 & 8.211 & 9.670 \\ 
H & 7.823 & 7.299 & 9.101 \\ 
O & 6.033 & 9.626 & 8.798 \\ 
H & 6.712 & 9.949 & 8.186 \\ 
H & 5.876 & 10.350 & 9.440 \\ 
O & 8.848 & 7.954 & 7.743 \\ 
H & 9.819 & 7.862 & 7.808 \\ 
H & 8.665 & 8.903 & 7.658 \\ 
O & 8.744 & 13.222 & 7.805 \\ 
H & 8.452 & 13.743 & 7.035 \\ 
H & 9.710 & 13.364 & 7.831 \\ 
O & 11.704 & 11.815 & 9.557 \\ 
H & 12.034 & 12.510 & 10.156 \\ 
H & 10.945 & 11.420 & 10.026 \\ 
O & 5.548 & 11.484 & 10.620 \\ 
H & 5.780 & 10.852 & 11.330 \\ 
H & 4.584 & 11.612 & 10.670 \\ 
O & 11.580 & 7.979 & 7.659 \\ 
H & 11.805 & 8.848 & 7.279 \\ 
H & 11.827 & 7.314 & 6.990 \\ 
O & 10.286 & 10.297 & 5.640 \\ 
H & 11.136 & 10.382 & 6.110 \\ 
H & 9.594 & 10.372 & 6.321 \\ 
O & 12.365 & 10.507 & 7.297 \\ 
H & 13.170 & 10.946 & 6.964 \\ 
H & 12.132 & 10.986 & 8.117 \\ 
O & 6.451 & 9.615 & 12.307 \\ 
H & 5.968 & 9.509 & 13.147 \\ 
H & 7.375 & 9.373 & 12.491 \\ 
O & 10.739 & 11.608 & 13.537 \\ 
H & 10.390 & 11.329 & 12.673 \\ 
H & 10.700 & 10.825 & 14.123 \\ 
O & 9.005 & 8.835 & 12.880 \\ 
H & 9.025 & 7.957 & 12.450 \\ 
H & 8.963 & 8.663 & 13.836 \\ 
O & 8.661 & 14.481 & 12.104 \\ 
H & 8.167 & 14.253 & 12.912 \\ 
H & 8.399 & 13.796 & 11.456 \\ 
O & 11.160 & 14.450 & 10.701 \\ 
H & 10.303 & 14.491 & 11.153 \\ 
H & 11.426 & 15.367 & 10.503 \\ 
O & 8.933 & 6.410 & 11.706 \\ 
H & 8.365 & 6.653 & 10.951 \\ 
H & 8.691 & 5.491 & 11.940 \\ 
O & 11.415 & 13.635 & 7.505 \\ 
H & 11.640 & 13.030 & 8.234 \\ 
H & 11.849 & 14.484 & 7.715 \\ 
O & 3.906 & 8.054 & 9.312 \\ 
H & 3.765 & 8.140 & 10.273 \\ 
H & 4.667 & 8.621 & 9.086 \\ 
O & 6.749 & 11.629 & 5.198 \\ 
H & 5.904 & 11.223 & 4.950 \\ 
H & 6.494 & 12.563 & 5.330 \\ 
O & 13.764 & 8.586 & 9.298 \\ 
\end{tabular}
\end{minipage}\hfill
\hspace{0.02\textwidth}
\begin{minipage}{0.3\textwidth}
\begin{tabular}{lrrr}
H & 13.040 & 8.675 & 8.648 \\ 
H & 13.716 & 7.662 & 9.598 \\ 
O & 7.925 & 6.915 & 5.397 \\ 
H & 8.142 & 7.246 & 6.287 \\ 
H & 8.451 & 7.462 & 4.789 \\ 
O & 9.129 & 8.702 & 3.745 \\ 
H & 8.646 & 9.263 & 3.112 \\ 
H & 9.610 & 9.297 & 4.349 \\ 
O & 3.374 & 8.007 & 12.047 \\ 
H & 2.435 & 7.813 & 11.885 \\ 
H & 3.394 & 8.863 & 12.513 \\ 
O & 11.531 & 6.361 & 12.193 \\ 
H & 10.597 & 6.437 & 11.915 \\ 
H & 11.510 & 5.789 & 12.983 \\ 
O & 12.444 & 13.473 & 4.904 \\ 
H & 13.027 & 12.703 & 4.787 \\ 
H & 12.068 & 13.396 & 5.800 \\ 
O & 11.016 & 5.011 & 8.761 \\ 
H & 11.210 & 5.321 & 7.857 \\ 
H & 11.718 & 5.366 & 9.324 \\ 
O & 13.632 & 10.253 & 11.431 \\ 
H & 13.310 & 11.155 & 11.254 \\ 
H & 13.607 & 9.769 & 10.584 \\ 
O & 7.057 & 16.495 & 11.113 \\ 
H & 7.569 & 17.277 & 11.395 \\ 
H & 7.636 & 15.744 & 11.335 \\ 
O & 8.364 & 4.187 & 8.942 \\ 
H & 8.121 & 4.144 & 7.996 \\ 
H & 9.264 & 4.560 & 8.966 \\ 
O & 5.747 & 6.499 & 12.066 \\ 
H & 4.899 & 6.972 & 12.130 \\ 
H & 6.299 & 7.023 & 11.459 \\ 
O & 8.661 & 14.447 & 5.375 \\ 
H & 9.163 & 13.749 & 4.917 \\ 
H & 9.194 & 15.263 & 5.347 \\ 
O & 4.815 & 15.179 & 10.055 \\ 
H & 4.865 & 15.279 & 9.088 \\ 
H & 5.554 & 15.675 & 10.439 \\ 
O & 12.952 & 8.364 & 13.241 \\ 
H & 13.057 & 9.074 & 12.582 \\ 
H & 12.361 & 7.696 & 12.845 \\ 
O & 12.999 & 12.848 & 14.168 \\ 
H & 13.289 & 13.091 & 13.271 \\ 
H & 12.187 & 12.324 & 14.005 \\ 
O & 4.037 & 11.666 & 4.761 \\ 
H & 3.917 & 10.738 & 4.492 \\ 
H & 3.243 & 11.892 & 5.278 \\ 
O & 5.452 & 14.006 & 5.077 \\ 
H & 4.790 & 13.346 & 4.801 \\ 
H & 5.886 & 14.354 & 4.280 \\ 
O & 13.293 & 5.991 & 10.214 \\ 
H & 13.966 & 5.499 & 10.710 \\ 
H & 12.615 & 6.206 & 10.883 \\ 
O & 9.321 & 12.770 & 15.554 \\ 
H & 9.828 & 12.200 & 14.943 \\ 
H & 9.938 & 13.480 & 15.786 \\ 
O & 2.847 & 11.491 & 10.815 \\ 
\end{tabular}
\end{minipage}\hfill
\hspace{0.02\textwidth}
\begin{minipage}{0.3\textwidth}
\begin{tabular}{lrrr}
H & 2.618 & 12.441 & 10.872 \\ 
H & 2.117 & 11.039 & 10.357 \\ 
O & 13.300 & 13.053 & 11.462 \\ 
H & 12.637 & 13.753 & 11.309 \\ 
H & 14.151 & 13.395 & 11.129 \\ 
O & 13.036 & 7.714 & 4.963 \\ 
H & 12.937 & 8.274 & 4.177 \\ 
H & 13.989 & 7.603 & 5.127 \\ 
O & 9.218 & 12.123 & 4.050 \\ 
H & 8.308 & 11.880 & 4.296 \\ 
H & 9.789 & 11.534 & 4.576 \\ 
O & 7.640 & 4.315 & 6.322 \\ 
H & 6.667 & 4.233 & 6.367 \\ 
H & 7.812 & 5.217 & 6.003 \\ 
O & 7.210 & 13.949 & 14.375 \\ 
H & 8.000 & 13.618 & 14.838 \\ 
H & 6.473 & 13.499 & 14.816 \\ 
O & 4.901 & 15.499 & 7.322 \\ 
H & 3.940 & 15.419 & 7.208 \\ 
H & 5.274 & 14.969 & 6.598 \\ 
O & 15.985 & 9.677 & 5.821 \\ 
H & 15.907 & 8.716 & 5.702 \\ 
H & 16.173 & 9.808 & 6.764 \\ 
O & 5.867 & 4.450 & 10.166 \\ 
H & 6.743 & 4.385 & 9.737 \\ 
H & 5.948 & 5.128 & 10.861 \\ 
O & 1.852 & 8.398 & 7.562 \\ 
H & 2.686 & 8.388 & 8.067 \\ 
H & 1.700 & 7.462 & 7.330 \\ 
O & 7.458 & 10.670 & 16.049 \\ 
H & 7.854 & 9.782 & 16.040 \\ 
H & 8.202 & 11.282 & 15.913 \\ 
O & 7.023 & 16.947 & 8.355 \\ 
H & 6.977 & 16.793 & 9.314 \\ 
H & 6.271 & 16.474 & 7.954 \\ 
O & 3.729 & 5.382 & 8.842 \\ 
H & 4.409 & 4.898 & 9.351 \\ 
H & 3.919 & 6.326 & 9.000 \\ 
O & 7.205 & 14.716 & 3.076 \\ 
H & 7.703 & 14.509 & 3.889 \\ 
H & 7.699 & 15.478 & 2.726 \\ 
O & 3.725 & 6.358 & 4.777 \\ 
H & 4.487 & 6.109 & 4.226 \\ 
H & 3.676 & 7.325 & 4.699 \\ 
O & 11.897 & 5.623 & 6.229 \\ 
H & 11.804 & 4.724 & 5.845 \\ 
H & 12.218 & 6.195 & 5.510 \\ 
O & 3.234 & 10.488 & 13.291 \\ 
H & 3.026 & 11.202 & 13.923 \\ 
H & 3.107 & 10.893 & 12.413 \\ 
O & 8.479 & 3.862 & 12.506 \\ 
H & 8.499 & 3.930 & 13.474 \\ 
H & 9.241 & 3.296 & 12.259 \\ 
O & 8.605 & 8.105 & 15.553 \\ 
H & 9.344 & 7.568 & 15.902 \\ 
H & 7.926 & 7.464 & 15.277 \\ 
O & 12.083 & 16.942 & 10.106 \\ 
\end{tabular}
\end{minipage}
\end{table}

\begin{table}[H]
\footnotesize
\setlength{\tabcolsep}{6pt} 
\renewcommand{\arraystretch}{0.95} 
\begin{minipage}{0.3\textwidth}
\begin{tabular}{lrrr}
H & 12.443 & 16.662 & 9.242 \\ 
H & 11.467 & 17.670 & 9.930 \\ 
O & 7.954 & 10.034 & 1.619 \\ 
H & 8.691 & 9.907 & 0.987 \\ 
H & 7.679 & 10.959 & 1.531 \\ 
O & 13.955 & 11.073 & 4.651 \\ 
H & 14.750 & 10.550 & 4.871 \\ 
H & 13.447 & 10.511 & 4.044 \\ 
O & 11.423 & 14.618 & 15.538 \\ 
H & 11.942 & 15.051 & 16.218 \\ 
H & 12.037 & 14.035 & 15.050 \\ 
O & 0.899 & 9.851 & 9.703 \\ 
H & 0.727 & 9.170 & 10.369 \\ 
H & 1.222 & 9.372 & 8.919 \\ 
O & 12.896 & 15.949 & 7.767 \\ 
H & 12.901 & 16.197 & 6.828 \\ 
H & 13.761 & 15.510 & 7.927 \\ 
O & 6.149 & 6.071 & 3.530 \\ 
H & 6.241 & 6.712 & 2.803 \\ 
H & 6.737 & 6.381 & 4.248 \\ 
O & 5.252 & 9.436 & 14.753 \\ 
H & 4.514 & 9.958 & 14.384 \\ 
H & 5.845 & 10.029 & 15.242 \\ 
O & 15.945 & 10.147 & 8.572 \\ 
H & 15.237 & 9.558 & 8.894 \\ 
H & 16.341 & 10.573 & 9.347 \\ 
O & 3.568 & 9.058 & 4.117 \\ 
H & 2.629 & 9.190 & 4.335 \\ 
H & 3.642 & 8.892 & 3.159 \\ 
O & 14.282 & 5.589 & 7.597 \\ 
H & 13.448 & 5.437 & 7.117 \\ 
H & 14.035 & 5.690 & 8.527 \\ 
O & 5.589 & 16.026 & 13.466 \\ 
H & 6.199 & 15.375 & 13.849 \\ 
H & 5.984 & 16.311 & 12.627 \\ 
O & 12.697 & 9.346 & 2.724 \\ 
H & 12.084 & 8.650 & 2.429 \\ 
H & 12.664 & 10.014 & 2.022 \\ 
O & 1.651 & 12.192 & 6.038 \\ 
H & 1.116 & 12.163 & 6.858 \\ 
H & 1.396 & 11.376 & 5.583 \\ 
O & 10.740 & 7.500 & 1.939 \\ 
H & 10.174 & 7.806 & 2.672 \\ 
H & 10.258 & 6.765 & 1.537 \\ 
O & 15.111 & 14.560 & 8.138 \\ 
H & 15.425 & 14.376 & 9.039 \\ 
H & 14.977 & 13.695 & 7.713 \\ 
O & 4.944 & 4.535 & 6.494 \\ 
H & 4.516 & 5.135 & 5.861 \\ 
H & 4.566 & 4.728 & 7.366 \\ 
O & 11.287 & 4.865 & 14.464 \\ 
H & 11.846 & 4.161 & 14.096 \\ 
H & 10.409 & 4.484 & 14.635 \\ 
O & 0.731 & 7.587 & 11.349 \\ 
H & 0.000 & 7.377 & 11.932 \\ 
H & 0.830 & 6.825 & 10.743 \\ 
O & 2.623 & 14.142 & 11.185 \\ 
\end{tabular}
\end{minipage}\hfill
\hspace{0.02\textwidth}
\begin{minipage}{0.3\textwidth}
\begin{tabular}{lrrr}
H & 2.652 & 14.439 & 12.108 \\ 
H & 3.429 & 14.508 & 10.765 \\ 
O & 9.028 & 18.157 & 11.962 \\ 
H & 9.589 & 17.611 & 12.538 \\ 
H & 9.570 & 18.448 & 11.214 \\ 
O & 6.843 & 6.220 & 14.614 \\ 
H & 6.561 & 6.246 & 13.679 \\ 
H & 6.034 & 6.425 & 15.124 \\ 
O & 10.052 & 16.800 & 5.333 \\ 
H & 11.017 & 16.685 & 5.376 \\ 
H & 9.782 & 17.367 & 6.074 \\ 
O & 0.080 & 12.030 & 8.168 \\ 
H & 0.183 & 12.886 & 8.626 \\ 
H & 0.244 & 11.325 & 8.814 \\ 
O & 10.982 & 7.025 & 16.213 \\ 
H & 11.157 & 6.247 & 15.657 \\ 
H & 11.814 & 7.524 & 16.262 \\ 
O & 8.638 & 4.273 & 15.197 \\ 
H & 7.997 & 5.000 & 15.072 \\ 
H & 8.239 & 3.658 & 15.815 \\ 
O & 11.515 & 14.674 & 2.561 \\ 
H & 11.843 & 14.243 & 3.365 \\ 
H & 11.044 & 13.985 & 2.064 \\ 
O & 13.107 & 16.350 & 12.558 \\ 
H & 12.903 & 16.634 & 11.649 \\ 
H & 12.246 & 16.313 & 13.003 \\ 
O & 1.088 & 5.534 & 9.654 \\ 
H & 0.699 & 5.451 & 8.776 \\ 
H & 2.037 & 5.369 & 9.522 \\ 
O & 15.436 & 7.265 & 13.044 \\ 
H & 14.516 & 7.524 & 13.230 \\ 
H & 15.907 & 8.108 & 12.965 \\ 
O & 13.098 & 8.878 & 15.851 \\ 
H & 13.114 & 8.617 & 14.909 \\ 
H & 13.728 & 9.607 & 15.914 \\ 
O & 1.122 & 9.597 & 5.247 \\ 
H & 1.361 & 9.211 & 6.114 \\ 
H & 0.207 & 9.360 & 5.092 \\ 
O & 15.874 & 10.042 & 12.897 \\ 
H & 15.044 & 10.085 & 12.380 \\ 
H & 15.640 & 10.382 & 13.775 \\ 
O & 5.595 & 12.630 & 16.206 \\ 
H & 5.558 & 12.978 & 17.099 \\ 
H & 6.218 & 11.882 & 16.227 \\ 
O & 8.926 & 18.420 & 7.290 \\ 
H & 8.411 & 19.052 & 6.785 \\ 
H & 8.266 & 17.804 & 7.677 \\ 
O & 11.717 & 2.374 & 8.077 \\ 
H & 11.012 & 1.865 & 8.509 \\ 
H & 11.569 & 3.289 & 8.361 \\ 
O & 15.162 & 4.859 & 11.896 \\ 
H & 15.298 & 5.697 & 12.387 \\ 
H & 16.013 & 4.635 & 11.514 \\ 
O & 10.083 & 12.602 & 1.545 \\ 
H & 9.218 & 12.774 & 1.145 \\ 
H & 9.872 & 12.372 & 2.470 \\ 
O & 6.686 & 7.653 & 1.336 \\ 
\end{tabular}
\end{minipage}\hfill
\hspace{0.02\textwidth}
\begin{minipage}{0.3\textwidth}
\begin{tabular}{lrrr}
H & 7.107 & 8.525 & 1.458 \\ 
H & 7.390 & 7.060 & 1.029 \\ 
O & 14.608 & 12.064 & 7.108 \\ 
H & 15.187 & 11.434 & 7.574 \\ 
H & 14.653 & 11.831 & 6.165 \\ 
O & 10.660 & 2.424 & 12.038 \\ 
H & 11.598 & 2.648 & 12.073 \\ 
H & 10.449 & 2.098 & 11.140 \\ 
O & 11.685 & 3.134 & 5.418 \\ 
H & 10.811 & 2.898 & 5.058 \\ 
H & 11.769 & 2.677 & 6.276 \\ 
O & 9.548 & 1.744 & 9.644 \\ 
H & 9.051 & 2.553 & 9.424 \\ 
H & 8.880 & 1.062 & 9.841 \\ 
O & 14.788 & 11.108 & 15.247 \\ 
H & 15.303 & 11.591 & 15.895 \\ 
H & 14.143 & 11.742 & 14.877 \\ 
O & 15.602 & 6.965 & 5.685 \\ 
H & 15.217 & 6.498 & 6.454 \\ 
H & 16.018 & 6.291 & 5.146 \\ 
O & 10.440 & 16.165 & 13.425 \\ 
H & 10.662 & 15.750 & 14.270 \\ 
H & 9.867 & 15.533 & 12.962 \\ 
O & 12.208 & 11.194 & 0.711 \\ 
H & 11.514 & 11.807 & 1.020 \\ 
H & 12.733 & 11.670 & 0.066 \\ 
O & 6.338 & 2.250 & 11.762 \\ 
H & 5.950 & 2.904 & 11.156 \\ 
H & 7.109 & 2.701 & 12.142 \\ 
O & 15.679 & 15.508 & 12.795 \\ 
H & 14.739 & 15.778 & 12.763 \\ 
H & 16.185 & 16.317 & 12.868 \\ 
O & 9.294 & 16.393 & 2.678 \\ 
H & 10.084 & 15.877 & 2.444 \\ 
H & 9.476 & 16.755 & 3.559 \\ 
O & 0.833 & 14.461 & 9.150 \\ 
H & 1.398 & 14.397 & 9.940 \\ 
H & 1.407 & 14.716 & 8.412 \\ 
O & 1.901 & 5.790 & 6.800 \\ 
H & 2.374 & 5.880 & 5.955 \\ 
H & 2.564 & 5.464 & 7.431 \\ 
O & 2.263 & 14.830 & 6.715 \\ 
H & 2.131 & 13.897 & 6.465 \\ 
H & 1.634 & 15.340 & 6.202 \\ 
O & 4.014 & 8.356 & 1.512 \\ 
H & 4.957 & 8.134 & 1.421 \\ 
H & 3.564 & 7.971 & 0.759 \\ 
O & 7.357 & 12.784 & 1.164 \\ 
H & 7.233 & 13.435 & 1.882 \\ 
H & 6.818 & 13.088 & 0.431 \\ 
O & 10.605 & 9.977 & 15.592 \\ 
H & 9.985 & 9.241 & 15.713 \\ 
H & 11.485 & 9.604 & 15.780 \\ 
O & 17.130 & 11.322 & 10.877 \\ 
H & 16.711 & 10.843 & 11.618 \\ 
H & 18.073 & 11.170 & 10.947 \\ 
O & 7.379 & 0.223 & 10.271 \\ 
\end{tabular}
\end{minipage}
\end{table}

\begin{table}[H]
\footnotesize
\setlength{\tabcolsep}{6pt} 
\renewcommand{\arraystretch}{0.95} 
\begin{minipage}{0.3\textwidth}
\begin{tabular}{lrrr}
H & 6.710 & 0.000 & 9.625 \\ 
H & 6.971 & 0.849 & 10.900 \\ 
O & 8.767 & 5.802 & 0.842 \\ 
H & 8.943 & 5.378 & 0.000 \\ 
H & 8.430 & 5.108 & 1.439 \\ 
O & 15.794 & 13.783 & 10.695 \\ 
H & 15.914 & 14.372 & 11.464 \\ 
H & 16.356 & 13.002 & 10.816 \\ 
O & 10.135 & 9.453 & 0.108 \\ 
H & 10.849 & 10.101 & 0.216 \\ 
H & 10.401 & 8.671 & 0.623 \\ 
O & 7.510 & 3.991 & 2.432 \\ 
H & 6.980 & 4.643 & 2.930 \\ 
H & 6.978 & 3.195 & 2.376 \\ 
O & 3.054 & 15.147 & 13.724 \\ 
H & 2.525 & 15.891 & 14.015 \\ 
H & 3.970 & 15.479 & 13.629 \\ 
O & 10.307 & 18.977 & 9.584 \\ 
H & 10.468 & 19.920 & 9.545 \\ 
H & 9.807 & 18.740 & 8.783 \\ 
O & 3.137 & 12.651 & 14.902 \\ 
H & 3.974 & 12.686 & 15.389 \\ 
H & 3.027 & 13.514 & 14.469 \\ 
O & 9.082 & 2.835 & 4.531 \\ 
H & 8.603 & 3.235 & 5.279 \\ 
H & 8.741 & 3.255 & 3.733 \\ 
O & 13.182 & 3.300 & 13.080 \\ 
H & 13.643 & 2.545 & 13.449 \\ 
H & 13.846 & 3.849 & 12.633 \\ 
O & 4.529 & 7.004 & 15.737 \\ 
H & 3.674 & 6.771 & 15.375 \\ 
H & 4.717 & 7.918 & 15.455 \\ 
O & 12.743 & 16.203 & 5.054 \\ 
H & 13.102 & 16.569 & 4.244 \\ 
H & 12.688 & 15.239 & 4.912 \\ 
\multicolumn{4}{l}{450}\\ 
\multicolumn{4}{l}{w150 optmized with ChargeNN}\\ 
O & 11.458 & 10.624 & 11.662 \\ 
H & 10.990 & 9.783 & 11.506 \\ 
H & 10.767 & 11.304 & 11.787 \\ 
O & 9.640 & 12.605 & 11.897 \\ 
H & 9.759 & 12.993 & 11.009 \\ 
H & 8.719 & 12.282 & 11.951 \\ 
O & 9.702 & 11.277 & 8.280 \\ 
H & 9.919 & 12.094 & 8.771 \\ 
H & 9.228 & 11.568 & 7.482 \\ 
O & 9.976 & 8.310 & 11.231 \\ 
H & 9.193 & 8.749 & 10.853 \\ 
H & 10.472 & 7.967 & 10.463 \\ 
O & 7.990 & 9.846 & 10.051 \\ 
H & 8.460 & 10.323 & 9.349 \\ 
H & 7.715 & 10.510 & 10.712 \\ 
O & 10.734 & 8.738 & 8.610 \\ 
H & 11.704 & 8.817 & 8.663 \\ 
H & 10.386 & 9.638 & 8.482 \\ 
O & 10.211 & 13.633 & 9.476 \\ 
H & 9.669 & 14.444 & 9.475 \\ 
\end{tabular}
\end{minipage}\hfill
\hspace{0.02\textwidth}
\begin{minipage}{0.3\textwidth}
\begin{tabular}{lrrr}
H & 11.107 & 13.937 & 9.242 \\ 
O & 13.471 & 12.099 & 10.616 \\ 
H & 13.881 & 12.486 & 11.410 \\ 
H & 12.742 & 11.540 & 10.950 \\ 
O & 7.197 & 11.519 & 12.033 \\ 
H & 7.344 & 11.022 & 12.863 \\ 
H & 6.254 & 11.756 & 12.041 \\ 
O & 13.408 & 9.129 & 8.678 \\ 
H & 13.500 & 10.067 & 8.427 \\ 
H & 13.755 & 8.616 & 7.926 \\ 
O & 11.715 & 11.280 & 6.190 \\ 
H & 12.472 & 11.461 & 6.775 \\ 
H & 10.917 & 11.533 & 6.690 \\ 
O & 13.760 & 11.700 & 7.957 \\ 
H & 14.706 & 11.538 & 7.786 \\ 
H & 13.680 & 11.964 & 8.896 \\ 
O & 7.562 & 10.066 & 14.256 \\ 
H & 7.032 & 10.209 & 15.062 \\ 
H & 8.485 & 9.955 & 14.560 \\ 
O & 12.081 & 10.607 & 14.335 \\ 
H & 11.959 & 10.628 & 13.370 \\ 
H & 11.308 & 10.123 & 14.679 \\ 
O & 9.996 & 8.948 & 14.929 \\ 
H & 10.035 & 8.176 & 14.342 \\ 
H & 9.932 & 8.608 & 15.841 \\ 
O & 10.195 & 14.741 & 13.431 \\ 
H & 9.872 & 14.652 & 14.344 \\ 
H & 9.984 & 13.901 & 12.978 \\ 
O & 12.653 & 15.071 & 12.270 \\ 
H & 11.772 & 14.941 & 12.667 \\ 
H & 12.761 & 16.031 & 12.158 \\ 
O & 10.217 & 6.692 & 13.349 \\ 
H & 10.194 & 7.095 & 12.455 \\ 
H & 10.297 & 5.723 & 13.239 \\ 
O & 12.884 & 14.420 & 9.309 \\ 
H & 13.078 & 13.691 & 9.924 \\ 
H & 13.154 & 15.235 & 9.771 \\ 
O & 5.793 & 8.233 & 10.353 \\ 
H & 5.947 & 8.285 & 11.317 \\ 
H & 6.538 & 8.726 & 9.969 \\ 
O & 8.957 & 12.318 & 5.948 \\ 
H & 8.369 & 11.930 & 5.276 \\ 
H & 9.198 & 13.199 & 5.599 \\ 
O & 14.677 & 8.947 & 11.073 \\ 
H & 14.098 & 9.043 & 10.294 \\ 
H & 14.516 & 8.050 & 11.415 \\ 
O & 9.939 & 7.126 & 6.640 \\ 
H & 10.156 & 7.597 & 7.469 \\ 
H & 10.348 & 7.669 & 5.948 \\ 
O & 10.673 & 9.125 & 4.903 \\ 
H & 9.984 & 9.587 & 4.394 \\ 
H & 11.101 & 9.816 & 5.441 \\ 
O & 6.184 & 8.175 & 13.007 \\ 
H & 5.291 & 8.149 & 13.401 \\ 
H & 6.663 & 8.914 & 13.435 \\ 
O & 12.754 & 7.016 & 14.258 \\ 
H & 11.816 & 7.043 & 14.001 \\ 
\end{tabular}
\end{minipage}\hfill
\hspace{0.02\textwidth}
\begin{minipage}{0.3\textwidth}
\begin{tabular}{lrrr}
H & 12.816 & 6.372 & 14.990 \\ 
O & 13.843 & 14.561 & 6.813 \\ 
H & 14.447 & 13.843 & 6.564 \\ 
H & 13.598 & 14.371 & 7.740 \\ 
O & 11.814 & 6.192 & 10.401 \\ 
H & 12.174 & 6.185 & 9.496 \\ 
H & 12.570 & 6.106 & 10.997 \\ 
O & 15.177 & 10.541 & 13.208 \\ 
H & 14.904 & 11.469 & 13.123 \\ 
H & 14.846 & 10.069 & 12.422 \\ 
O & 8.488 & 16.581 & 12.491 \\ 
H & 8.822 & 17.491 & 12.600 \\ 
H & 9.181 & 15.979 & 12.834 \\ 
O & 9.444 & 4.735 & 10.297 \\ 
H & 9.192 & 4.749 & 9.351 \\ 
H & 10.299 & 5.194 & 10.358 \\ 
O & 7.668 & 6.176 & 14.059 \\ 
H & 7.087 & 6.915 & 13.797 \\ 
H & 8.578 & 6.446 & 13.834 \\ 
O & 11.031 & 14.989 & 6.941 \\ 
H & 11.959 & 14.719 & 7.008 \\ 
H & 11.043 & 15.962 & 6.876 \\ 
O & 6.024 & 15.431 & 11.889 \\ 
H & 6.078 & 15.360 & 10.923 \\ 
H & 6.866 & 15.805 & 12.193 \\ 
O & 14.324 & 8.967 & 15.348 \\ 
H & 14.385 & 9.684 & 14.692 \\ 
H & 13.687 & 8.332 & 14.976 \\ 
O & 13.334 & 12.718 & 15.354 \\ 
H & 13.784 & 13.029 & 14.553 \\ 
H & 12.853 & 11.911 & 15.070 \\ 
O & 5.508 & 11.198 & 7.450 \\ 
H & 5.280 & 10.739 & 6.627 \\ 
H & 4.701 & 11.629 & 7.781 \\ 
O & 6.979 & 13.390 & 7.541 \\ 
H & 6.417 & 12.590 & 7.612 \\ 
H & 7.818 & 13.076 & 7.164 \\ 
O & 14.224 & 6.431 & 12.097 \\ 
H & 15.022 & 6.125 & 12.565 \\ 
H & 13.617 & 6.674 & 12.823 \\ 
O & 9.276 & 11.880 & 15.908 \\ 
H & 10.021 & 11.711 & 16.514 \\ 
H & 9.269 & 12.835 & 15.736 \\ 
O & 4.459 & 11.708 & 12.270 \\ 
H & 3.972 & 12.426 & 11.832 \\ 
H & 3.942 & 10.903 & 12.095 \\ 
O & 14.474 & 13.222 & 12.887 \\ 
H & 13.860 & 13.964 & 12.704 \\ 
H & 15.378 & 13.585 & 12.812 \\ 
O & 14.434 & 7.911 & 6.474 \\ 
H & 14.376 & 8.165 & 5.539 \\ 
H & 15.383 & 7.974 & 6.689 \\ 
O & 11.465 & 11.926 & 3.541 \\ 
H & 10.520 & 11.722 & 3.429 \\ 
H & 11.639 & 11.762 & 4.486 \\ 
O & 8.632 & 4.898 & 7.722 \\ 
H & 7.669 & 5.061 & 7.779 \\ 
\end{tabular}
\end{minipage}
\end{table}

\begin{table}[H]
\footnotesize
\setlength{\tabcolsep}{6pt} 
\renewcommand{\arraystretch}{0.95} 
\begin{minipage}{0.3\textwidth}
\begin{tabular}{lrrr}
H & 9.044 & 5.690 & 7.336 \\ 
O & 9.013 & 14.657 & 15.912 \\ 
H & 9.774 & 14.639 & 16.516 \\ 
H & 8.206 & 14.490 & 16.441 \\ 
O & 6.205 & 15.406 & 9.130 \\ 
H & 5.265 & 15.542 & 8.910 \\ 
H & 6.470 & 14.624 & 8.608 \\ 
O & 16.326 & 10.806 & 7.597 \\ 
H & 16.556 & 9.959 & 7.173 \\ 
H & 16.596 & 10.673 & 8.522 \\ 
O & 7.212 & 4.723 & 11.754 \\ 
H & 8.021 & 4.738 & 11.200 \\ 
H & 7.384 & 5.303 & 12.520 \\ 
O & 3.296 & 8.552 & 9.311 \\ 
H & 4.256 & 8.491 & 9.475 \\ 
H & 2.979 & 7.631 & 9.235 \\ 
O & 7.447 & 11.665 & 17.867 \\ 
H & 7.874 & 11.081 & 18.516 \\ 
H & 8.094 & 11.718 & 17.135 \\ 
O & 8.762 & 15.889 & 9.878 \\ 
H & 8.698 & 16.137 & 10.818 \\ 
H & 7.849 & 15.718 & 9.585 \\ 
O & 4.972 & 5.658 & 10.652 \\ 
H & 5.731 & 5.153 & 11.003 \\ 
H & 5.272 & 6.587 & 10.636 \\ 
O & 9.241 & 14.803 & 4.974 \\ 
H & 9.791 & 14.931 & 5.772 \\ 
H & 9.686 & 15.381 & 4.327 \\ 
O & 5.318 & 7.582 & 6.368 \\ 
H & 6.227 & 7.505 & 6.011 \\ 
H & 5.053 & 8.490 & 6.161 \\ 
O & 13.125 & 6.012 & 7.986 \\ 
H & 12.968 & 5.154 & 7.543 \\ 
H & 13.448 & 6.623 & 7.305 \\ 
O & 4.136 & 11.827 & 15.014 \\ 
H & 4.118 & 12.759 & 15.301 \\ 
H & 4.296 & 11.847 & 14.053 \\ 
O & 10.312 & 3.981 & 13.705 \\ 
H & 10.470 & 3.919 & 14.661 \\ 
H & 11.117 & 3.626 & 13.275 \\ 
O & 9.793 & 7.772 & 17.362 \\ 
H & 10.679 & 7.529 & 17.707 \\ 
H & 9.309 & 6.945 & 17.210 \\ 
O & 12.862 & 17.821 & 12.077 \\ 
H & 13.449 & 17.573 & 11.339 \\ 
H & 12.148 & 18.320 & 11.634 \\ 
O & 9.030 & 10.718 & 3.309 \\ 
H & 8.974 & 10.290 & 2.439 \\ 
H & 8.144 & 11.073 & 3.507 \\ 
O & 15.819 & 12.789 & 5.881 \\ 
H & 16.081 & 12.054 & 6.467 \\ 
H & 15.640 & 12.387 & 5.018 \\ 
O & 11.511 & 13.959 & 16.918 \\ 
H & 11.342 & 13.272 & 17.586 \\ 
H & 12.147 & 13.532 & 16.309 \\ 
O & 2.781 & 9.552 & 11.777 \\ 
H & 3.044 & 8.970 & 12.512 \\ 
\end{tabular}
\end{minipage}\hfill
\hspace{0.02\textwidth}
\begin{minipage}{0.3\textwidth}
\begin{tabular}{lrrr}
H & 3.044 & 9.104 & 10.949 \\ 
O & 14.145 & 16.835 & 9.937 \\ 
H & 14.060 & 17.161 & 9.026 \\ 
H & 15.104 & 16.797 & 10.127 \\ 
O & 7.696 & 7.277 & 5.206 \\ 
H & 7.999 & 7.503 & 4.310 \\ 
H & 8.495 & 7.204 & 5.762 \\ 
O & 5.809 & 10.216 & 16.345 \\ 
H & 5.166 & 10.827 & 15.937 \\ 
H & 6.315 & 10.733 & 17.001 \\ 
O & 17.024 & 9.796 & 10.003 \\ 
H & 16.195 & 9.658 & 10.499 \\ 
H & 17.304 & 8.898 & 9.745 \\ 
O & 4.901 & 9.999 & 4.994 \\ 
H & 3.935 & 10.138 & 5.013 \\ 
H & 5.089 & 9.437 & 4.216 \\ 
O & 15.249 & 5.834 & 9.611 \\ 
H & 14.496 & 5.989 & 9.008 \\ 
H & 14.886 & 5.910 & 10.509 \\ 
O & 7.501 & 16.881 & 15.070 \\ 
H & 7.931 & 16.120 & 15.495 \\ 
H & 7.802 & 16.874 & 14.147 \\ 
O & 14.800 & 8.744 & 3.864 \\ 
H & 13.940 & 8.463 & 3.499 \\ 
H & 14.878 & 9.701 & 3.725 \\ 
O & 3.418 & 12.756 & 8.372 \\ 
H & 3.314 & 12.903 & 9.330 \\ 
H & 2.776 & 12.060 & 8.152 \\ 
O & 12.306 & 7.907 & 3.160 \\ 
H & 11.744 & 8.303 & 3.855 \\ 
H & 11.867 & 7.074 & 2.929 \\ 
O & 16.767 & 16.694 & 10.589 \\ 
H & 16.887 & 16.749 & 11.550 \\ 
H & 17.174 & 15.862 & 10.294 \\ 
O & 5.954 & 5.451 & 8.010 \\ 
H & 5.661 & 6.231 & 7.510 \\ 
H & 5.583 & 5.518 & 8.903 \\ 
O & 12.939 & 5.178 & 16.224 \\ 
H & 13.602 & 4.627 & 15.778 \\ 
H & 12.130 & 4.646 & 16.315 \\ 
O & 3.649 & 8.004 & 13.903 \\ 
H & 2.994 & 8.440 & 14.480 \\ 
H & 3.402 & 7.064 & 13.771 \\ 
O & 3.735 & 16.618 & 12.674 \\ 
H & 3.954 & 17.004 & 13.536 \\ 
H & 4.563 & 16.188 & 12.378 \\ 
O & 11.420 & 18.195 & 14.442 \\ 
H & 11.724 & 17.570 & 15.118 \\ 
H & 12.015 & 18.095 & 13.679 \\ 
O & 8.316 & 5.530 & 16.644 \\ 
H & 7.998 & 5.710 & 15.737 \\ 
H & 7.561 & 5.709 & 17.231 \\ 
O & 11.220 & 17.679 & 6.564 \\ 
H & 12.119 & 17.776 & 6.913 \\ 
H & 10.602 & 17.995 & 7.251 \\ 
O & 2.605 & 13.256 & 10.979 \\ 
H & 2.366 & 14.186 & 10.775 \\ 
\end{tabular}
\end{minipage}\hfill
\hspace{0.02\textwidth}
\begin{minipage}{0.3\textwidth}
\begin{tabular}{lrrr}
H & 1.801 & 12.716 & 10.892 \\ 
O & 12.271 & 7.195 & 18.168 \\ 
H & 12.718 & 6.577 & 17.570 \\ 
H & 12.706 & 8.063 & 18.088 \\ 
O & 10.481 & 3.915 & 16.461 \\ 
H & 9.731 & 4.529 & 16.581 \\ 
H & 10.259 & 3.121 & 16.948 \\ 
O & 13.550 & 15.489 & 4.202 \\ 
H & 13.615 & 15.153 & 5.110 \\ 
H & 13.596 & 14.723 & 3.611 \\ 
O & 14.476 & 17.107 & 14.251 \\ 
H & 13.921 & 17.414 & 13.514 \\ 
H & 13.875 & 16.824 & 14.958 \\ 
O & 3.707 & 5.395 & 13.232 \\ 
H & 4.241 & 4.774 & 13.759 \\ 
H & 4.074 & 5.385 & 12.333 \\ 
O & 16.761 & 7.759 & 15.294 \\ 
H & 15.889 & 8.149 & 15.498 \\ 
H & 17.228 & 8.427 & 14.765 \\ 
O & 13.171 & 9.793 & 17.680 \\ 
H & 13.469 & 9.532 & 16.788 \\ 
H & 13.786 & 10.501 & 17.938 \\ 
O & 1.832 & 10.530 & 8.208 \\ 
H & 2.380 & 9.784 & 8.524 \\ 
H & 1.289 & 10.778 & 8.972 \\ 
O & 17.687 & 9.517 & 13.393 \\ 
H & 16.773 & 9.851 & 13.310 \\ 
H & 18.238 & 10.321 & 13.394 \\ 
O & 6.653 & 14.185 & 17.116 \\ 
H & 5.778 & 14.207 & 16.696 \\ 
H & 6.764 & 13.304 & 17.515 \\ 
O & 9.435 & 18.175 & 8.548 \\ 
H & 8.561 & 18.359 & 8.164 \\ 
H & 9.325 & 17.308 & 8.977 \\ 
O & 11.896 & 1.924 & 8.627 \\ 
H & 10.995 & 1.882 & 9.005 \\ 
H & 12.494 & 1.935 & 9.389 \\ 
O & 16.452 & 5.794 & 13.542 \\ 
H & 16.523 & 6.440 & 14.280 \\ 
H & 17.142 & 6.073 & 12.921 \\ 
O & 13.499 & 13.269 & 2.490 \\ 
H & 13.241 & 13.623 & 1.637 \\ 
H & 12.699 & 12.851 & 2.865 \\ 
O & 8.459 & 7.520 & 2.584 \\ 
H & 8.764 & 8.192 & 1.955 \\ 
H & 9.140 & 6.829 & 2.597 \\ 
O & 17.479 & 14.565 & 7.177 \\ 
H & 17.743 & 15.248 & 6.561 \\ 
H & 16.894 & 13.955 & 6.696 \\ 
O & 12.765 & 3.215 & 12.857 \\ 
H & 13.530 & 3.423 & 13.413 \\ 
H & 13.099 & 2.795 & 12.046 \\ 
O & 12.890 & 3.606 & 6.838 \\ 
H & 12.360 & 3.662 & 6.031 \\ 
H & 12.433 & 2.963 & 7.425 \\ 
O & 9.666 & 1.940 & 10.109 \\ 
H & 9.522 & 2.882 & 10.293 \\ 
\end{tabular}
\end{minipage}
\end{table}

\begin{table}[H]
\footnotesize
\setlength{\tabcolsep}{6pt} 
\renewcommand{\arraystretch}{0.95} 
\begin{minipage}{0.3\textwidth}
\begin{tabular}{lrrr}
H & 9.477 & 1.442 & 10.921 \\ 
O & 14.853 & 11.963 & 17.536 \\ 
H & 15.735 & 11.781 & 17.211 \\ 
H & 14.340 & 12.326 & 16.792 \\ 
O & 17.082 & 8.293 & 6.955 \\ 
H & 17.420 & 7.739 & 7.682 \\ 
H & 17.340 & 7.885 & 6.108 \\ 
O & 12.591 & 16.434 & 16.258 \\ 
H & 12.209 & 15.548 & 16.390 \\ 
H & 12.790 & 16.769 & 17.134 \\ 
O & 15.325 & 11.502 & 3.443 \\ 
H & 14.626 & 12.095 & 3.114 \\ 
H & 16.060 & 11.585 & 2.833 \\ 
O & 7.803 & 2.802 & 13.516 \\ 
H & 7.505 & 3.357 & 12.769 \\ 
H & 8.670 & 3.177 & 13.751 \\ 
O & 16.965 & 16.656 & 13.361 \\ 
H & 16.059 & 16.817 & 13.696 \\ 
H & 17.539 & 17.249 & 13.847 \\ 
O & 10.969 & 16.616 & 3.960 \\ 
H & 11.878 & 16.280 & 3.895 \\ 
H & 10.956 & 17.179 & 4.752 \\ 
O & 2.218 & 15.839 & 10.467 \\ 
H & 2.699 & 16.298 & 11.180 \\ 
H & 2.709 & 15.979 & 9.641 \\ 
O & 2.643 & 5.915 & 9.292 \\ 
H & 2.440 & 5.154 & 8.750 \\ 
H & 3.409 & 5.674 & 9.845 \\ 
O & 3.707 & 15.523 & 8.085 \\ 
H & 3.558 & 14.570 & 7.952 \\ 
H & 3.967 & 15.908 & 7.232 \\ 
O & 5.585 & 8.231 & 2.973 \\ 
H & 6.507 & 8.013 & 2.772 \\ 
H & 5.169 & 7.407 & 3.277 \\ 
O & 6.798 & 11.852 & 4.224 \\ 
H & 6.512 & 12.768 & 4.388 \\ 
H & 6.087 & 11.261 & 4.533 \\ 
O & 10.992 & 11.474 & 18.035 \\ 
H & 10.334 & 10.963 & 18.540 \\ 
H & 11.749 & 10.871 & 17.906 \\ 
O & 18.374 & 7.238 & 12.124 \\ 
H & 19.265 & 7.072 & 12.437 \\ 
H & 18.099 & 8.081 & 12.541 \\ 
O & 8.785 & 0.507 & 12.350 \\ 
H & 8.050 & 0.000 & 12.005 \\ 
H & 8.418 & 1.258 & 12.846 \\ 
O & 10.666 & 5.697 & 2.727 \\ 
H & 10.598 & 5.132 & 1.956 \\ 
H & 10.824 & 5.093 & 3.477 \\ 
O & 17.014 & 14.058 & 12.554 \\ 
H & 17.089 & 14.958 & 12.917 \\ 
H & 17.230 & 14.124 & 11.605 \\ 
O & 11.977 & 10.067 & 1.531 \\ 
H & 11.881 & 10.770 & 2.194 \\ 
H & 12.211 & 9.257 & 2.020 \\ 
O & 6.952 & 4.792 & 4.195 \\ 
H & 7.274 & 5.641 & 4.540 \\ 
\end{tabular}%
\end{minipage}
\end{table}

\bibliography{manuscript}